\pdfoutput=1
\documentclass[letterpaper,twocolumn,10pt]{article}
\usepackage{usenix2019_v3}
\usepackage{tikz}
\usepackage{amsmath}
\usepackage{graphicx}
\usepackage{subfigure} 
\usepackage{caption}
\usepackage{dsfont}
\usepackage{multirow}
\usepackage{lipsum}
\usepackage[inline, shortlabels]{enumitem}

\usepackage{xspace}
\usepackage{filecontents}
\usepackage{bbm}
\usepackage[linesnumbered,ruled]{algorithm2e}
\usepackage{epstopdf}
\usepackage{array}

\usepackage[numbers,sort&compress]{natbib}
\usepackage{pifont}

\usepackage{amsthm}

\newcommand{\CR}[1]{{\color{black}{#1}}}

\newcommand{\cut}[1]{}

\newcommand{\etc}{\textit{etc.}\xspace}
\newcommand{\ie}{\textit{i.e.,}\xspace}
\newcommand{\eg}{\textit{e.g.,}\xspace}
\newcommand{\etal}{\textit{et al.}\xspace}

\iffalse

\newcommand{\nsection}[1]{\section{#1}}
\newcommand{\nsubsection}[1]{\subsection{#1}}
\newcommand{\nsubsubsection}[1]{\subsubsection{#1}}

\else

\newcommand{\nsection}[1]{\vspace{-0.40cm}\section{#1}\vspace{-0.3cm}}
\newcommand{\nsubsection}[1]{\vspace{-0.4cm}\subsection{#1}\vspace{-0.15cm}}
\newcommand{\nsubsubsection}[1]{\vspace{-0.3cm}\subsubsection{#1}\vspace{-0.15cm}}

\fi

\usepackage{amsmath,amsfonts,bm}

\def\eqref#1{equation~\ref{#1}}

\def\1{\bm{1}}

\DeclareMathAlphabet{\mathsfit}{\encodingdefault}{\sfdefault}{m}{sl}
\SetMathAlphabet{\mathsfit}{bold}{\encodingdefault}{\sfdefault}{bx}{n}

\def\gL{{\mathcal{L}}}

\def\gV{{\mathcal{V}}}

\newcommand{\advdelta}{T'}

\newcommand{\wx}{\mathbf{w_x}}
\newcommand{\wy}{\mathbf{w_y}}
\newcommand{\wz}{\mathbf{w_z}} 
\newcommand{\wi}{\mathbf{w_i}} 

\newcommand{\advdeltad}{t'}

\newcommand{\V}[1]{\mathbf{#1}}

\hypersetup{                    
 colorlinks,
  linkcolor={green!80!black},
  citecolor={red!70!black},
  urlcolor={blue!70!black}
}

\begin{document}
\date{}
%

\title{\Large \bf Towards Robust LiDAR-based Perception in Autonomous Driving: General Black-box Adversarial Sensor Attack and Countermeasures}




\author{
{\rm Jiachen Sun}\\
University of Michigan\\
\href{mailto:jiachens@umich.edu}{jiachens@umich.edu}
\and
{\rm Yulong Cao}\\
University of Michigan\\
\href{mailto:yulongc@umich.edu}{yulongc@umich.edu}
\and
{\rm Qi Alfred Chen} \\
UC Irvine\\
\href{mailto:alfchen@uci.edu}{alfchen@uci.edu}
\and
{\rm Z. Morley Mao}\\
University of Michigan\\
\href{mailto:zmao@umich.edu}{zmao@umich.edu} 
}


\maketitle

\begin{abstract}
\emph{Perception} plays a pivotal role in autonomous driving systems, which utilizes onboard sensors like cameras and LiDARs (Light Detection and Ranging) to assess surroundings. Recent studies have demonstrated that LiDAR-based perception is vulnerable to spoofing attacks, in which adversaries spoof a fake vehicle in front of a victim self-driving car by strategically transmitting laser signals to the victim's LiDAR sensor. However, existing attacks suffer from effectiveness and generality limitations. In this work, we perform the first study to explore the general vulnerability of current LiDAR-based perception architectures and discover that the ignored occlusion patterns in LiDAR point clouds make self-driving cars vulnerable to spoofing attacks. We construct the first black-box spoofing attack based on our identified vulnerability, which universally achieves around 80\% mean success rates on all target models. We perform the first defense study, proposing CARLO to mitigate LiDAR spoofing attacks. CARLO detects spoofed data by treating ignored occlusion patterns as invariant physical features, which reduces the mean attack success rate to 5.5\%. Meanwhile, we take the first step towards exploring a general architecture for robust LiDAR-based perception, and propose SVF that embeds the neglected physical features into end-to-end learning. SVF further reduces the mean attack success rate to around 2.3\%.\footnote{Project website: \url{https://sites.google.com/view/cav-sec/adv-lidar-defense}}  



\end{abstract}

\nsection{Introduction}
\label{introduction}


Today, self-driving cars, or autonomous vehicles (AV), are undergoing rapid development, and some are already operating on public roads, \eg self-driving taxis from Google's Waymo One~\cite{waymo-one} and Baidu's Apollo Go~\cite{baidu-apollo-go}, and self-driving trucks from TuSimple~\cite{tusimple-truck}  used by UPS. To enable self-driving, AVs rely on autonomous driving (AD) software, in which \textit{perception} is a fundamental pillar that detects surrounding obstacles using sensors like cameras and LiDARs (Light Detection and Ranging). Since perception directly impacts safety-critical driving decisions such as collision avoidance, it is imperative to ensure its security under potential attacks. 


In AD perception, 3D object detection is indispensable for ensuring safe and correct autonomous driving. To achieve this, almost all AV makers~\cite{apollo,waymo_lidar,gm_lidar} adopt LiDAR sensors, since they capture high-resolution 360$^\circ$ 3D information called point clouds and are more reliable in challenging weather and lighting conditions than other sensors such as cameras. 
Due to such heavy reliance on LiDAR, a few prior studies have explored the security of LiDAR and its usage in AD systems~\cite{petit2015remote,shin2017illusion, cao2019adversarial}. Among them, Cao \etal are the first to discover that the deep learning model for LiDAR-based perception used in a real-world AD system can be fooled to detect a fake vehicle by strategically injecting a small number of spoofed LiDAR points~\cite{cao2019adversarial}. \CR{Such LiDAR spoofing attacks could lead to severe safety consequences (\eg emergency brake operations that may injure passengers).} However, the attack proposed was evaluated on only one specific model (\ie~Baidu Apollo 2.5), assuming white-box access, which may be unrealistic. Moreover, it is unclear 1) whether the attack generalizes to other machine learning models, and 2) how to mitigate such spoofing attacks.

In this work, we perform the first study to systematically explore, discover, and defend against a \textit{general} vulnerability existing among three state-of-the-art LiDAR-based 3D object detection model designs: bird’s-eye view-based, voxel-based, and point-wise (introduced in~\S\ref{background}).
More specifically, we first demonstrate that existing LiDAR spoofing attacks~\cite{shin2017illusion,cao2019adversarial} cannot directly generalize to all three model designs (\S\ref{limitations_existing}). Meanwhile, we find that in these prior works the required sensor attack capabilities to succeed in fooling AD perception are quite intriguing: Cao~\etal~\cite{cao2019adversarial} found that an attack trace with merely \textit{60 points} is sufficient to spoof a front-near vehicle in Apollo 2.5, while a valid one should have \textit{$\sim$2000 points}~\cite{Geiger2013IJRR}, which is almost two magnitudes more. Thus, there must exist certain LiDAR-related physical invariants that are not correctly learned in the model, which could also be generalizable to other state-of-the-art 3D object detection model designs.


To explore the cause, we perform experiments based on hypotheses formed by empirical observations of deep learning models and unique physical features of LiDAR, and discover that \textit{all} the three state-of-the-art 3D object detection model designs above generally ignore the \textit{occlusion patterns} in LiDAR point clouds, a set of physical invariants for LiDAR~(\S\ref{hypothesis_validation}). For example, when a vehicle is behind another vehicle, its point cloud can legitimately have much fewer points due to the front vehicle's occlusion of the LiDAR beams. \textit{However, such point cloud with much fewer points should not be detected as a vehicle at front-near locations with no occlusions, due to the physical law.} Unfortunately, all three model designs today fail to differentiate these two cases. This allows an adversary to spoof almost two magnitudes fewer points into the victim's LiDAR but can still fool the perception model into detecting a fake front-near vehicle (\S\ref{vulnerability_id}). 




Based on this general vulnerability, we design the first black-box (i.e., without any knowledge about the models) adversarial sensor attack on LiDAR-based perception models to spoof a front-near vehicle to a victim AV that can alter its driving decisions (\S\ref{bb_attack}). To realize this, we enumerate different occlusion patterns of a 3D vehicle mesh (\eg different occluded postures) to fit the sensor attack capability, and leverage ray-casting techniques~\cite{cao2019adversarial2} to render the attack traces. We perform large-scale experiments on the three target model designs with around 15,000 point cloud samples from the KITTI~\cite{Geiger2013IJRR} dataset. Evaluations show that with the same sensor attack capability as prior works~\cite{cao2019adversarial} (\ie 60 spoofed points), adversaries can generally achieve over 80\% success rates on all three model designs.



Since these spoofed point clouds directly violate the physical laws of the LiDAR occlusion patterns mentioned above, we then leverage them as physical invariants to defend against this class of LiDAR spoofing attacks. First, we design a model-agnostic defense solution, CARLO: o\underline{C}clusion-\underline{A}ware hie\underline{R}archy anoma\underline{L}y detecti\underline{O}n, which can be applied to LiDAR-based perception immediately without changing the existing models. CARLO exploits two occlusion-related characteristics: 1) the free space inside a detected bounding box, and 2) the locations of points inside the frustum corresponding to a detected bounding box. Large-scale evaluations of CARLO show that it can efficiently and effectively defend both white- and black-box LiDAR spoofing attacks~\cite{cao2019adversarial}. 
CARLO is also found to have high resilience to adaptive attacks since it exploits physical invariants that are highly difficult, if not impossible, for attackers to break.

While the model-agnostic defense is already useful, it is also beneficial if we can improve the robustness of the model designs themselves. Thus, we further design a general architecture for robust LiDAR-based perception in AVs. We observe that LiDAR measures range data by nature; hence the front view (FV) of the LiDAR sensor preserves the physical features as well as the occlusion information~\cite{li2016vehicle}. Recent studies present view fusion-based models that combines the FV and 3D representations~\cite{chen2017multi,zhou2019end,ku2018joint}. However, our experiment results show that current designs are still vulnerable to LiDAR spoofing attacks since features from the 3D representation dominate the fusion process.
To address such limitations, we propose sequential view fusion (SVF), a novel view fusion-based model design that sequentially fuses the FV and 3D representations to ensure that the end-to-end learning makes sufficient use of the features from FV (\S\ref{svf}). Evaluations show that SVF can effectively reduce the attack success rate to 2.3\% without sacrificing the original performance, which is a 2.2$\times$ improvement compared to CARLO.
We find that SVF is also resilient to white-box attacks and adaptive attacks.







Overall, our key contributions are summarized as follows:
\vspace{-\topsep}
\begin{itemize}
\setlength{\itemsep}{0pt}
\setlength{\parskip}{0pt}
\item We perform the first study to explore the general vulnerability of current LiDAR-based perception architectures. We discover that current LiDAR-based perception models do not learn occlusion information in the LiDAR point clouds, which enables a class of spoofing attacks. We construct the first black-box spoofing attack based on this vulnerability. Large-scale evaluations show that attackers can achieve around 80\% mean success rates on all target models.
\item To defend against LiDAR spoofing attacks, we design a model-agnostic defense CARLO that leverages the ignored occlusion patterns as invariant physical features to detect spoofed fake vehicles. We also perform large-scale evaluations on CARLO, and demonstrate that CARLO can effectively reduce the mean attack success rate to 5.5\% on all target models without sacrificing the original performance. 
\item We design a general architecture for robust LiDAR-based perception in AVs by embedding the front view (FV) representation of LiDAR point clouds. We find that existing view fusion-based models are dominated by features from the 3D representation, meaning they are still vulnerable to LiDAR spoofing attacks. To address such limitations, we propose sequential view fusion (SVF). SVF leverages a semantic segmentation module to better utilize FV features. Evaluations show that SVF can further reduce the mean attack success rate to 2.3\%. 
\vspace{-\topsep}
\end{itemize}

\nsection{Background}
\label{background}
\vspace{0.2in}

\nsubsection{LiDAR-based Perception in AVs}
\label{background_percpetion}



\begin{figure}[t!]
\centering
  \vspace{-0.2cm}
  \includegraphics[width=\linewidth]{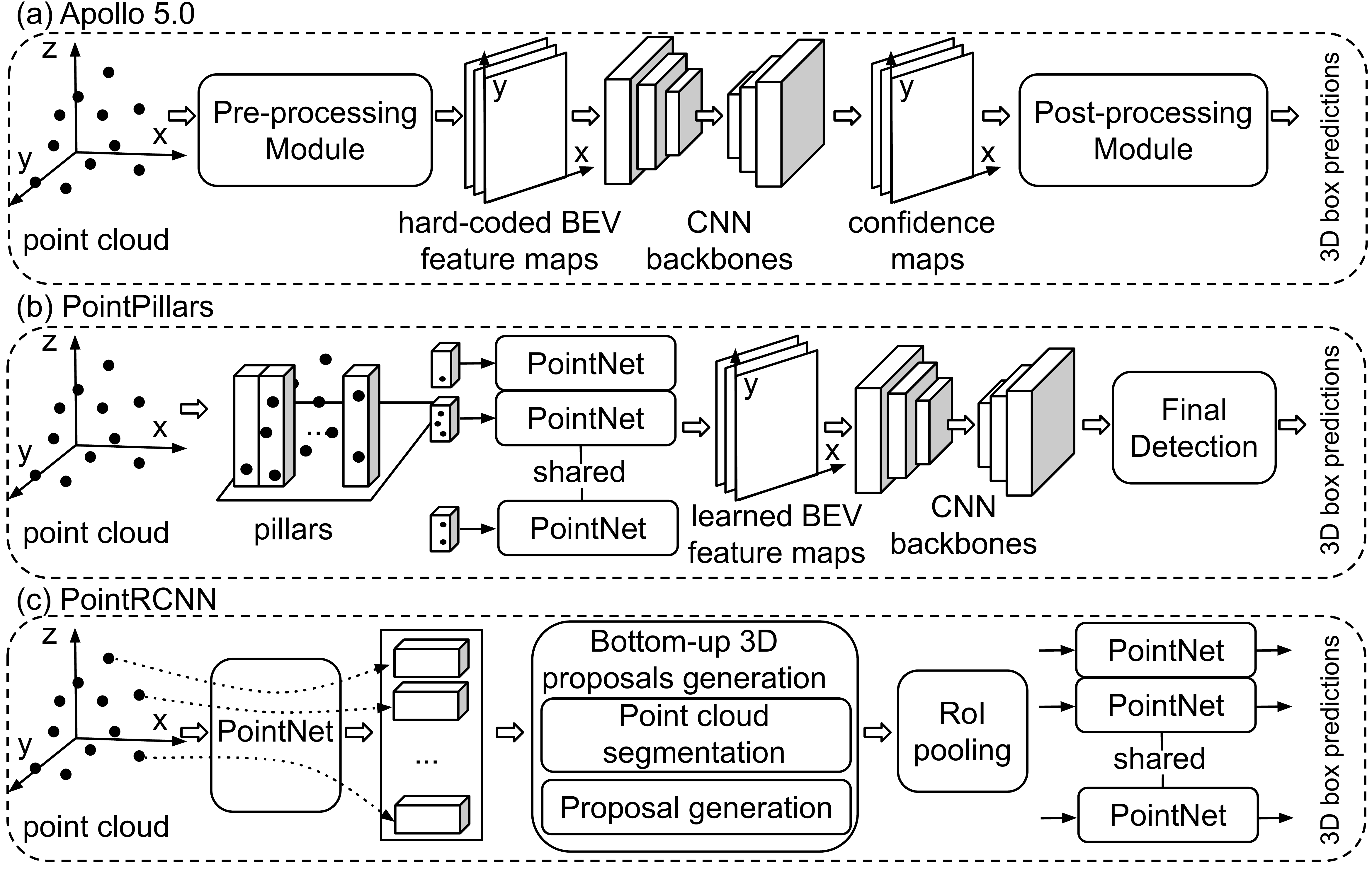}
  \vspace{-0.3cm}
  \vspace{-0.15in}
  \caption{State-of-the-art LiDAR-based perception models.}
  \label{fig:sota_models}
  \vspace{-0.5cm}
\end{figure}

LiDAR-based perception leverages 3D object detection models to understand driving environments, in which the models output 3D bounding boxes for detected objects. Deep learning has achieved great success in computer vision tasks for 2D images. However, standard convolutional pipelines cannot digest point clouds due to their sparsity and irregularity. To this end, significant research efforts have been made for 3D object detection recently~\cite{shi2019part,shi2019pointrcnn,lang2019pointpillars,zhou2018voxelnet}, among which the state-of-the-art models can be grouped into three classes:


\textbf{1. Bird's-eye view (BEV)-based 3D object detection.} Due to the remarkable progress made in 2D image recognition tasks, a large number of existing works~\cite{apollo,yang2018pixor,liang2018deep,meyer2019lasernet} attempt to transform LiDAR point clouds into the 2D structure for 3D object detection in AD systems. Most state-of-the-art methods~\cite{apollo,yang2018pixor,liang2018deep} conduct the transformation by projecting point clouds into the top-down view, also known as the BEV, and utilize 
convolutional neural networks (CNNs) to perform the final detection. Figure \ref{fig:sota_models} (a) shows the architecture of \emph{Apollo 5.0}\footnote{In this paper, we use ``Apollo 5.0'' to denote the Baidu Apollo 5.0 model.}, \CR{an industry-level BEV-based model,} that has six hard-coded feature maps in the BEV and follows a UNet-like~\cite{ronneberger2015u} pipeline to output the grid-level confidence score. The final stage heuristically clusters the grids that belong to the same object. 

\textbf{2. Voxel-based 3D object detection.} VoxelNet~\cite{zhou2018voxelnet} is the first model that slices the point clouds into voxels and extracts learnable features by applying a PointNet~\cite{qi2017pointnet} to each voxel, after which a 2D convolutional detection layer is applied in the final stage. Many recent works~\cite{lang2019pointpillars,yan2018second,wang2019voxel,lehner2019patch} adopt this voxel-based architecture and achieve state-of-the-art performance~\cite{kitti_3d}. Figure \ref{fig:sota_models} (b) shows the architecture of \emph{PointPillars} that creatively voxelizes the point cloud into pillars (a representation of point clouds organized in vertical columns) to enhance the efficiency and follows the general design of voxel-based detection architectures. \CR{Notably, PointPillars is adopted by Autoware~\cite{autoware}, an industry-level AV platform.}



\textbf{3. Point-wise 3D object detection.} Instead of transforming point clouds to regular 2D structures or voxels for feature extraction, recent studies propose to directly operate on point clouds for 3D object detection~\cite{shi2019pointrcnn,yang2019std,chen2019fast,shi2019part} and achieve the state-of-the-art performance. Most existing works in this category use a classic two-stage architecture similar to Faster RCNN~\cite{ren2015faster} in 2D object detection. The first stage is responsible for generating high-quality region proposals in the 3D space. Based on these proposals, the second stage regresses the bounding box parameters and classifies the detected objects. As shown in Figure \ref{fig:sota_models} (c), \emph{PointRCNN} adopts a bottom-up method that generates point-wise region proposals in the first stage and regresses these proposals in the later stage. 

\nsubsubsection{KITTI Benchmark}
\label{background_kitti}

KITTI~\cite{Geiger2013IJRR} is a popular dataset for benchmarking AD research, of which the point cloud data are by design divided into a \textbf{trainval set} containing 7481 samples and a \textbf{test set} containing 7518 samples. We follow the methodology by Chen \etal to split the trainval set to a \textbf{training set} (3712 samples) and a \textbf{validation set} (3769 samples) for better experimental studies~\cite{chen20153d}. KITTI evaluates 3D object detection performance by average precision (AP) using the PASCAL~\cite{Everingham10} criteria and requires a 3D bounding box overlap (IoU) over 70\% for car detection. KITTI also defines objects into three difficulty classes: Easy, Moderate, and Hard~\cite{kitti_3d}. The difficulties correspond to different occlusion and truncation levels. We train PointPillars and PointRCNN on the training set, and Table \ref{tb:model_og_performance} shows their APs evaluated on the validation set. We utilize the publicly released Apollo 5.0 model since it has its own labeling, which is incompatible with KITTI. \CR{In this work, we target car detection on the KITTI benchmark as the APs of pedestrian and cyclist detection are not yet satisfactory. However, our methodology can be generalized to other categories.} 


\nsubsection{LiDAR Sensor and Spoofing Attacks}
\label{background_spoof}

\textbf{LiDAR sensor}. A LiDAR instrument measures the distance to surroundings by firing rapid laser pulses and obtaining the reflected light with a sensor. Since the speed of light is constant, the accurate distance measurements can be derived from the time difference between laser fires and returns. By firing laser pulses at many predetermined vertical and horizontal angles, a LiDAR generates a point cloud that can be used to make digital 3D representations of surroundings. Each point in a point cloud contains its $xyz$-$i$ information, corresponding to its location and the intensity of the captured laser return.

\nsubsubsection{Sensor-level LiDAR Spoofing Attack}
\label{background_direct_spoof}

In the context of sensors, a spoofing attack is the injection of a deceiving physical signal into a victim sensor~\cite{198480}. Since they share the same physical channels, the victim sensor accepts the malicious signal, trusting it as legitimate. Prior works~\cite{petit2015remote,shin2017illusion} have shown that LiDAR is vulnerable to laser spoofing attacks. Specifically, Petit~\etal showed the feasibility to relay LiDAR laser pulses from other locations to inject fake points into the point cloud~\cite{petit2015remote}. Shin~\etal further improved the attack to control fake points at different locations in the point cloud, even very close to the victim vehicles~\cite{shin2017illusion}. 



\nsubsubsection{Adv-LiDAR: Model-level LiDAR Spoofing Attack}
\label{background_adv_spoof}
Besides directly spoofing fake points into LiDAR point clouds, a recent study proposes Adv-LiDAR that uses adversarial machine learning to not only spoof a set of fake points into the point cloud but also manage to deceive the LiDAR-based perception model~\cite{cao2019adversarial}. The authors formulate the attack on Apollo 2.5\footnote{Apollo 2.5 was the latest version when Adv-LiDAR~\cite{cao2019adversarial} was published. In this work, we target Apollo 5.0, the latest version at the time of writing.} as an optimization problem:
\begin{equation}
\begin{split}
    & \min \qquad \gL (x \oplus \advdeltad;\mathcal{M}) \qquad
    \\ & \text{s.t.} \qquad \advdeltad \in  \{ \Phi(\advdelta) \,|\, \advdelta \in \mathcal{A} \} \And x = \Phi(X)
\end{split}
\label{eq:adv_lidar}
\end{equation}
where $X$ is the pristine point cloud and $x$ represents the hard-coded feature maps in Apollo (\S\ref{background_percpetion}). $\Phi(\cdot)$ is the pre-processing function for crafting the feature maps. $T'$ and $t'$ are the spoofed point cloud and its corresponding feature maps, respectively. $\mathcal{A}$ stands for the sensor attack capability, and $\oplus$ merges the pristine and adversarial feature maps. 

The attack goal is to spoof a fake vehicle right in front of the victim AV that leads to safety issues, \CR{and the success condition is that the confidence score of the optimized spoofed points ($T'$) exceeds the default threshold so that Apollo 2.5 ($\mathcal{M}$) will detect $T'$ as a valid vehicle.} The authors formulate the sensor attack capability ($\mathcal{A}$) for general LiDAR spoofing attacks and design a specific loss function ($\gL$) and a merging function ($\oplus$) for Apollo 2.5 ($\mathcal{M}$). By strategically controlling the spoofed points, Adv-LiDAR achieves around 75\% attack success rate towards Apollo 2.5 and is considered as the state-of-the-art LiDAR spoofing attack. 

\nsection{Threat Model}
\label{threat}

\textbf{Sensor attack capability.} We perform the sensor-level spoofing attack experiments towards a Velodyne VLP-16 PUCK LiDAR~\cite{vpl16}. The attack setup is the same as Cao \etal~\cite{cao2019adversarial}, and the utilized devices are detailed in Appendix \ref{ap:Spoof}. 


We adopt the formulation in Adv-LiDAR~\cite{cao2019adversarial} to describe the sensor attack capability ($\mathcal{A}$): 
\vspace{-\topsep}
\begin{itemize}
\setlength{\itemsep}{0pt}
\setlength{\parskip}{0pt}
\item \emph{Number of spoofed points.} Compared to Adv-LiDAR, \CR{we fine-tune the comparator circuit that bridges the photodiode and delay components to calculate the time delay more accurately. Moreover, we also use a better COTS lens put in front of the attack laser to refract the laser beams to a slightly wider azimuth range. Based on these improvements, we can stably spoof at most 200 points.} Thus, we assume that attackers can spoof at most 200 points in the pristine point cloud. \CR{Such a capability is constrained by the attack hardware devices.}

\item \emph{Location of spoofed points.} Similar to Adv-LiDAR, we assume that attackers are able to modify the \emph{distance}, \emph{altitude}, and \emph{azimuth} of a spoofed point to the victim LiDAR by changing the delay intervals of the attack devices. Especially, the \emph{azimuth} of a spoofed point can be modified within a horizontal viewing angle of 10$^\circ$.
\vspace{-\topsep}
\end{itemize}

\textbf{Black-box model-level spoofing attack.} We consider LiDAR spoofing attacks as our threat model, which has been shown as a practical attack vector for LiDAR sensors~\cite{shin2017illusion,petit2015remote}. We adopt the attack goal of Adv-LiDAR: to spoof a \emph{front-near} vehicle located 5-8 meters in front of the victim AV. To perform the attack, adversaries can place an attack device at roadsides to shoot malicious laser pulses to AVs passing by, or launch attacks in another vehicle in front of the victim car (\eg on the adjacent lane)~\cite{cao2019adversarial}. \CR{LiDAR spoofing attack has been demonstrated to cause severe safety consequences in \textit{Sim-control}, an AV simulator provided by Baidu Apollo~\cite{apollo}. For example, spoofing a front-near vehicle to a high-speed AV will make it trigger a hard brake, which may injure the passengers. Adversaries can also launch a spoofing attack on an AV waiting for the traffic lights to freeze the local transportation system~\cite{cao2019adversarial}.} We assume that attackers can control the spoofed points within the observed sensor attack capability ($\mathcal{A}$). Note that attackers do not have access to the machine learning model nor the perception system. We deem such an attack model realistic since we adopt the demonstrated sensor attack settings by Shin~\etal~\cite{shin2017illusion} and relax the white-box assumptions in Adv-LiDAR~\cite{cao2019adversarial}. 


\textbf{Defense against general spoofing attacks.} We also consider defending such LiDAR spoofing attacks. We assume a stronger attack model that adversaries have white-box access to the machine learning model and the perception systems. We also assume that defenders can only strengthen the software-level design, but cannot modify the AV hardware (\eg sensors) due to cost concerns. We deem it a realistic setting since we propose to defend state-of-the-art spoofing attacks, and software-level countermeasures can be easily adopted in current AD systems.


\begin{figure*}[!t]
\centering
    \vspace{-0.2cm}
  \begin{minipage}[b]{0.48\linewidth}
  \includegraphics[width=\linewidth,height=4.75cm]{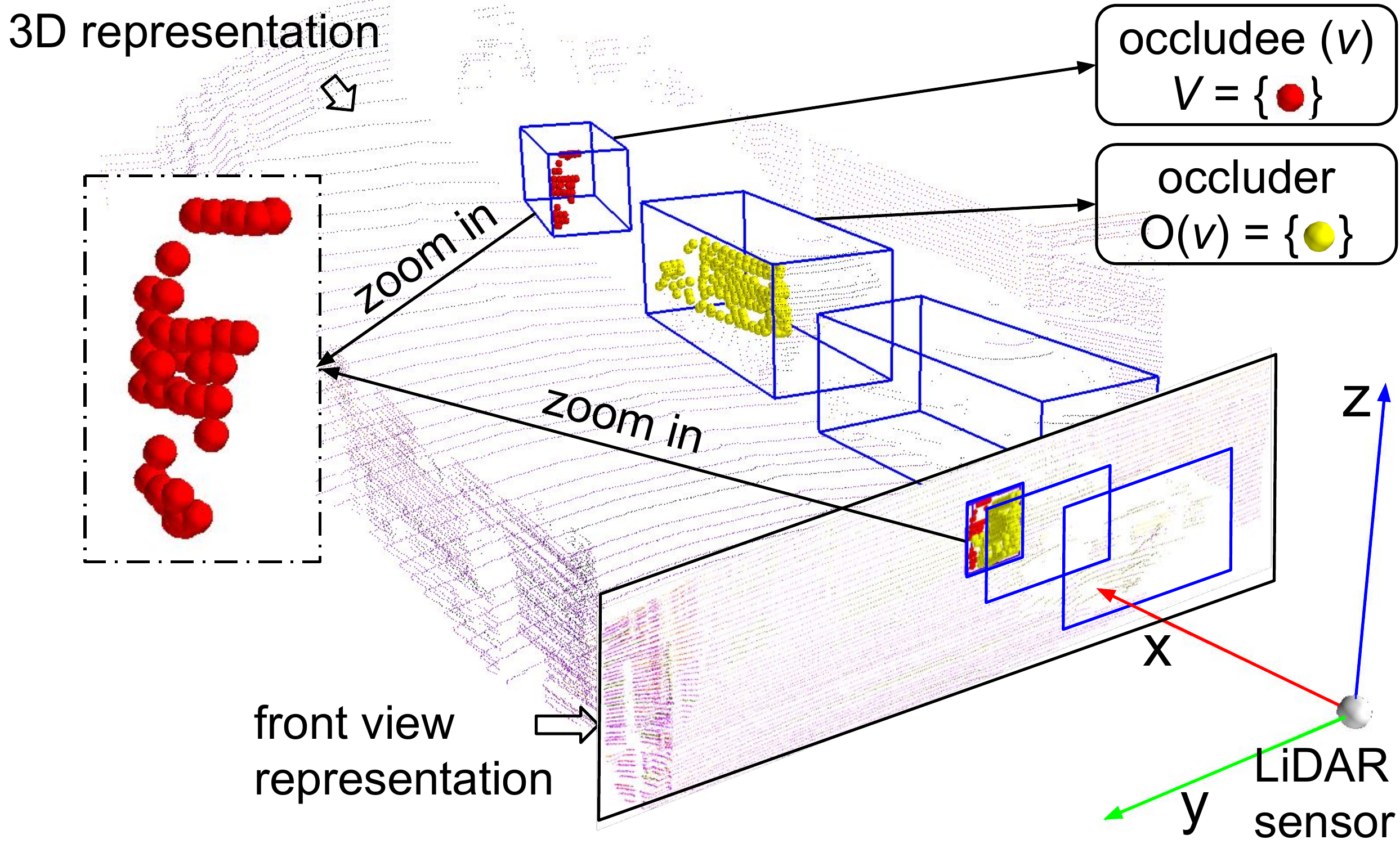}
  \vspace{-0.3cm}
  \caption{Illustration of an occluded vehicle (\textbf{C1}) in LiDAR point clouds. The yellow points from another vehicle occlude the vehicle $v$ from the perspective of the LiDAR sensor. The blue 3D cubes are the bounding boxes of detected vehicles.}
  \label{fig:c1-3}
  \end{minipage}
  \hfill
  \begin{minipage}[b]{.5\linewidth}
  \includegraphics[width=\linewidth,height=4.75cm]{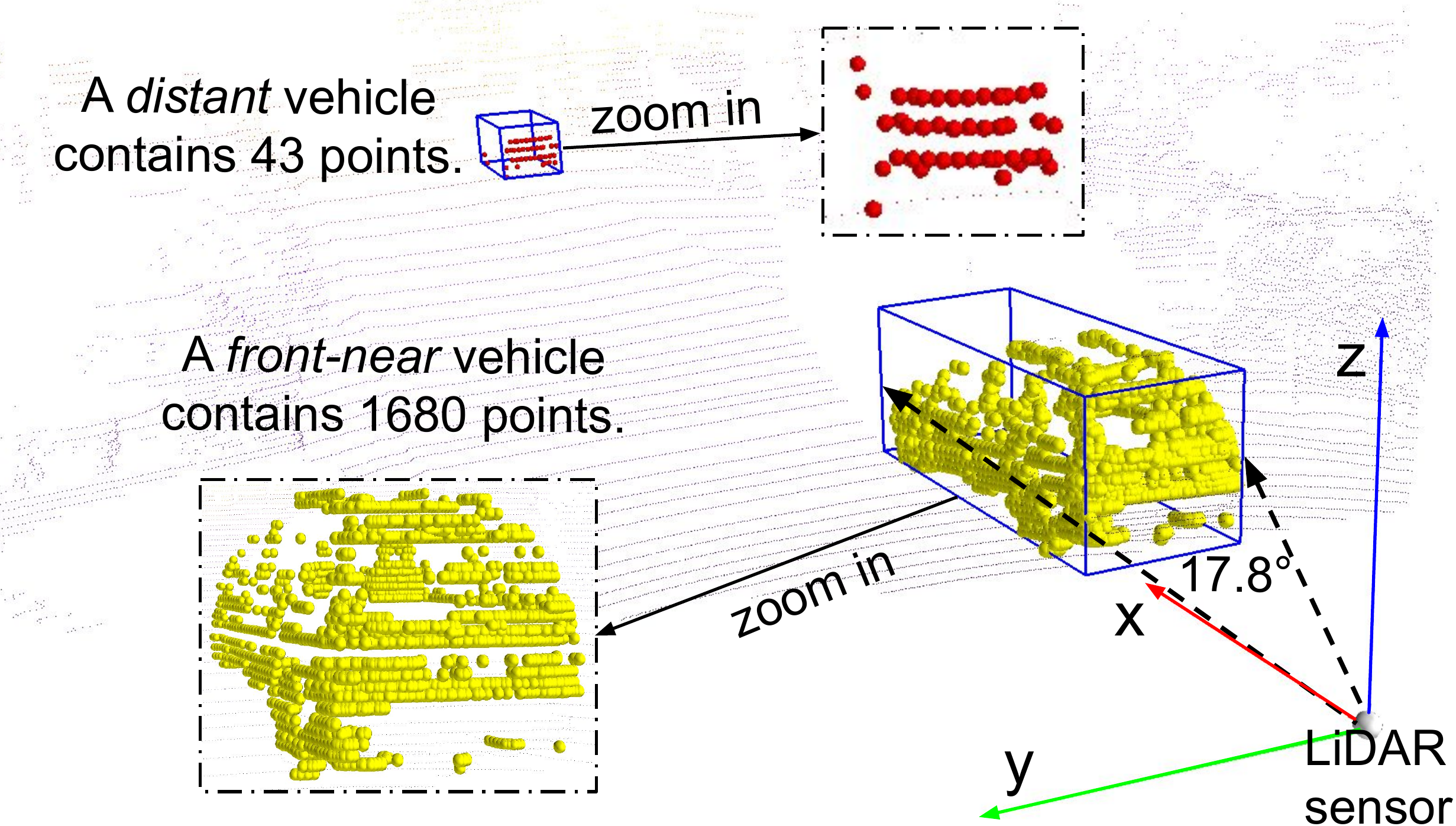}
  \vspace{-0.3cm}
  \caption{Illustration of a distant vehicle (\textbf{C2}) and a front-near vehicle in LiDAR point clouds, where the front-vehicle occupies 17.8$^\circ$ in azimuth from the perspective of the LiDAR sensor. The blue 3D cubes are the bounding boxes of detected vehicles.}
  \label{fig:distant}
  \end{minipage}
  \vspace{-0.3cm}
\end{figure*}
\nsection{Limitations of Existing Attacks}
\label{limitations_existing}
In this section, we first study whether existing LiDAR spoofing attacks can realize the attack goal on three target models, and further discuss their limitations accordingly.  

\textbf{Limitations of sensor-level LiDAR spoofing attacks:}

\emph{1. Blind attack limitation.} The sensor-level spoofing attack suffers from the effectiveness issue due to no control strategies for the spoofed points. We apply the reproduced sensor attack traces to three target models and further explore whether they will be detected as vehicles at target locations. The results (detailed in \S\ref{attack_effe}) show that blindly spoofing cannot effectively achieve the attack goal other than Apollo 5.0, which also confirms the findings by Cao \etal~\cite{cao2019adversarial}. 

\textbf{Limitations of Adv-LiDAR:}

\emph{1. White-box attack limitation.} Adv-LiDAR, the state-of-the-art spoofing attack by Cao \etal, demonstrates the feasibility of leveraging adversarial machine learning techniques to enhance its effectiveness~\cite{cao2019adversarial}. However, it suffers from the white-box limitation. Adv-LiDAR assumes that attackers have access to the deep learning model parameters and its pre- and post-processing modules. However, very few AV companies publicly release their perception systems, making Adv-LiDAR challenging to launch in the real world.

\emph{2. Attack generality limitation.} Adv-LiDAR cannot be easily generalized. First, as introduced in \S\ref{background_adv_spoof}, Adv-LiDAR only targets Apollo 2.5 and utilizes a specific pre-processing function ($\Phi(\cdot)$) and merging function ($\oplus$) which are not applicable to other models. Constructing such functions is non-trivial since they need to be differentiable so that the optimization problem can be solved by gradient descent-based methods~\cite{carlini2017adversarial}. For example, the $\Phi(\cdot)$ and $\oplus$ correspond to the voxelization and stacking processes, respectively, in PointPillars. It is still unknown whether such processes can be properly approximated differentiablely. Second, adversarial examples generated by Adv-LiDAR cannot transfer between models. We construct 20 optimized attack traces using Adv-LiDAR that successfully fool Apollo 5.0, and apply them to the other two models. However, none can achieve the attack goal in either PointPillars or PointRCNN. Third, the attack trace $T'$ is optimized with one specific point cloud at a time (Equation \ref{eq:adv_lidar}), which indicates that $T'$ may not succeed in attacking other point cloud samples. The robustness analysis by Cao \etal also validates that the attack success rate consistently drops with the change of the pristine point cloud~\cite{cao2019adversarial}. 

Overall, existing spoofing attacks cannot easily achieve the attack goal on all target models. Though Adv-LiDAR shows the feasibility to attack Apollo 2.5, more work is needed to understand the potential reasons that lead to its success. 

\nsection{A General Design-level Vulnerability}
\label{vulnerability}

Motivated by the limitations of existing attacks, in this section, we leverage an in-depth understanding of the intrinsic physical nature of LiDAR to identify a general design-level vulnerability for LiDAR-based perception in AD systems.


\nsubsection{Behind the Scenes of Adv-LiDAR}
\label{hypothesis}

Despite a lack of generality, Adv-LiDAR was able to spoof a fake front-near vehicle by injecting much fewer points than required for a valid vehicle representation. For example, Cao~\etal have demonstrated that an attack trace with merely 60 points and 8$^\circ$ of horizontal angles is sufficient to deceive Apollo 2.5~\cite{cao2019adversarial}. However, a valid front-near vehicle (\S\ref{threat}) contains around 2000 points and occupies about 15$^\circ$ of horizontal angles in KITTI point clouds~\cite{Geiger2013IJRR}. It remains unclear why such spoofing attacks can succeed despite a massive gap in the number of points between that of a fake and a valid vehicle. To answer this question and comprehend the general vulnerability exposed by Adv-LiDAR, it is necessary to consider the distinct physical features of LiDAR. In particular, we identify two situations where a valid vehicle contains a small number of points in LiDAR point clouds: 1) \textbf{an occluded vehicle} and 2) \textbf{a distant vehicle}, each corresponding to a unique characteristic (\textbf{C}) of LiDAR. 

\textbf{C1}: Occlusions between objects will make occluded objects partially visible in the LiDAR point cloud. As introduced in \S\ref{background_spoof}, a LiDAR sensor functions by firing laser pulses and capturing their returns. As a result, each point in a point cloud represents the distance to the \emph{nearest} solid object along the laser ray. Similar to human eyes, a LiDAR sensor can only perceive parts of an object (\eg a vehicle) if other obstacles, that obstruct the laser beams, are standing between the LiDAR and the object. Consequently, an occluded vehicle contains significantly fewer points than a fully exposed one since only a portion of it is visible.

In this paper, we name such occluded objects as \emph{occludees} and the obstacles that occlude others as \emph{occluders}. Particularly, as shown in Figure~\ref{fig:c1-3}, we use $\mathsf{O}(v)$ to represent the point set that occludes a vehicle $v$, and $V$ to denote the point set that belongs to the vehicle $v$ in a point cloud $F$.



\textbf{C2}: The density of data decreases with increasing distance from the LiDAR sensor, due to the working principles of LiDAR sensors (\S\ref{background_spoof}). Since the generated point clouds are collected uniformly in vertical and horizontal angles, the density of point clouds varies in the 3D space. Similar to human eyes in which a far object occupies much fewer pixels than a near one with identical size, a distant vehicle contains significantly fewer points since its point set is much sparser than that of a front-near vehicle in LiDAR point clouds (Figure~\ref{fig:distant}).  

Based upon these observations, we propose two hypotheses of potential false positive (\textbf{FP}) conditions for current LiDAR-based perception models, which could contribute to the success of Adv-LiDAR:



\vspace{0.15cm}


\noindent \textbf{FP1}: \emph{If an \underline{occluded vehicle} can be detected in the pristine point cloud by the model, its \textbf{point set} will still be detected as a vehicle when directly moved to a front-near location.}


\vspace{0.15cm}

\noindent \textbf{FP2}: \emph{If a \underline{distant vehicle} can be detected in the pristine point cloud by the model, its \textbf{point set} will still be detected as a vehicle when directly moved to a front-near location.}

\nsubsection{Experimental Validation}
\label{hypothesis_validation}

We design experiments (\textbf{E}) to test the existence of such potential erroneous predictions (\ie \textbf{FP}) on three target models using the KITTI validation set. 



\textbf{E1}: To validate \textbf{FP1}, we first randomly pick 100 point cloud samples $\mathcal{F}=\{F_{i}\}_{i=1}^{100}$ that contain 100 target \emph{occluded} vehicles $\{v_{i}\}_{i=1}^{100}$ with their point sets $V_i \subseteq F_i$. We then feed $\mathcal{F}$ into three target models and record the confidence scores (\ie outputs of models to represent the confidence of detection) of the occluded vehicles as $s_i$ for each $v_i$. 


Second, we leverage a global translation matrix $H(\theta,\tau)$ (Equation \ref{eq:trans}) to move every $V_{i}$ to a front-near location (\ie 5-8 meters in front of the victim AV) in the point cloud $F_{i}$ as $V_{i}'$, where $\theta$ and $\tau$ correspond to the \emph{azimuth} and \emph{distance} of the translation, respectively:
\CR{
\vspace{-0.2cm}
\begin{equation}
\begin{split}
& \quad {V_{i}'}_{\wi} = {V_{i}}_{\wi} \\
\begin{bmatrix}
  {V_{i}'}_{\wx}\\
  {V_{i}'}_{\wy}\\
  {V_{i}'}_{\wz}\\
  1 
\end{bmatrix} & = \begin{bmatrix}
  \cos{\theta} & -\sin{\theta} & 0 & \tau\cos{(\theta+\alpha)}\\
  \sin{\theta} & \cos{\theta} & 0 & \tau\sin{(\theta+\alpha)}\\
  0 & 0 & 1 & 0 \\
  0 & 0 & 0 & 1 \\
\end{bmatrix} 
\cdot 
\begin{bmatrix}
  {V_{i}}_{\wx}\\
  {V_{i}}_{\wy}\\
  {V_{i}}_{\wz}\\
  1
\end{bmatrix}
\end{split}
\label{eq:trans}
\end{equation}
}
$({V_{i}}_{\wx},{V_{i}}_{\wy},{V_{i}}_{\wz},{V_{i}}_{\wi})$ denotes the $xyz$-$i$ feature vectors (introduced in \S\ref{background_spoof}) of all points in $V_{i}$, and \CR{$\alpha=arctan ({V_{i}}_{\wy} / {V_{i}}_{\wx})$}. We make sure that there are no other objects standing between the LiDAR and each $V_{i}'$. By doing so, we construct a new set, where the points belong to the target occluded vehicles are moved to a front-near location by Equation \ref{eq:trans}. We further feed the new point cloud set $\mathcal{F}'$ into three target models and record the confidence scores of the translated points $V_{i}'$ as $s_{i}'$.

\begin{figure}[t!]
\centering
\vspace{-0.1cm}
  \includegraphics[width=1\linewidth, height=0.4\linewidth]{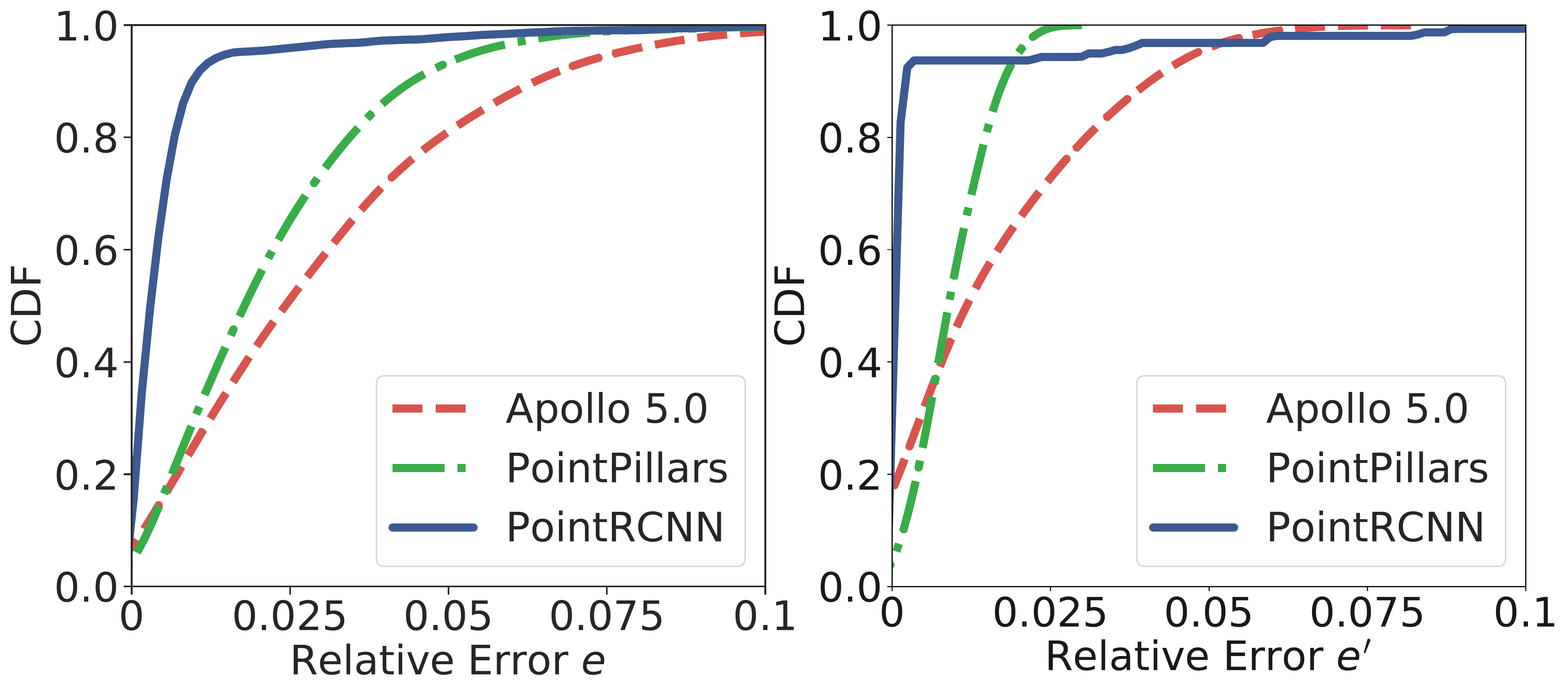}
  \vspace{-0.5cm}
  \caption{Left: CDF of the relative errors $e$ in \textbf{E1}. Right: CDF of the relative errors $e'$ in \textbf{E2}.}
  \label{fig:cdf}
  \vspace{-0.5cm}
\end{figure}

Experimental results show that all of the translated points $V_{i}'$ are detected by three target models, and we calculate the relative errors $e = \frac{|{s_i}'-{s_i}|}{{s_i}}$. Figure \ref{fig:cdf} shows the CDF of $e$ for three target models. As shown, 99.5\% of the picked occluded vehicles only have below 10\% fluctuations of their confidence scores, which successfully validate \textbf{FP1}. 

The success of \textbf{E1} comes from the fact that LiDAR-based 3D object detection models perform \emph{amodal} perception, where given only the visible portions of a vehicle $v$, the model attempts to reason about occlusions and predict the bounding box for the complete vehicle (Figure \ref{fig:c1-3}). However, \CR{convolutional operations exploit spatial locality by enforcing a local connectivity pattern between neurons of adjacent layers. Such architecture thus ensures to produce the strongest response to a spatially local input pattern.} Since the \emph{occludee}'s and \emph{occluder}'s point sets $V$ and $\mathsf{O}(v)$ stand apart from each other in the 3D space, deep learning models may fail to identify the causality between $V$ and $\mathsf{O}(v)$ and thus learns to regress the bounding box for $v$ by $V$ only.

\textbf{E2}: To validate \textbf{FP2}, similarly, we first randomly pick 100 point cloud samples that contain 100 target \emph{distant} vehicles $\{v_{i}\}_{i=1}^{100}$ that locate farther than 30 meters away from the AV, and follow the same procedure with \textbf{E1} to record the confidence score changes. Experimental results show that all of the translated points $V_{i}'$ are detected by three target models, and we calculate the relative errors $e' = \frac{|{s_i}'-{s_i}|}{{s_i}}$. Figure \ref{fig:cdf} shows the CDF of $e'$ for three target models. As shown, 99.5\% of the picked distant vehicles only have below 7.5\% fluctuations of their confidence scores, which successfully validate \textbf{FP2}.

The success of \textbf{E2} comes from that 3D object detection models are designed to be non-sensitive to the locations of objects. For example, Apollo 5.0 does not incorporate location information in its hard-coded feature maps, and PointRCNN regards the centers of each bounding box as the origins of their coordinates. Hence the global locations of objects are not valued by the 3D object detection models in AD systems.

\nsubsection{Vulnerability Identification}
\label{vulnerability_id}



As mentioned earlier, the sensor attack capability $\mathcal{A}$ is far from spoofing a fully exposed front-near vehicle's point set. However, \textbf{E1} and \textbf{E2} provide two strategies for adversaries to launch spoofing attacks with fewer points and horizontal angles. As a result, attackers can directly spoof a vehicle imitating various occlusion and sparsity patterns that satisfy the sensor attack capability $\mathcal{A}$ to fool the state-of-the-art models. For example, the $V$ (red points) in Figure \ref{fig:c1-3} only contains 38 points and occupies 4.92$^\circ$ horizontally when translated to 6 meters in front of the AV. We confirm that it can deceive all three target models successfully, as visualized in Appendix~\ref{ap:figure}.

\begin{figure*}[!t]
\centering
\vspace{-0.25cm}
\begin{minipage}[b]{.72\linewidth}
\subfigure[$ASR$ of Apollo 5.0.]{
\centering
\includegraphics[width=0.3333\linewidth]{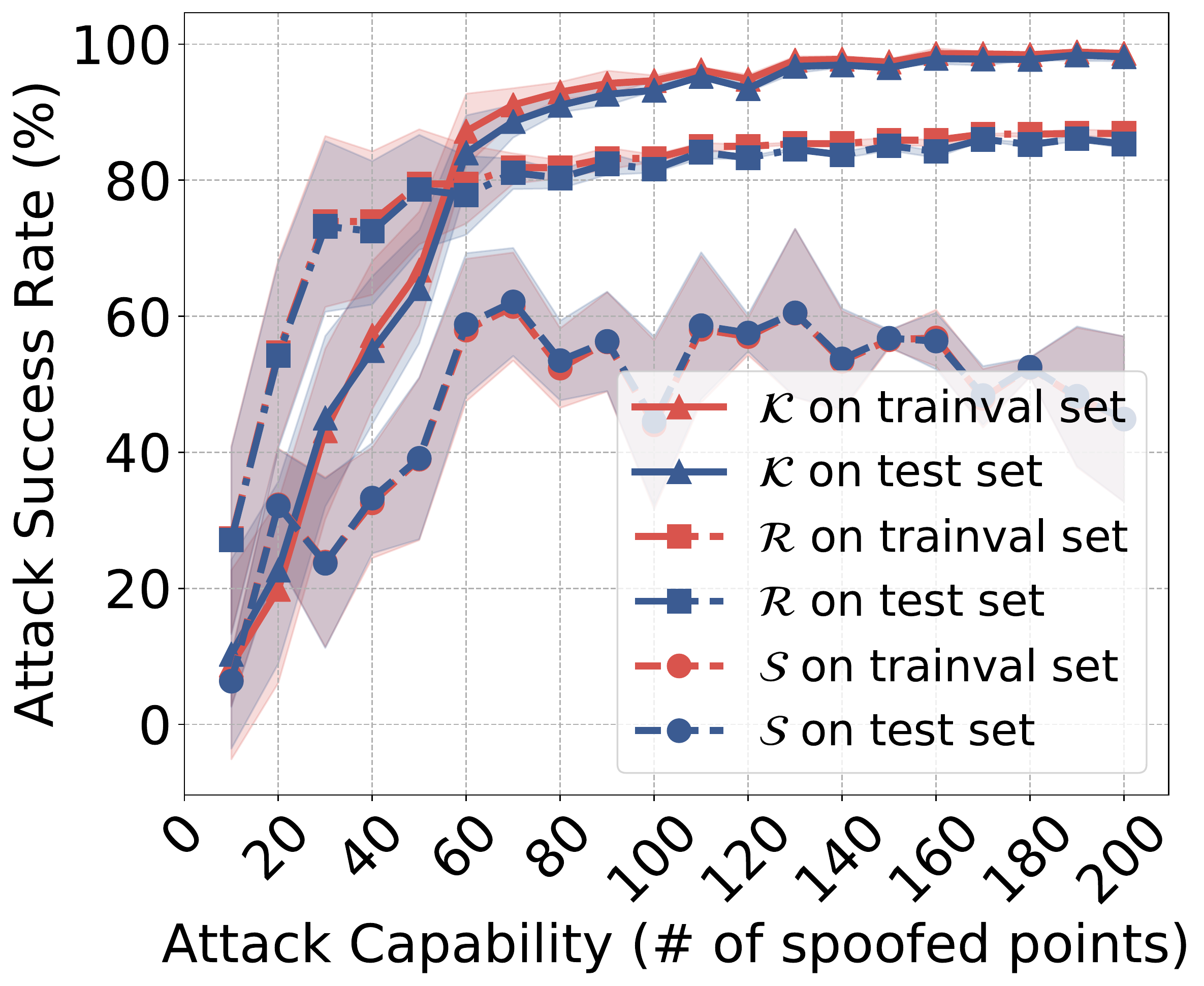}
}%
\subfigure[$ASR$ of PointPillars.]{
\centering
\includegraphics[width=0.3333\linewidth]{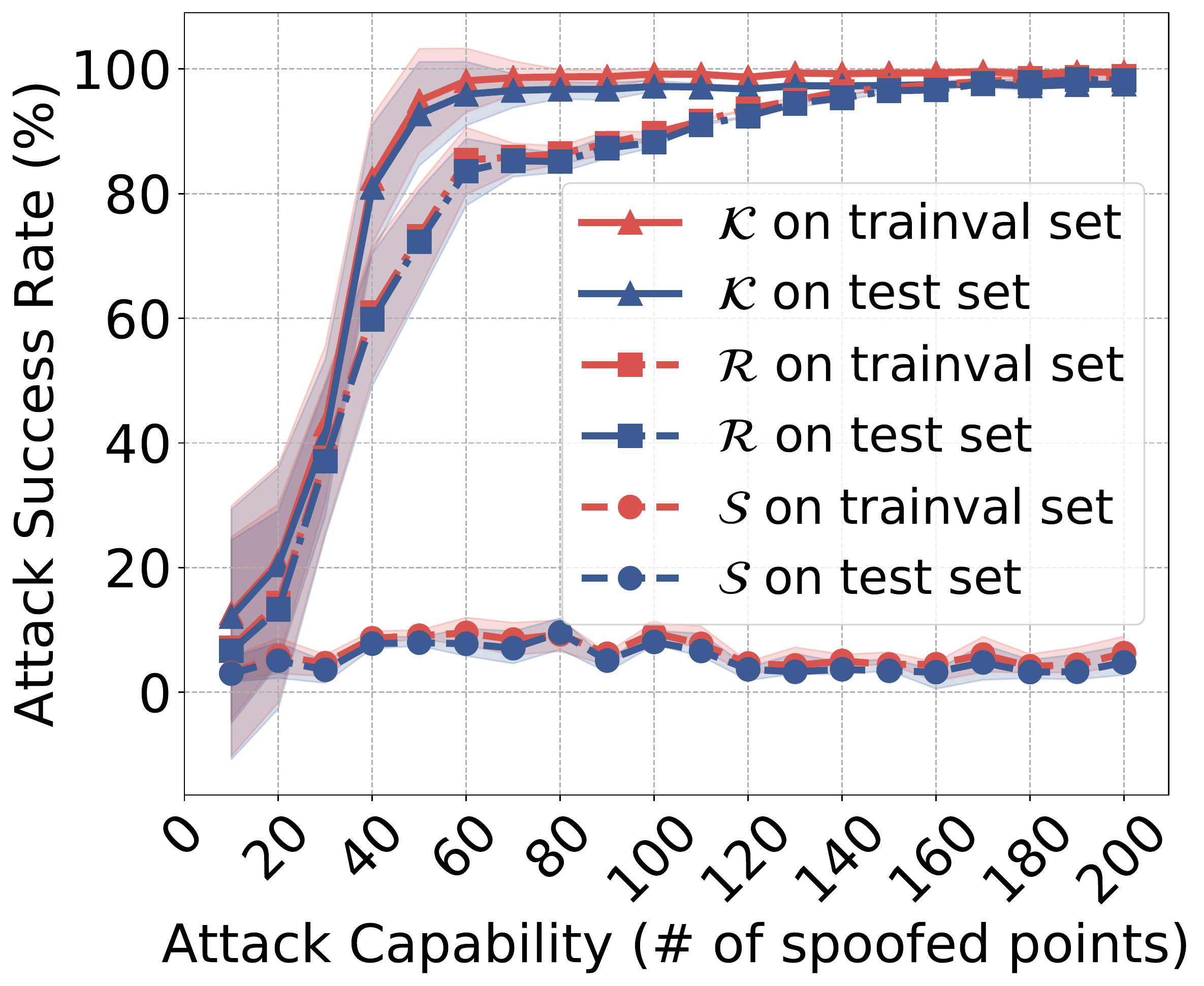}
}%
\subfigure[$ASR$ of PointRCNN.]{
\centering
\includegraphics[width=0.3333\linewidth]{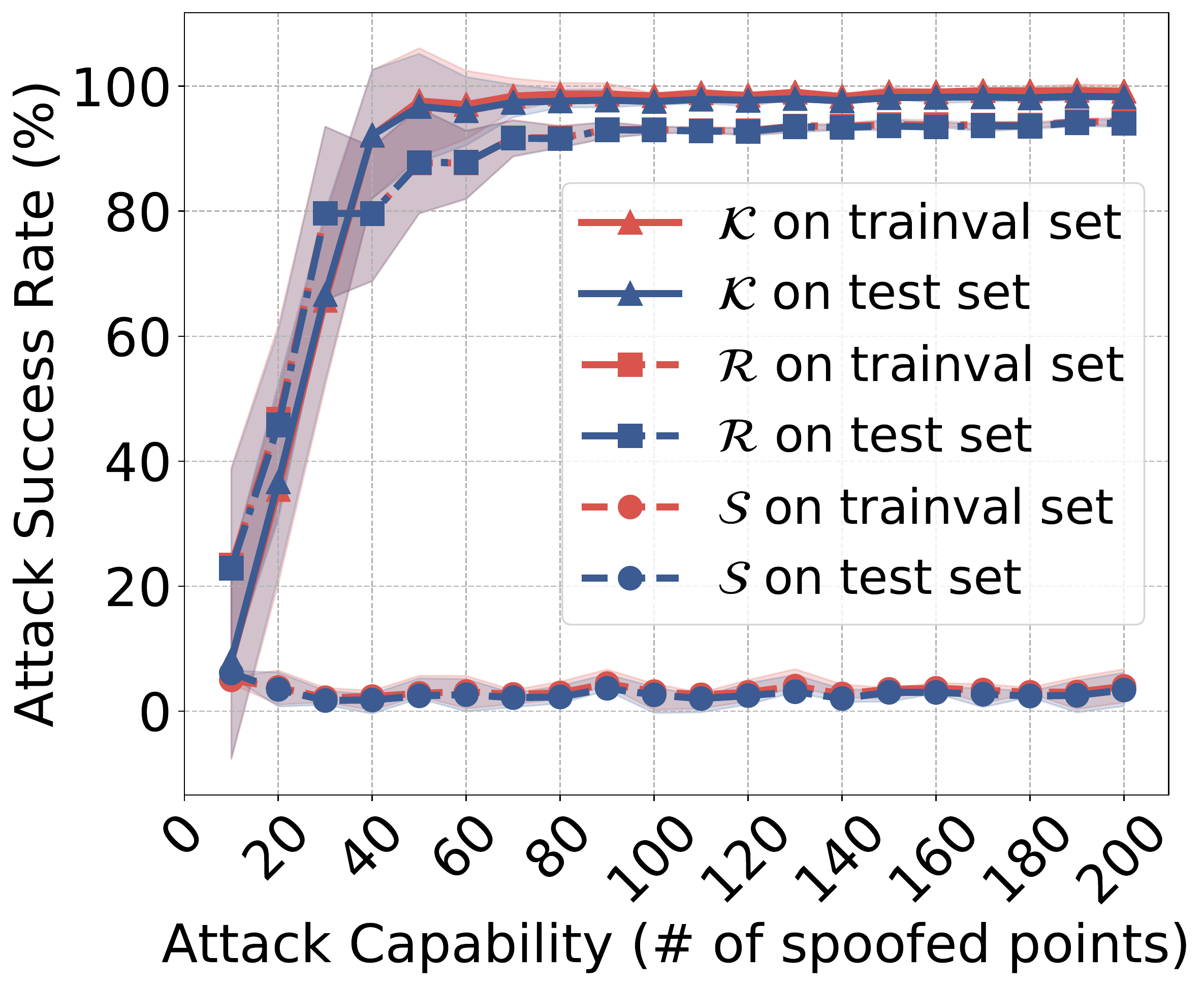}
}%
\end{minipage}
\hfill
\begin{minipage}[b]{.25\linewidth}
	\centering
	\includegraphics[width=\linewidth]{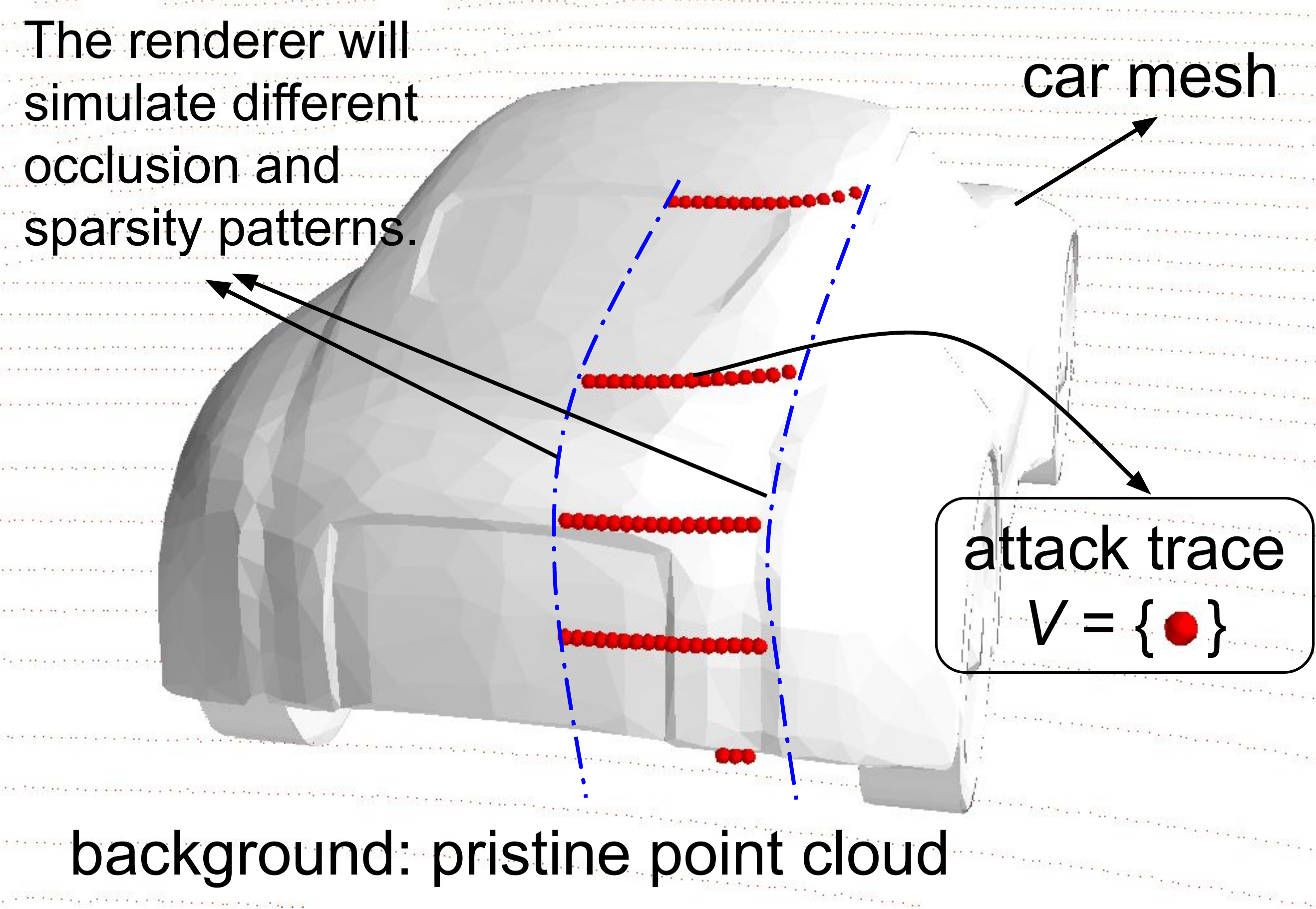}
\end{minipage}

\centering
\vspace{-0.2cm}

\begin{minipage}[b]{.70\linewidth}
\caption{Attack success rates ($ASR$s) of proposed black-box spoofing attacks on target state-of-the-art models.}
\label{fig:asr_all}
\end{minipage}
\hfill
\begin{minipage}[b]{.26\linewidth}
\centering
\caption{The process of generating attack traces for $\mathcal{R}$ from the implemented renderer.}
\label{fig:render}
\end{minipage}
\vspace{-0.8cm}
\end{figure*}



The vulnerability comes from the observation that the state-of-the-art 3D object detection architectures ignore the distinct physical features of LiDAR. Therefore, they leave a gap, as well as an attack surface, between the model capacity and LiDAR point clouds. We further abstract the neglected physical features as two occlusion patterns inside the LiDAR point clouds, described below.

\textbf{Inter-occlusion}. We abstract the typical occlusion introduced in \S\ref{hypothesis} as inter-occlusion. As its name indicates, inter-occlusion describes a causal relationship between \emph{occludee} and the corresponding \emph{occluders} (\ie the \emph{occluders} cause the \emph{occludee} partially visible). \textbf{FP1} violates the physical law of inter-occlusion since a translated ``occluded'' vehicle's point set $V'$ no longer has its valid \emph{occluder} $\mathsf{O}(v)$. However, \textbf{E1} demonstrates that state-of-the-art LiDAR-based perception models overlook such inter-occlusions in the point clouds.  



\begin{figure}[t!]
\centering
  \vspace{-0.2cm}
  \includegraphics[width=\linewidth]{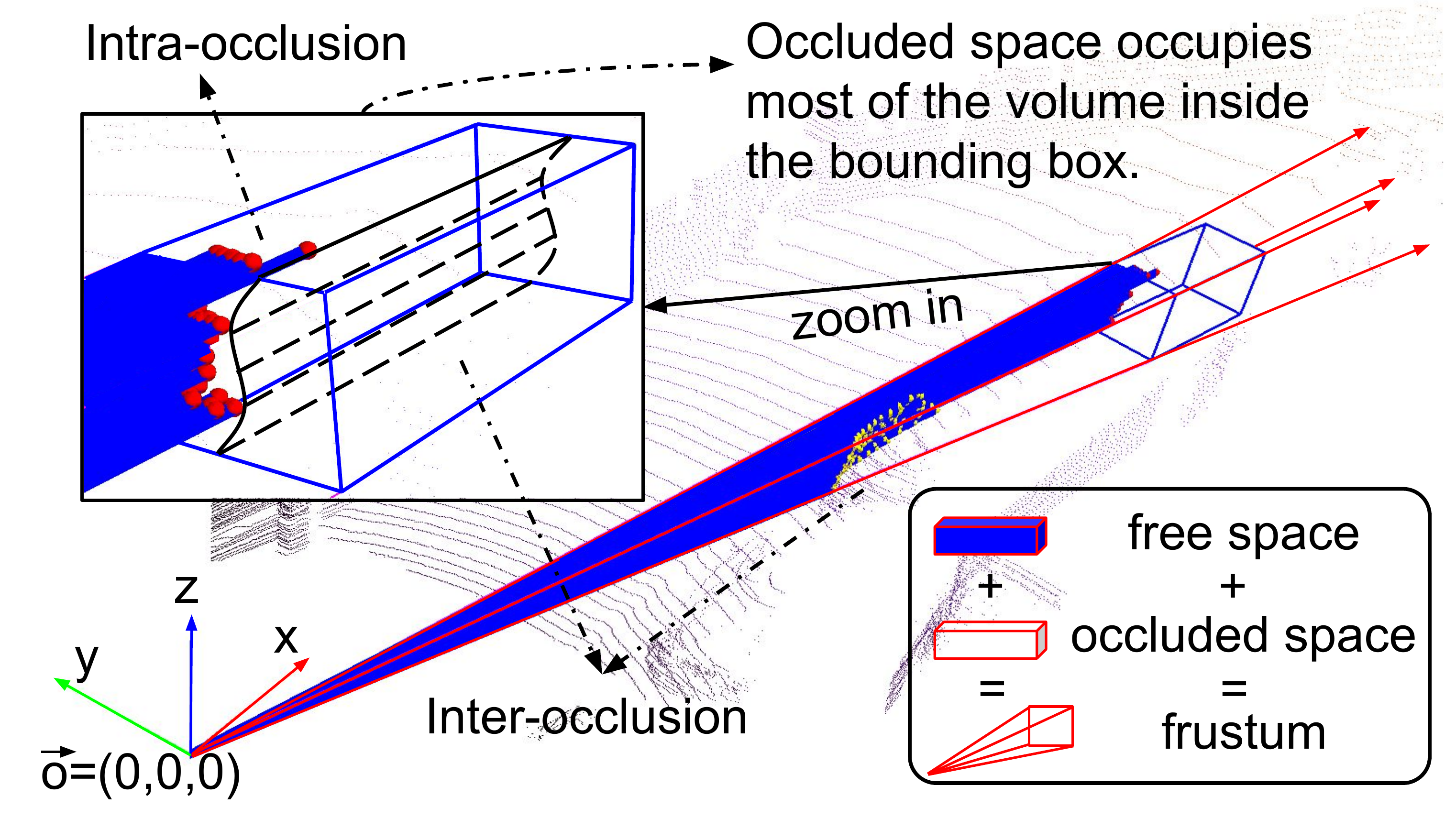}
  \vspace{-0.5cm}
  \caption{Illustration of free space (\textbf{FS}) and occluded space (\textbf{OS}) in a frustum corresponding to a detected bounding box.}
  \label{fig:fsd}
  \vspace{-0.5cm}
\end{figure}

\textbf{Intra-occlusion}. We abstract the other occlusion pattern hidden inside an object as intra-occlusion. The facing surface of a solid object (\eg~a vehicle) occludes itself in the point cloud, which indicates that the LiDAR cannot perceive the interior of the object (Figure \ref{fig:fsd}). \textbf{FP2} violates the physical law of intra-occlusion since the abnormal sparseness of a translated ``distant'' vehicle's point set $V'$ can no longer fully occlude a valid vehicle since other laser pulses could penetrate its ``surface''. However, \textbf{E2} demonstrates that state-of-the-art LiDAR-based perception models \CR{are unable to differentiate reflected points of real solid objects from sparse injected points of the same overall shape so that they also} overlook the intra-occlusions in the LiDAR point clouds. 

To demonstrate the potential real-world impacts of this identified vulnerability, we construct the first black-box spoofing attack on state-of-the-art LiDAR-based perception models in \S\ref{bb_attack}. We find the violations of the physical law of occlusion generally enable LiDAR spoofing attacks. Therefore, we perform the first defense study, exploiting the occlusion patterns as physical invariants to detect spoofing attacks in \S\ref{piad}. Lastly, in \S\ref{ml_defense}, we present a general architecture for robust LiDAR-based perception that embeds occlusion patterns as robust features into end-to-end learning.



\nsection{Black-box Spoofing Attack}
\label{bb_attack}
Constructing black-box attacks on deep learning models is non-trivial. Prior works have studied black-box attacks on image classification~\cite{papernot2017practical} and speech recognition models~\cite{247642}. However, none explored LiDAR-based perception models, and their approaches usually suffer from efficiency limitations (\eg building a local substitute model). In this section, we present the first black-box LiDAR spoofing attack based on our identified vulnerability (\S\ref{vulnerability_id}) that achieves both high efficiency and success rates. 



\emph{1. Constructing original attack traces.} As demonstrated in \S\ref{vulnerability_id}, occluded or distant vehicles' point sets that meet the sensor attack capability can be utilized to spoof front-near vehicles. Therefore, our methodology attempts to closely represent realistic physical attacks using traces from real-world datasets (\eg KITTI). In order to test different sensor attack capability, we extract occluded vehicles' point sets with varying numbers of points (5-200 points) from the KITTI validation set. Furthermore, we take 10 points as interval, and divide the extracted point sets into 20 groups per their number of points (The first group contains traces with the number of points from 0 to 10, and the second group contains traces with the number of points from 10 to 20, \etc). We then randomly pick five traces in each group forming a small dataset $\mathcal{K}$ containing 100 point sets. 

Besides collecting existing real-world traces, the identified vulnerability also supports adversaries in generating customized attack traces, which are more efficient for pipelining the attack process. We leverage ray-casting techniques to generate customized attack traces. More specifically, we utilize a 3D car mesh and implement a renderer~\cite{cao2019adversarial2} simulating the function of a LiDAR sensor that probes the car mesh by casting lasers. By doing so, we can render the car mesh's point cloud. We further simulate different occlusion and sparsity patterns on the car mesh to fit the sensor attack capability, as shown in Figure \ref{fig:render}. Similar to $\mathcal{K}$, we collect rendered point clouds with different numbers of points by using different postures and occlusion patterns. We also follow the same procedure to build a small dataset $\mathcal{R}$ containing 100 rendered point sets. More figures of $\mathcal{R}$ are shown in Appendix \ref{ap:figure}.


\emph{2. Spoofing original attack traces at target locations.} To trigger severe security and safety consequences, adversaries need to inject the constructed attack traces at target locations in the point cloud.  We consider spoofing $\mathcal{K}$ and $\mathcal{R}$ in both digital and physical environments. For digital spoofing, we make sure the injection of attack traces meets the sensor attack capability $\mathcal{A}$ and real-world requirements. We follow the high-level formulation in Adv-LiDAR~\cite{cao2019adversarial} utilizing a global transformation matrix $H(\theta,\tau)$ (Equation \ref{eq:trans}) to translate the attack traces (\ie ${V'}^{T} = H(\theta,\tau)\cdot V^{T}$, where $V \in \mathcal{K} \cup \mathcal{R}$). Here the translation interprets the attack capability ($\mathcal{A}$) in terms of modifying the \emph{azimuth} and \emph{distance} of attack traces. We further calibrate each point in the translated attack trace to its nearest laser ray's direction and prune the translated attack trace to fit the attack capability (\ie $V' \in \mathcal{A}$). Finally, we merge the attack trace with the pristine point cloud according to the physics of LiDAR. We feed the modified point cloud samples containing the attack traces into three target models. For physical spoofing, we program attack traces from $\mathcal{R}$ as input to the function generator so that we can control the spoofed points and launch the spoofing attack~\cite{shin2017illusion} in our lab. We further collect the physical attack traces and feed them into target models. Due to the limitation of our attack devices, we only conduct preliminary physical spoofing experiments. More details of physical spoofing can be found in \S\ref{practi}. It is worth noting that such limitations do not hurt the validity of our attack model (\S\ref{threat}) since the attack capability $\mathcal{A}$ is adopted from Adv-LiDAR~\cite{cao2019adversarial}, in which has been demonstrated in the real world.




\nsubsection{Attack Evaluation and Analysis}
\label{attack_analysis}
We perform large-scale evaluations on our proposed black-box attack in terms of effectiveness and robustness.

\textbf{Experimental setup.} \CR{The evaluations are performed on the KITTI trainval and test sets (introduced in~\S\ref{background_kitti}), which are collected in the physical world. As mentioned before, limited by our attack devices, we leverage $\mathcal{K},\mathcal{R}$ to launch \textit{digital spoofing} attacks. We also utilize attack traces ($\mathcal{S}$) generated by the sensor-level spoofing attack (\S\ref{limitations_existing}) as a baseline. $\mathcal{S}$ is collected from blindly \textit{physical spoofing} attacks on a real Velodyne VLP-16 PUCK LiDAR~\cite{vpl16}. We further inject all the attack traces from above three constructed datasets into the KITTI point clouds at front-near locations (\ie 5-8 meters in front of the victim AV) to test their effectiveness.}  



\textbf{Evaluation metrics.} Object detection models often have default thresholds for confidence scores to filter out detected objects with low confidence (potential false positives). We leverage the default thresholds used by three target models to measure the attack success rate ($ASR$). We label an attack successful as long as the model detects a vehicle at the target location whose confidence score exceeds the default threshold:
\begin{equation}
    ASR = \frac{\mathrm{\#\; of\; successful\; attacks}}{\mathrm{\#\; of\; total\; point\; cloud\; samples}}
    \label{eq:asr}
\end{equation}

Besides the default threshold, we also define a new metric that leverages multiple thresholds to evaluate LiDAR spoofing attacks. The corresponding definitions and evaluations are described in Appendix \ref{ap:attack_eval}, which provide insights that point-wise features appear to be more robust than voxel-based features.


\nsubsubsection{Attack Effectiveness}
\label{attack_effe}

Figure \ref{fig:asr_all} shows the $ASR$ of the digital spoofing attack with different attack capabilities (\ie number of points). As expected, the $ASR$ increases with more spoofed points. The $ASR$s are able to universally achieve higher than 80\% in all target models with more than 60 points spoofed, and it also stabilizes to around 85\% with more than 80 points spoofed. Notably, the attack traces from $\mathcal{R}$ can achieve comparable $ASR$ with $\mathcal{K}$ on all target models, which demonstrates that adversaries can efficiently leverage a customized renderer to generate attack traces (Figure \ref{fig:render}). Such rendered traces can be directly programmed into hardware for physical spoofing attacks (Appendix \ref{ap:Spoof}). Interestingly, $\mathcal{S}$ achieves much higher $ASR$ on Apollo 5.0, indicating that BEV-based features are less robust to spoofing attacks than the other two categories, which could be attributed to the information loss of feature encoding from BEV. 






\nsubsubsection{Robustness Analysis}

We analyze the robustness of the proposed attack to variations of attack traces $V'$ and the average precision (AP) of target models $\mathcal{M}$. We also evaluate the attack robustness against state-of-the-art defense strategies~\cite{xu2017feature,yang2019menet} designed for image-based adversarial attacks. We find that spoofed traces with around 60 points to trigger major changes in $ASR$. Note that Cao \etal also utilized spoofed traces with 60 points for analysis~\cite{cao2019adversarial}. Therefore, we use attack traces with $(60,70]$ points from $\mathcal{R}$ for the robustness analysis. 

\begin{figure}[!t]
  \begin{minipage}[t]{0.478\linewidth}
    \centering 
    \includegraphics[width=\linewidth]{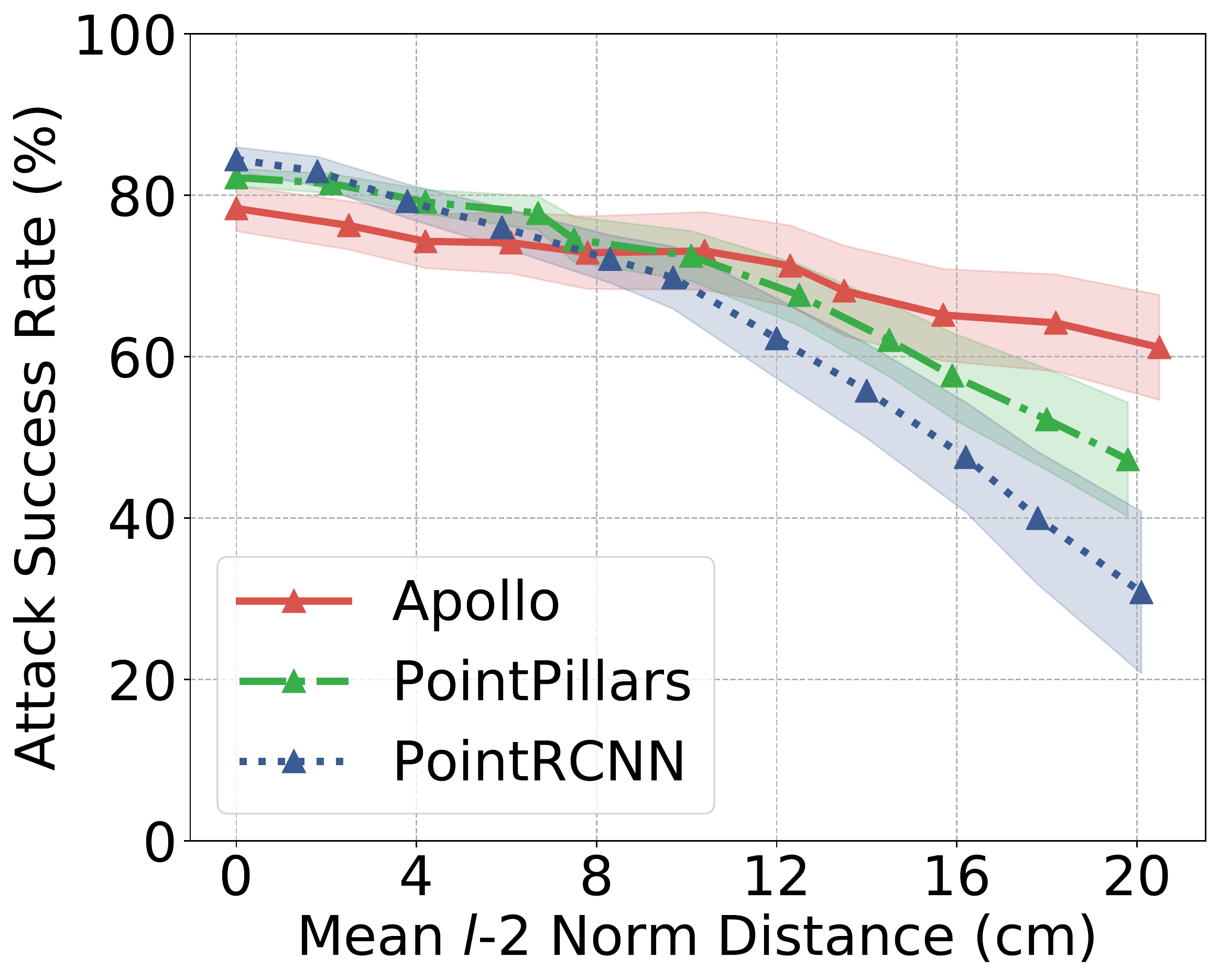} 
    \vspace{-0.5cm}
    \caption{Attack robustness to variations in generated attack traces from $\mathcal{R}$.} 
    \label{fig:robust_1}
  \end{minipage}%
  \hfill
  \begin{minipage}[t]{0.49\linewidth} 
    \centering 
    \includegraphics[width=\linewidth]{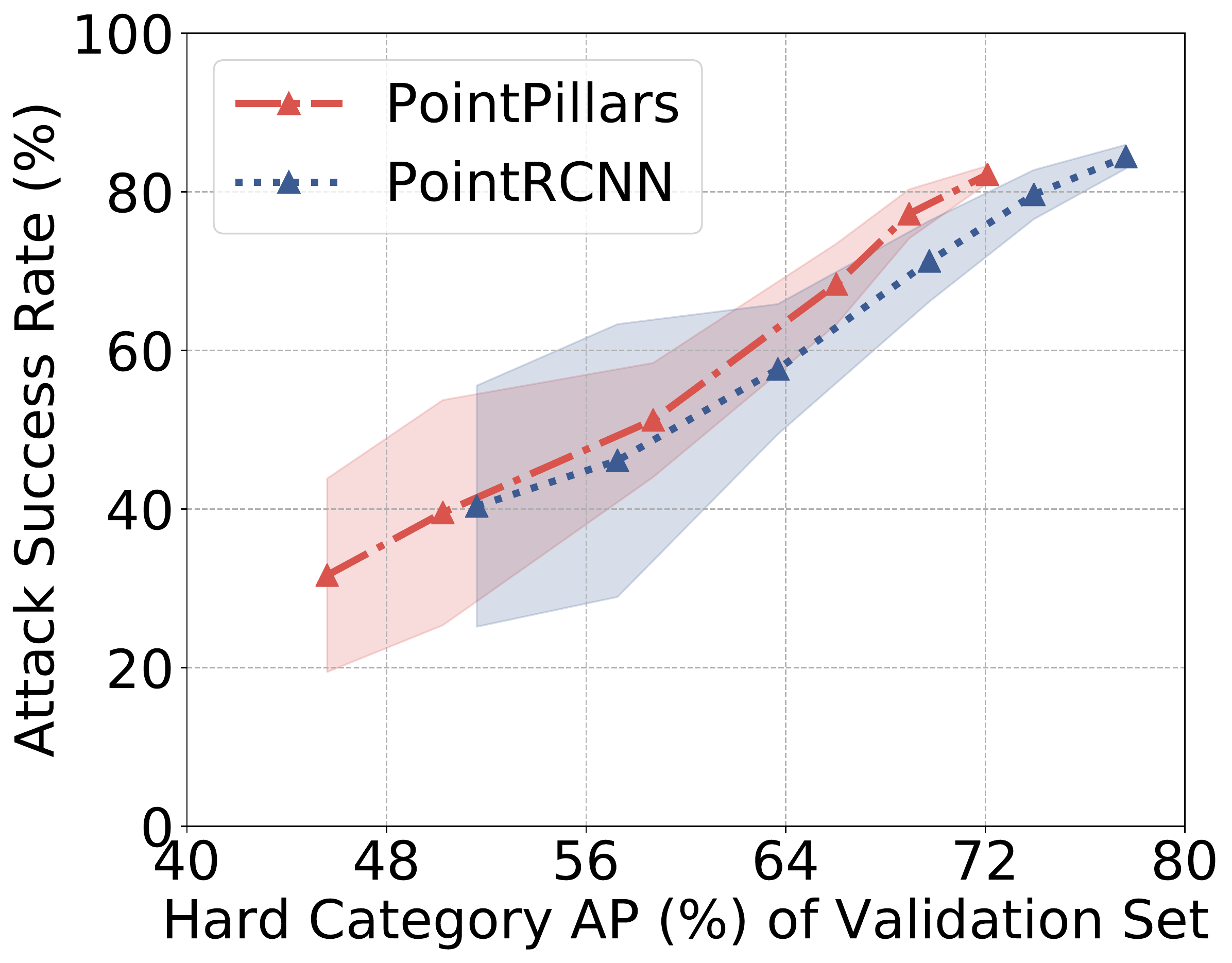} 
    \vspace{-0.5cm}
    \caption{Attack robustness to variations in target models' performance.}
    \label{fig:robust_2}
  \end{minipage} 
  \vspace{-0.3cm}
\end{figure}

\textbf{Robustness to variations in attack traces}. First, we apply a scaling matrix $S$ to the attack traces $V'$ with different-level randomness to simulate the inaccuracy of sensor attack:
\begin{equation}
\begin{split}
& \quad {V''}_{\wi} = {V'}_{\wi} \\
\begin{bmatrix}
  {V''}_{\wx}\\
  {V''}_{\wy}\\
  {V''}_{\wz}\\
\end{bmatrix} & = \begin{bmatrix}
  s & 0 & 0\\
  0 & s & 0\\
  0 & 0 & s\\
\end{bmatrix} 
\cdot 
\begin{bmatrix}
  {V'}_{\wx}\\
  {V'}_{\wy}\\
  {V'}_{\wz}\\
\end{bmatrix}
\end{split}
\end{equation}
where $s$ subjects to a uniform distribution $U(1-\epsilon,1+\epsilon)$.
We use the mean $l$-$2$ norm to measure the distance between $V''$ and $V'$. Figure \ref{fig:robust_1} shows the $ASR$ drops with larger $l$-$2$ distances which is expected. However, as shown, the $ASR$ still reaches around 70\% while the distance is around 10~cm. We also observe that the $ASR$ for PointRCNN drops faster than for PointPillars and Apollo 5.0, which also validates that point-wise features are arguably more robust than voxel-based and BEV-based features (detailed in Appendix \ref{ap:attack_eval}). 


\textbf{Robustness to variations in model performance}. To understand the relationship between $ASR$ and the original performance of models (\ie AP), we first extract the intermediate models when we trained PointPillars and PointRCNN. We then try to launch attacks on these models. Surprisingly, we find that the $ASR$ increases with higher AP\footnote{\CR{We evaluate the $ASR$s 
until the training procedures (\ie APs) converge on both models.}} (Figure \ref{fig:robust_2}), which implies that a model with better performance could be more vulnerable to such attacks. Our results indirectly demonstrate that the identified vulnerability could be attributed to an ignored dimension (\ie occlusion patterns) by current models. Since the models do not notice such a hidden dimension, they will be overfitted to be more vulnerable during training.

\CR{\textbf{Robustness to adversarial training}. Adversarial training is not rigorously applicable because it targets classification models, and requires norm-bounded perturbations to make the optimization problem tractable~\cite{madry2017towards}. In contrast, our study targets 3D object detection models, and the proposed attack is constrained by the sensor attack capability ($\mathcal{A}$), which does not fit any existing norm-bounded formulations. Thus, we perform this robust analysis in an empirical setting. Specifically, we generate another 100 attack traces with 60 points using the customized renderer and randomly inject two of them into each point cloud sample in the KITTI training set at areas without occlusions. We further train PointPillars and PointRCNN on this modified dataset and evaluate our proposed attack using the same 60-point attack traces with \S\ref{attack_effe} on them. We observe that the $ASR$s drop from 83.6\% to 70.1\% and 88.3\% to 79.7\% on PointPillars and PointRCNN, respectively, on the KITTI validation set. However, the ``Hard'' category's original detection performance has significant degradation of over 10\% on both models. Our results empirically show that current LiDAR-based perception model designs cannot learn the occlusion information correctly. The slight drop of the $ASR$s comes from the under-fitting effect of existing occluded vehicles (\ie significant AP degradation), which is not acceptable in real AD systems.}


\textbf{Robustness to randomization-based defenses}. We leverage state-of-the-art image-based defenses: feature squeezing~\cite{xu2017feature} and ME-Net~\cite{yang2019menet} to test the attack robustness on Apollo 5.0 since it has similar pipelines with image-based models. We demonstrate that none of them can defend the black-box spoofing attack without hurting the original AP. More details can be found in Appendix \ref{ap:attack_eval}.


\nsection{Physics-Informed Anomaly Detection }
\label{piad}

Our results show that a lack of awareness for occlusion patterns enables the proposed black-box attack in~\S\ref{bb_attack}. Since adversaries exploit an ignored hidden dimension, such attacks can succeed universally in target models and appear to be robust to existing defenses (\S\ref{attack_analysis}). Since anomaly detection methods have been widely adopted in different areas~\cite{chaman2018ghostbuster,li2019detecting}, one intuitive and immediate mitigation is to detect such violations of physics. \CR{We find that no existing open-source AV platforms enable such physical checking~\cite{apollo,autoware}.} In this section, we present CARLO:~o\underline{C}clusion-\underline{A}ware hie\underline{R}archy anoma\underline{L}y detecti\underline{O}n, that harnesses occlusion patterns as invariant physical features to accurately detect such spoofed fake vehicles.

\nsubsection{CARLO Design}
\label{3sd}

CARLO consists of two building blocks: free space detection and laser penetration detection. 

\nsubsubsection{Free Space Detection}

Free space detection (FSD) integrates both inter- and intra-occlusions (\S\ref{vulnerability_id}) to detect spoofed fake vehicles. As introduced in \S\ref{background_spoof}, each laser in a LiDAR sensor is responsible for perceiving a direction in the spherical coordinates. Due to resolution limits, such a laser direction actually corresponds to a thin frustum in the 3D space. As shown in Figure \ref{fig:fsd}, the frustum (as well as the straight-line path $\vec{p}-\vec{o}$) from the LiDAR sensor ($\vec{o}=(0,0,0)$) and any point in the point cloud ($\forall \vec{p}=(x,y,z)$) is considered as free space (drivable space occupied by air only).  Therefore, combined with all laser beams of the LiDAR, the entire 3D space is divided into free space (\textbf{FS}) and occluded space (\textbf{OS}) (\ie space behind the hit point from the LiDAR sensor's perspective). \textbf{FS} information is embedded at the point level. Occlusions, on the other hand, exist at the object level. \textbf{FS}, thus, is more fine-grained and incorporates occlusion information since the \textbf{OS} of an object directly reflects its occlusion status (Figure \ref{fig:fsd}).

Due to inter-occlusion and intra-occlusion, we observe that the ratio $f$ of the volume of \textbf{FS} over the volume of a detected bounding box should be subject to some distribution and upper-bounded by $\exists\, b \in (0,1)$, implying $f \in (0,b]$ (Figure \ref{fig:fsd}).
Nevertheless, since the fake vehicles do not obey the two occlusion patterns, their ratio $f'$ should be large enough and lower-bounded by $\exists\, a \in (0,1)$ such that $f' \in [a,1)$. Clearly, as long as $a > b$, we have opportunities to distinguish valid vehicles with the spoofed fake vehicles statistically. To estimate the ratio $f$, we grid the 3D space into cells and calculate: 
\begin{equation}
    f_{\mathsf{B}}=\frac{\sum_{\,c \in \mathsf{B}}\mathds{1} \cdot FS(c)}{|\mathsf{B}|}
    \label{eq:fsd}
\end{equation}
where $FS(c)$ indicates whether the cell $c$ is free or not, and $|\mathsf{B}|$ denotes the total number of cells in the bounding box $\mathsf{B}$. The algorithm to derive $FS(c)$ can be found in Appendix \ref{ap:algorithm}.

\begin{figure}[!t]
\vspace{-0.1cm}
  \begin{minipage}[t]{0.48\linewidth}
    \centering 
    \includegraphics[width=\linewidth,height=3.5cm]{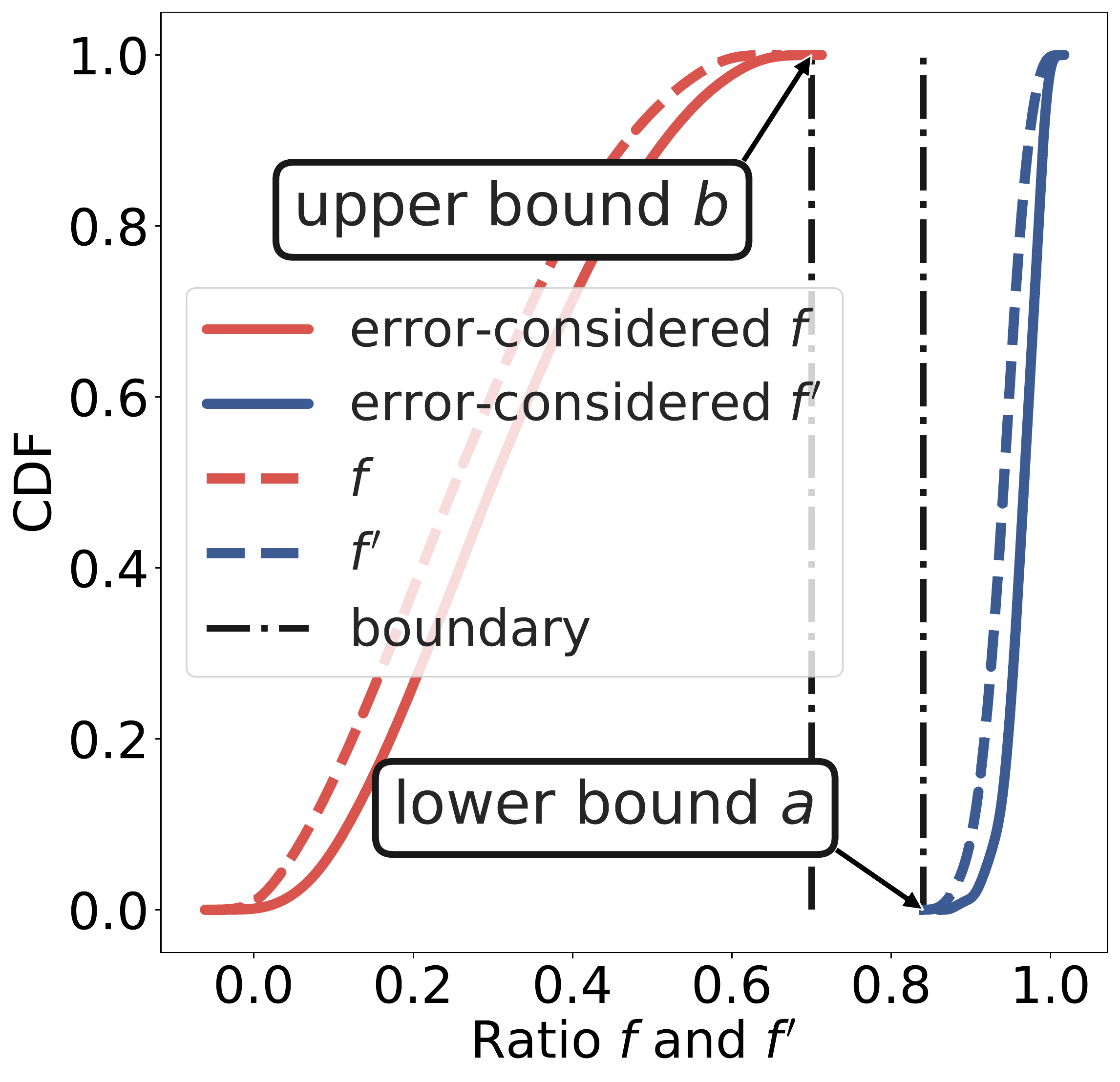} 
    \vspace{-0.5cm}
    \caption{CDF of $f$ and $f'$, and the two distributions are clearly separate.} 
    \label{fig:fsd_dis}
  \end{minipage}%
  \hfill
  \begin{minipage}[t]{0.48\linewidth} 
    \centering 
    \includegraphics[width=\linewidth,height=3.5cm]{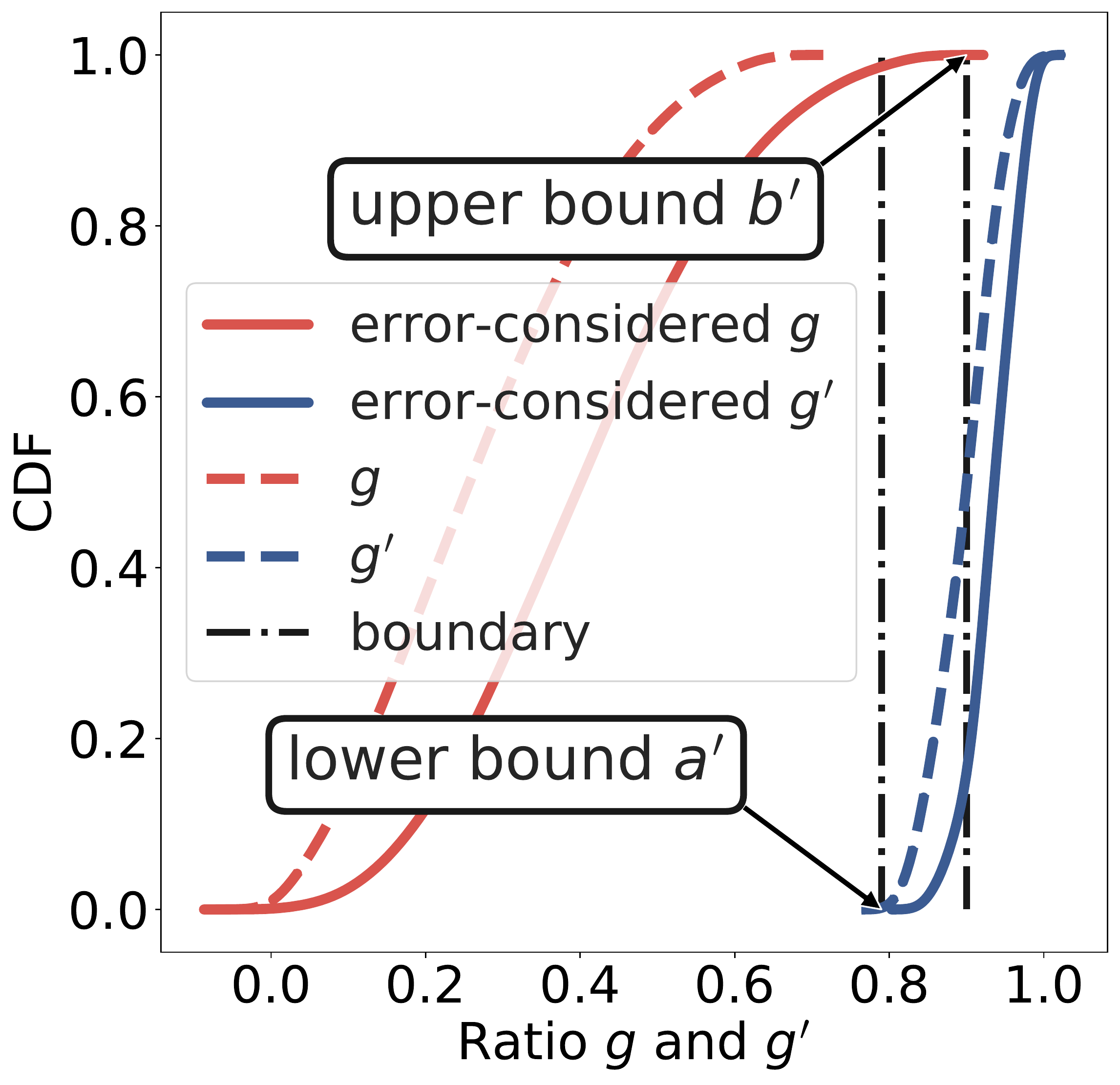} 
    \vspace{-0.5cm}
    \caption{CDF of $g$ and $g'$, but the two distributions overlap with each other.} 
    \label{fig:lpd_dis}
  \end{minipage} 
  \vspace{-0.5cm}
\end{figure}

We then estimate the distributions of valid and fake vehicles.We empirically set the cell size to $0.25^3$ $m^3$, and utilize the KITTI training set and 600 new attack traces generated by the implemented renderer (\S\ref{bb_attack}) for estimation. Figure \ref{fig:fsd_dis} shows that the CDF of $f$ and $f'$ clearly separate from each other. We further take the models' error into considerations (0.7 IoU), and estimate the distributions again. The two distributions still do not overlap with each other, as shown in Figure \ref{fig:fsd_dis}, which demonstrate the feasibility to leverage the ratio $f$ as an invariant indicator for detecting anomalies. 

However, though FSD provides a statistically significant method to detect adversarial examples, it is too time-consuming to perform ray-casting to all the detected bounding boxes in real-time. The mean processing time of one vehicle is around 100 ms in our implementation using C++ on a commodity Intel i7-6700K CPU @ 4.00GHz, which is already comparable to the inference time of deep learning models.

\nsubsubsection{Laser Penetration Detection}
\label{ltd}

Laser penetration detection (LPD) is a variant of FSD that aims to provide better efficiency for CARLO. As introduced in \S\ref{3sd}, each point in the point cloud represents one laser ray and the boundary between free space and occluded space. Given a vehicle's point set, its bounding box $\mathsf{B}$ also divides the corresponding frustum into three spaces which are: 1) the space between the LiDAR sensor and the bounding box $\mathsf{B}\uparrow$, 2) the space inside the bounding box $\mathsf{B}$, and 3) the space behind the bounding box $\mathsf{B}\downarrow$. Intuitively, only a small number of laser rays can \emph{penetrate} the bounding box (Figure \ref{fig:fsd}). As a result, from the perspective of the LiDAR sensor, the ratio $g$ of the number of points located in the space behind the bounding box $\mathsf{B}\downarrow$ over the total number of points in the whole frustum should be upper bounded by $\exists\, b' \in (0,1)$. For the same reason in \S\ref{3sd}, the ratio $g'$ of the spoofed vehicles is supposed to be large enough and lower bounded by $\exists\, a' \in (0,1)$.

Therefore, the ratio $g$ is derived from:
\begin{equation}
    g_{\mathsf{B}}=\frac{\sum_{\,\vec{p} \in \mathsf{B}\downarrow}\mathds{1}(\vec{p}) }{\sum_{\,\vec{p} \in \mathsf{B}\cup\mathsf{B}\downarrow\cup\mathsf{B}\uparrow}\mathds{1}(\vec{p})}
    \label{eq:lpd}
\end{equation}
Since LPD leverages information directly from the output of models, it is a good fit for parallel acceleration. The mean processing time of LPD is around 5 ms for each bounding box using Python on a commodity GeForce RTX 2080 GPU. 

Similarly, Figure \ref{fig:lpd_dis} shows the CDF of $g$ and $g'$ for valid vehicles from the KITTI training set and the 600 generated attack traces, respectively. As shown, though the distributions of ground-truth are separate, the error-considered distributions overlap with each other (\ie $b' > a'$). We verify that the overlap comes from the noise introduced by points of the ground plane. As a result, LPD will cause erroneous detection of potential anomalies. 





\nsubsubsection{Hierarchy Design}

To achieve both robustness and efficiency, CARLO hierarchically integrates FSD and LPD. In the first stage, CARLO accepts the detected bounding boxes and leverages LPD to filter the unquestionably fake and valid vehicles by two thresholds (\S\ref{ltd}). The rest bounding boxes are uncertain and will be further fed into FSD for final checking. CARLO achieves around 8.5 ms mean processing time for each vehicle. The entire algorithm of CARLO is detailed in Appendix \ref{ap:algorithm}. 


\nsubsection{CARLO Evaluation}

\textbf{Experimental setup.} We evaluate the defense performance of CARLO on the KITTI trainval and test sets. We apply all the attack traces from $\mathcal{K},\mathcal{R}$ to all point cloud samples at target locations (5-8 meters in front of the victim), and feed them into three CARLO-guarded models $\mathsf{CARLO}(\mathcal{M}(\cdot))$. We also evaluate CARLO against Adv-LiDAR~\cite{cao2019adversarial} on Apollo 5.0. The defense goal is to successfully detect the spoofed fake vehicles from the output bounding boxes without hurting the original performance (\ie AP) of the target models.

\begin{figure}[!tp]
\vspace{-0.4cm}
\centering
\subfigure[CARLO-guarded Apollo 5.0.]{
\begin{minipage}[t]{0.49\linewidth}
\centering
\includegraphics[width=4cm]{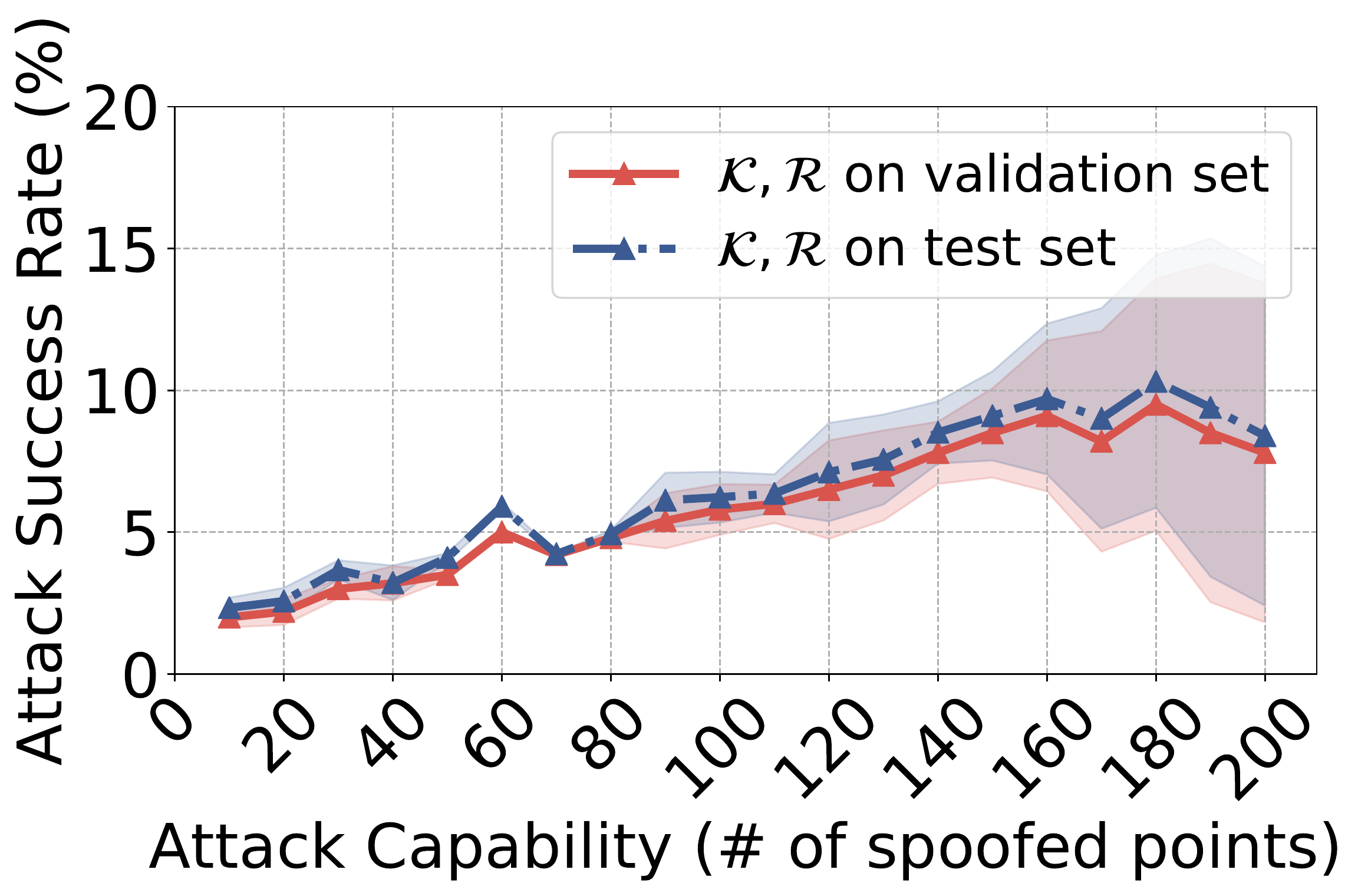}
\end{minipage}
}%
\subfigure[CARLO-guarded PointPillars.]{
\begin{minipage}[t]{0.49\linewidth}
\centering
\includegraphics[width=4cm]{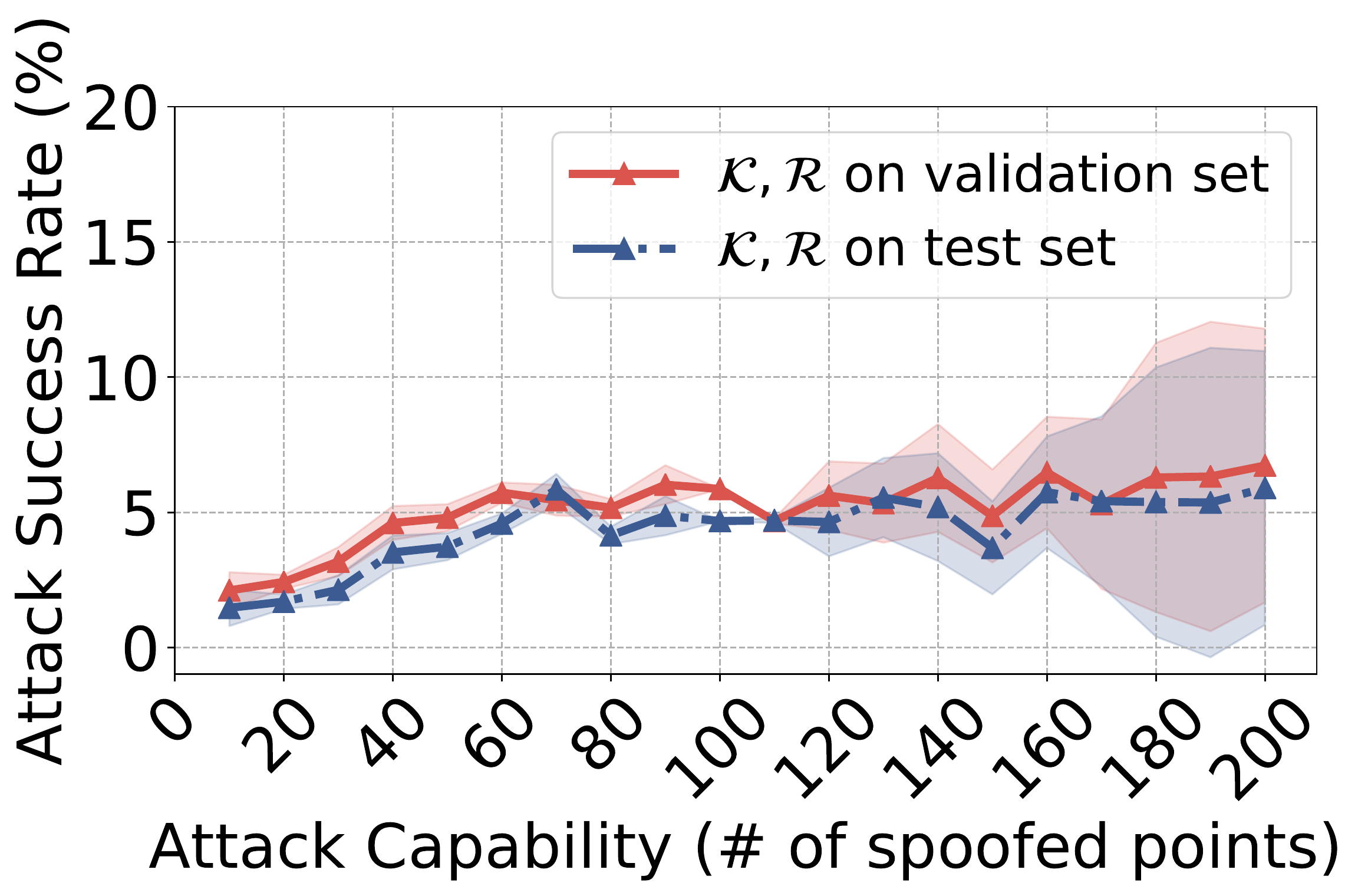}
\end{minipage}%
}%
\vskip 0pt
\vspace{-0.3cm}
\subfigure[CARLO-guarded PointRCNN.]{
\begin{minipage}[t]{0.49\linewidth}
\centering
\includegraphics[width=4cm]{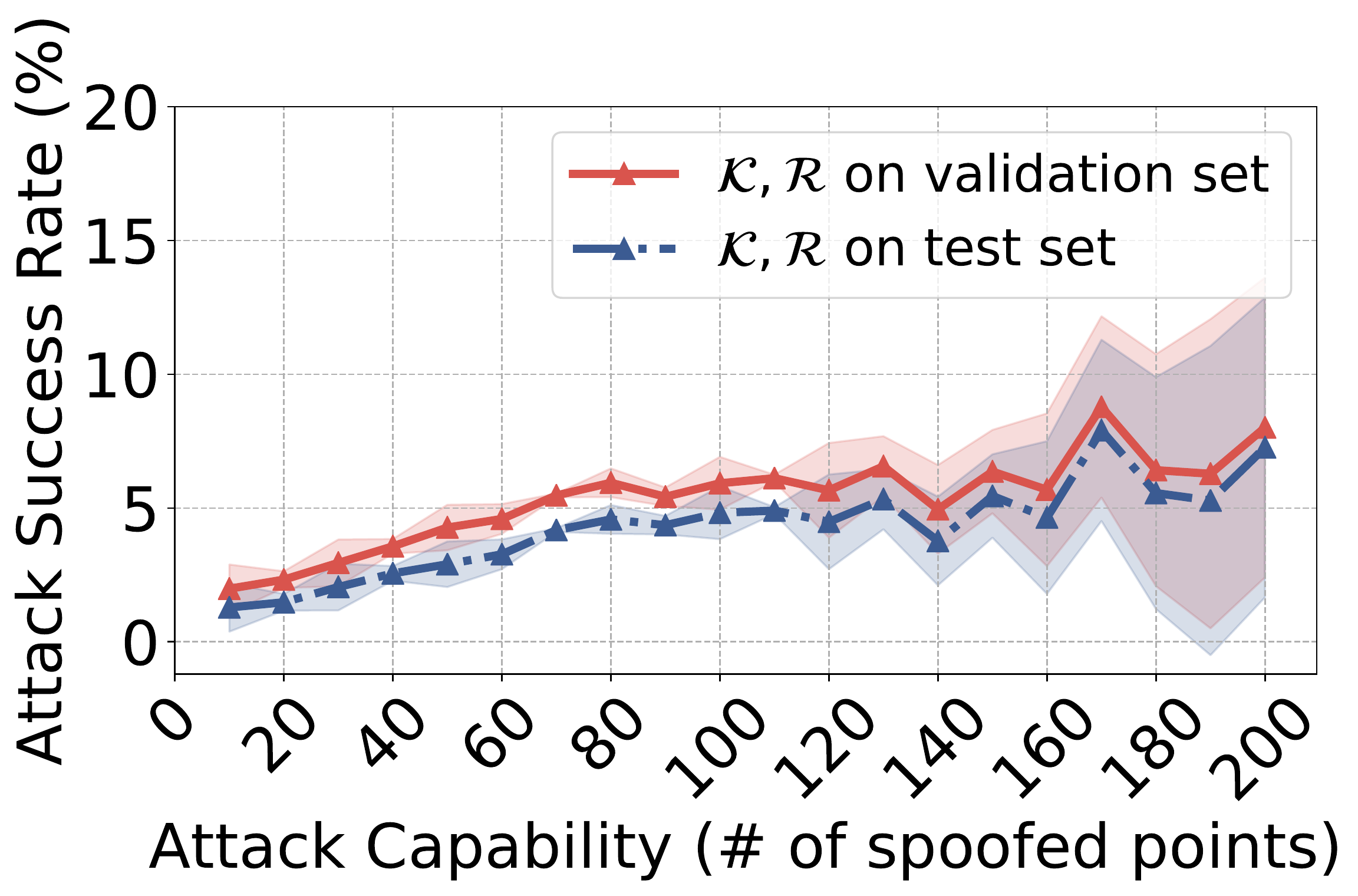}
\end{minipage}
}%
\subfigure[Precision and recall of CARLO.]{
\begin{minipage}[t]{0.49\linewidth}
\centering
\includegraphics[width=4cm,height=2.4cm]{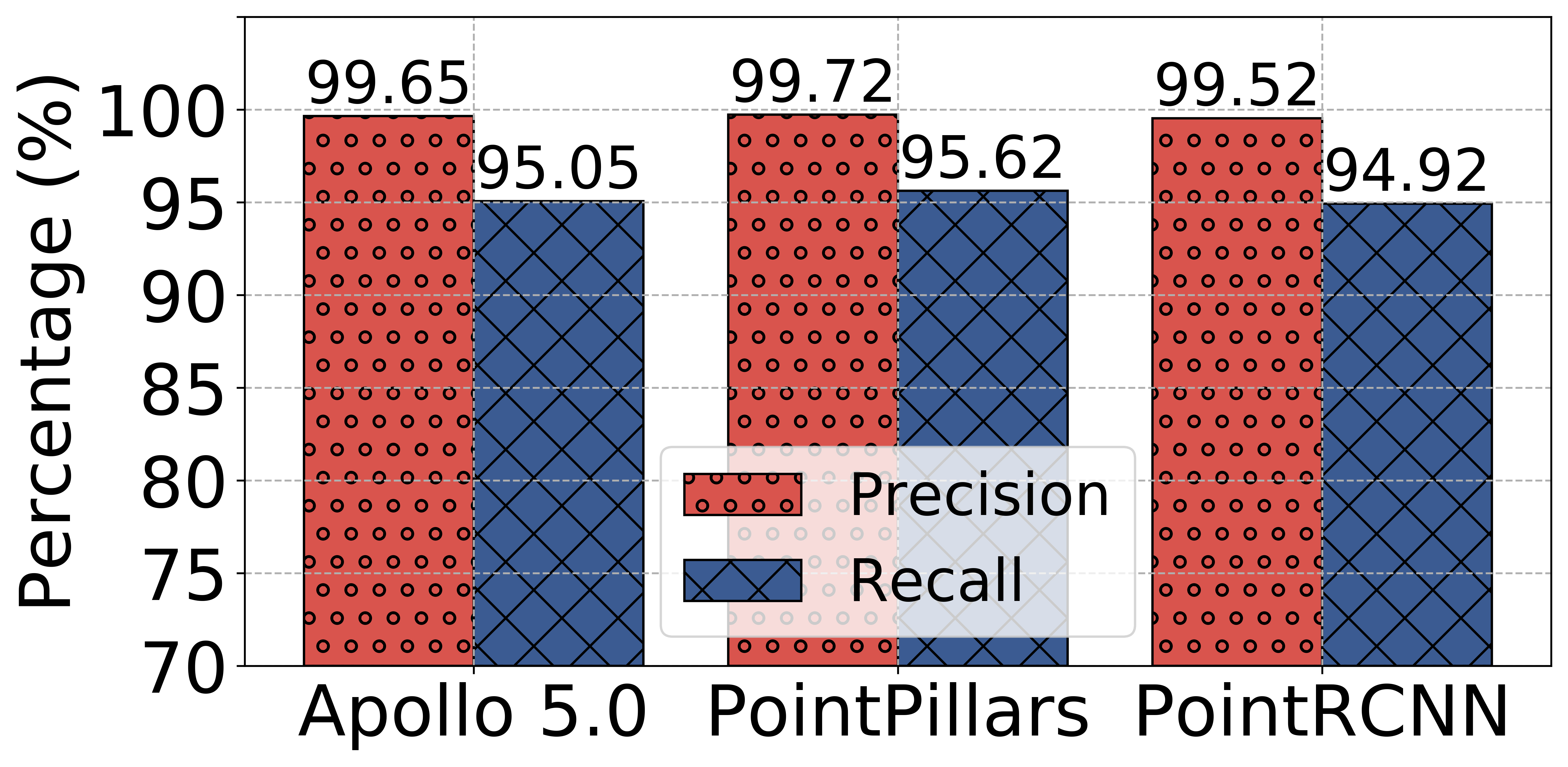}
\end{minipage}
}%

\centering
\vspace{-0.2cm}
\caption{Attack success rates ($ASR$s) of proposed black-box spoofing attacks on three CARLO-guarded models.}
\label{fig:asr_rhd}
\vspace{-0.5cm}
\end{figure}

\textbf{Evaluation metrics.} We evaluate the performance of CARLO in two aspects, which are the $ASR$ of the CARLO-guarded models, and the precision and recall of CARLO itself. $ASR$ directly relates to the defense performance, while the precision and recall of CARLO reflect whether it will harm the original AP of target models. We test the $ASR$ on the validation set and test set since the distributions are estimated from the training set. We only evaluate the precision and recall of CARLO on the validation set as we do not have the ground-truth for the test set.

Figure \ref{fig:asr_rhd} (a-c) shows the $ASR$ of three CARLO-guarded models. As shown, CARLO reduces the $ASR$ from more than 95\% to below 9.5\% with the maximum attack capability, and reduce the mean $ASR$ to around 5.5\%. \CR{We observe that the remaining 5.5\% comes from the detection errors (\ie the detected bounding box of the fake vehicle cannot match well with the ground-truth) that shift the $f'$ and $g'$ to the distribution of valid vehicles.} The errors occur randomly in the point clouds so that it is hard for adversaries to utilize. The recall in Figure \ref{fig:asr_rhd} (d) reaches around 95\% in all targets models which also validate the results in Figure \ref{fig:asr_rhd} (a-c).

\begin{table}[t]
    \captionsetup{font=small}
    \footnotesize
    \caption{PointPillars' and PointRCNN's APs (\%) of 3D car detection on the KITTI validation set. ``Mod.'' refers to the Moderate category introduced in \S\ref{background_kitti}; ``Original'' refers to the original performance of two models; ``Attack'' refers to the performance after spoofing attacks; ``CARLO'' refers to the performance after CARLO applied.}
    \vspace{-0.1in}
    \centering
    \begin{tabular}{|c|c|c|c|c|c|c|}
    \hline
    \multirow{2}*{Model} &  \multicolumn{3}{ c |}{PointPillars}  &  \multicolumn{3}{ c |}{PointRCNN} \\
    \cline{2-7}
     &  Easy & Mod. & Hard  &  Easy & Mod. & Hard \\
    \hline
    \hline
    Original & 86.56 & 76.87 & 72.09 &  88.80 & 78.58 & 77.64 \\
    \hline
    Attack &  74.06 & 56.69 & 53.98 & 84.51 & 71.17 & 68.06 \\
    \hline
    CARLO &  86.57 & 78.60 & 73.55 & 88.91 & 78.61 & 77.63 \\
    \hline
    \end{tabular}
    \label{tb:model_og_performance}
    \vspace{-0.5cm}
\end{table}

Besides delivering satisfactory defense performance, CARLO barely introduces misdetections (\ie false negatives) to the models. Figure \ref{fig:asr_rhd} (d) shows that the precision of anomaly detection reaches at least 99.5\% for all target models. We manually verify the 0.5\% misdetections, and find they are all vehicles at least 40 meters away from the AV, which will not affect its immediate driving behavior. Table \ref{tb:model_og_performance} also shows that the AP will slightly increase after CARLO being applied to the original model because the original model has internal false positives such as detecting a flower bed as a vehicle. CARLO detects some of those false positives generated by the original model.


\nsubsubsection{Defense against White-box and Adaptive Attacks}

\begin{figure}[t]
  \vspace{-0.3cm}
  \begin{minipage}[t]{0.49\linewidth}
    \centering 
    \includegraphics[width=\linewidth]{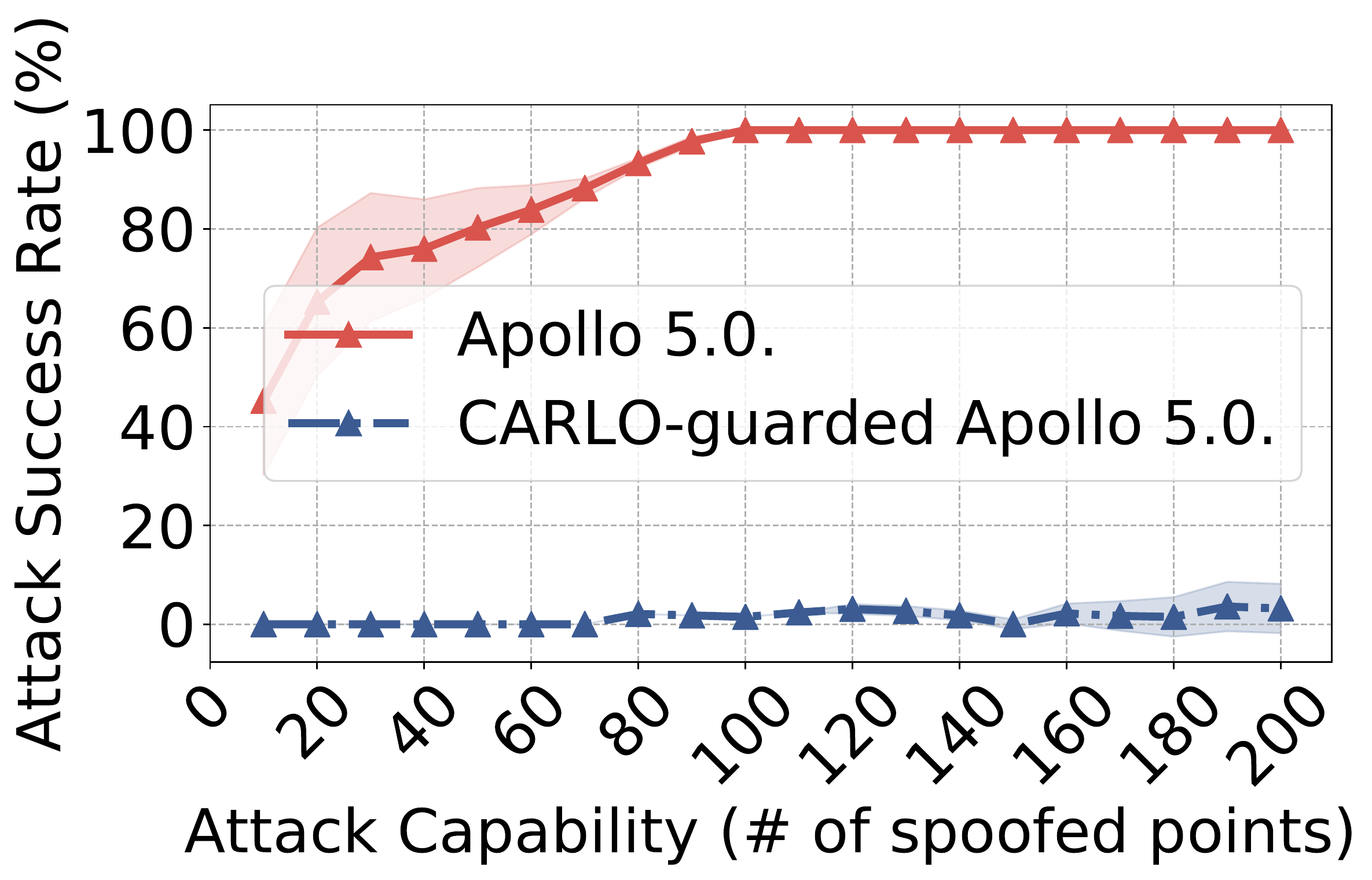} 
    \caption{Attack success rates ($ASR$s) of Adv-LiDAR on Apollo 5.0 and CARLO-guarded model.} 
    \label{fig:white_box_rhd}
  \end{minipage}%
  \hfill
  \begin{minipage}[t]{0.48\linewidth} 
    \centering 
    \includegraphics[width=\linewidth]{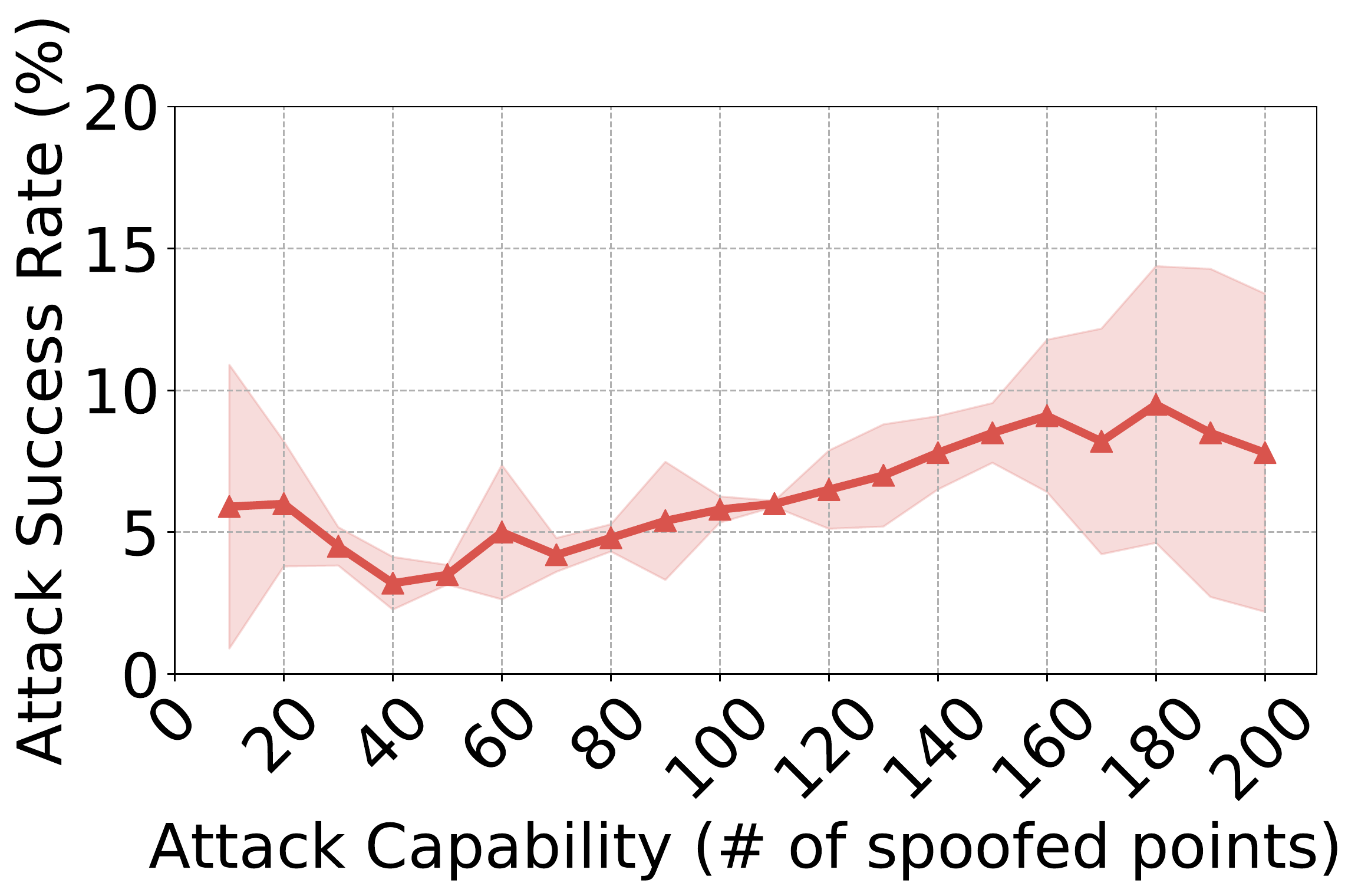} 
    \caption{Attack success rate ($ASR$) of the adaptive attack on CARLO-guarded Apollo 5.0.} 
    \label{fig:adaptive_rhd}
  \end{minipage} 
  \vspace{-0.5cm}
\end{figure}

To further evaluate CARLO against white-box attacks, we first leverage Adv-LiDAR to generate adversarial examples that fool Apollo 5.0 and test whether they can succeed in attacking CARLO-guarded Apollo 5.0. Figure \ref{fig:white_box_rhd} demonstrates that CARLO can effectively defend Adv-LiDAR, where the $ASR$ drops from more than 95\% to below 5\% consistently. We observe that the defense effects are better than the results shown in Figure~\ref{fig:asr_rhd} (a). We find out that Adv-LiDAR tends to translate the attack traces to a slightly higher place along $z$ axis. Such translations will isolate the adversarial examples in the point cloud, making themselves easier to be detected by CARLO.

We also try our best efforts to evaluate CARLO on the adaptive attacks. We assume attackers are aware of the CARLO pipelines and utilize Adv-LiDAR to break CARLO's defense. Due to the sensor attack capability, adversaries have limited ability to modify the absolute free space ($\sum_{\,c \in \mathsf{B}}\mathds{1} \cdot FS(c)$) in Equation \ref{eq:fsd}. However, attackers can try to shrink the volume of the bounding box ($|\mathsf{B}|$) to shift the distribution of $f'$, since it is controlled purely by models. Therefore, the attack goal is to spoof a vehicle at target locations at the same time, minimize the size of the bounding box. We formulate the loss function and follow Adv-LiDAR~\cite{cao2019adversarial} to utilize a global transformation matrix $H(\theta,\tau)$ (Equation~\ref{eq:trans}) for solving the optimization problem:
\CR{
\begin{equation}
\begin{split}
    & \min_{\theta,\tau} \qquad \gL (x \oplus  V \cdot H(\theta,\tau)^{T}) + \lambda \cdot \gV_{B} ( V \cdot H(\theta,\tau)^{T})
\end{split}
\label{eq:adaptive_1}
\end{equation}
where $\gL(\cdot)$ is the loss function defined in Adv-LiDAR~\cite{cao2019adversarial}, $\gV_{B}(\cdot)$ is the volume of the target bounding box $B$, and $\lambda$ is a hyper-parameter.} Figure~\ref{fig:adaptive_rhd} shows that such adaptive attacks cannot break CARLO, either. We attribute the reason to $H(\theta,\tau)$ that holistically modifies the spoofed points so that it can barely change the size of the bounding box.

\nsection{Physics-Embedded Perception Architecture}
\label{ml_defense}

In this section, we take a step further to explore the feasibility of embedding physical features into end-to-end learning that provides better robustness for AD systems. We find that, despite BEV or 3D representations, which are used by most models, the front view (FV) is a better representation for learning occlusion features by nature. However, prior works adopting FV are still vulnerable to the proposed attacks due to their model architecture designs' fundamental limitations. To improve the design and further enforce the learning of occlusion features, we propose sequential view fusion (SVF), a general architecture for robust LiDAR-based perception.




\nsubsection{Why should FV Representations help?}
\label{ml_defense_existing}

We observe that LiDAR natively measures range data (\S\ref{background_spoof}). Thus, projecting the LiDAR point cloud into the perspective of the LiDAR sensor will naturally preserve the physical features of LiDAR. Such projecting is also known as the FV of LiDAR point clouds~\cite{li2016vehicle}. Given a 3D point $\vec{p}=(x,y,z)$, we can compute its coordinates in FV $\vec{p}_{FV}=(r,c)$ by:
\begin{equation}
\begin{split}
    c & = \lfloor \arctan(y,x)/\Delta\theta\rfloor \\
    r & = \lfloor \arctan(z,\sqrt{x^2+y^2})/\Delta\phi\rfloor
\end{split}
\label{eq:mapping}
\end{equation}
where $\Delta \theta$ and $\Delta \phi$ are the horizontal and vertical fire angle intervals (\S\ref{background_spoof}). As shown in Figure \ref{fig:c1-3}, since the \emph{occluder} $\mathsf{O}(v)$ and \emph{occludee} $V$ neighbor with each other in the FV, deep learning models have opportunities to identify the inter-occlusion. The abnormal sparseness of a fake vehicle will also be exposed, as valid vehicles' points are clustered, while the spoofed points scatter in the FV (\S\ref{vulnerability_id}). Therefore, the FV representation of point clouds embeds both ignored occlusion patterns. 

Although prior works have utilized FV for object detection, little is known about its robustness to LiDAR spoofing attacks. LaserNet~\cite{meyer2019lasernet} is the latest model that only takes the FV representation of point clouds as input for 3D object detection. However, LaserNet cannot achieve state-of-the-art performance compared to models in the three classes introduced in \S\ref{background}. Other studies~\cite{li2016vehicle} also confirm that only by leveraging the FV representation, models cannot provide satisfactory detection results. The failure of FV-based models comes from the scale variation of objects as well as occlusions between objects in a cluttered scene~\cite{zhou2019end}. 
\begin{figure}[t!]
\centering
  \vspace{-0.3cm}
  \includegraphics[width=\linewidth]{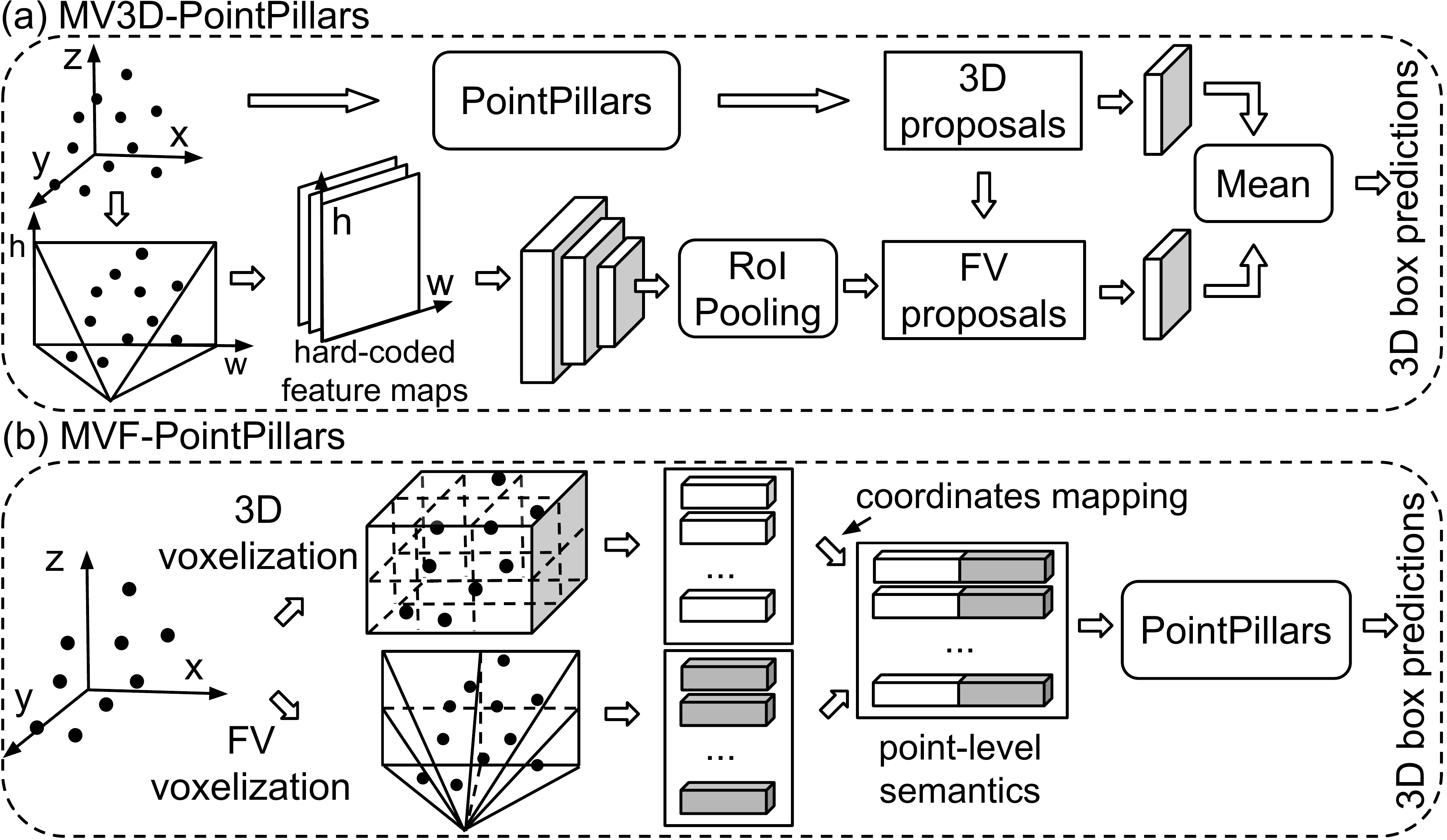}
  \caption{Existing view fusion-based architectures.}
  \label{fig:existing_fusion}
  \vspace{-0.6cm}
\end{figure}


Besides the models that only take FV as input, several studies~\cite{chen2017multi,zhou2019end,ku2018joint}  present fusion-based architectures for LiDAR-based perception that utilize the combinations of data from different sensors and views as input. MV3D~\cite{chen2017multi} is a classic fusion-based design that takes both LiDAR point clouds and RGB images as input and predicts 3D bounding boxes, where the point cloud is projected into multi-views (\ie FV and BEV) for feature encoding. Zhou \etal recently proposed multi-view fusion (MVF), which combines FV with 3D representations~\cite{zhou2019end}. MVF builds on top of PointPillars. Instead of only voxelizing points in 3D, MVF also voxelizes the point cloud into FV frustums and integrates the two voxels' features based on coordination mapping in the 3D space. 


To better understand the robustness of fusion-based architectures, we reproduce MV3D and MVF based on PointPillars. For MV3D, we ignore the RGB images, and take the FV and BEV as the model input since we focus on LiDAR-based perception. We use a VGG-16~\cite{Simonyan15} for FV feature learning in MV3D. Figure \ref{fig:existing_fusion} shows the architectures we adopt and reproduce. We train the two reproduced models on the KITTI training set and evaluate them on the KITTI validation set. As Table~\ref{tb:fusion_performance} shows, the FV-augmented models can achieve comparable performance than the original PointPillars. The reproduced results also align well with the evaluations in~\cite{zhou2019end,chen2017multi}. 

\begin{table}[thp!]
\captionsetup{font=small}
\vspace{-0.25cm}
    \footnotesize
    \caption{MV3D-PointPillars' and MVF-PointPillars' APs (\%) of 3D car detection on the KITTI validation set.}
    \vspace{-0.1in}
    \centering
    \begin{tabular}{|c|c|c|c|}
    \hline
    \multirow{2}*{Model} &  \multicolumn{3}{ c |}{Car Detection} \\
    \cline{2-4}
     &  Easy & Moderate & Hard \\
    \hline
    \hline
    MV3D-PointPillars & 85.67 & 77.12 & 71.65 \\
    \hline
    MVF-PointPillars &  86.77 & 79.15 & 75.72 \\
    \hline
    \end{tabular}
    \label{tb:fusion_performance}
    \vspace{-0.3cm}
\end{table}

\begin{figure}[tp]
\vspace{-0.3cm}
\centering

\subfigure[$ASR$ of the reproduced MV3D.]{
\begin{minipage}[t]{0.5\linewidth}
\centering
\includegraphics[width=0.95\linewidth]{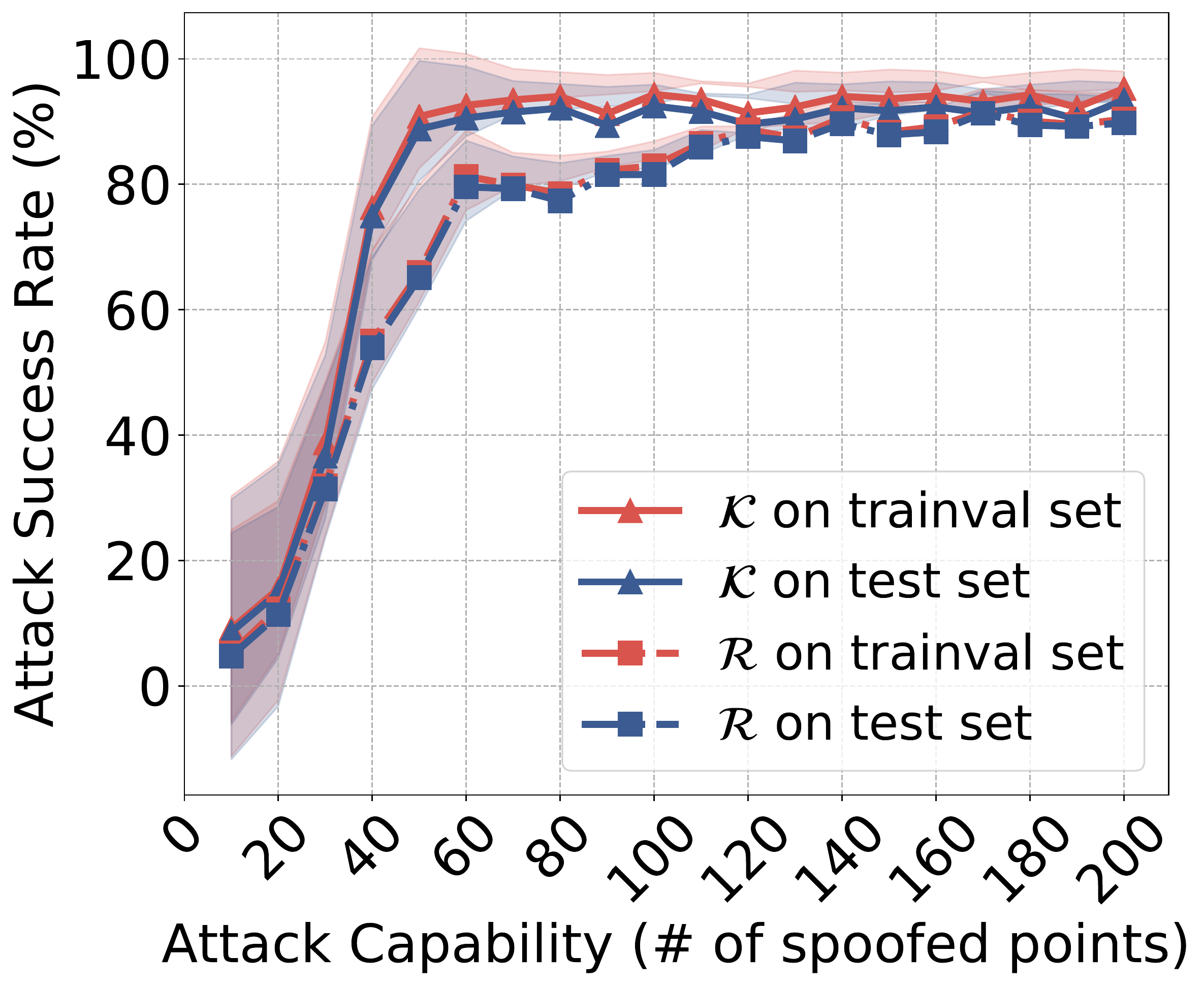}
\end{minipage}%
}%
\subfigure[$ASR$ of the reproduced MVF.]{
\begin{minipage}[t]{0.5\linewidth}
\centering
\includegraphics[width=0.95\linewidth]{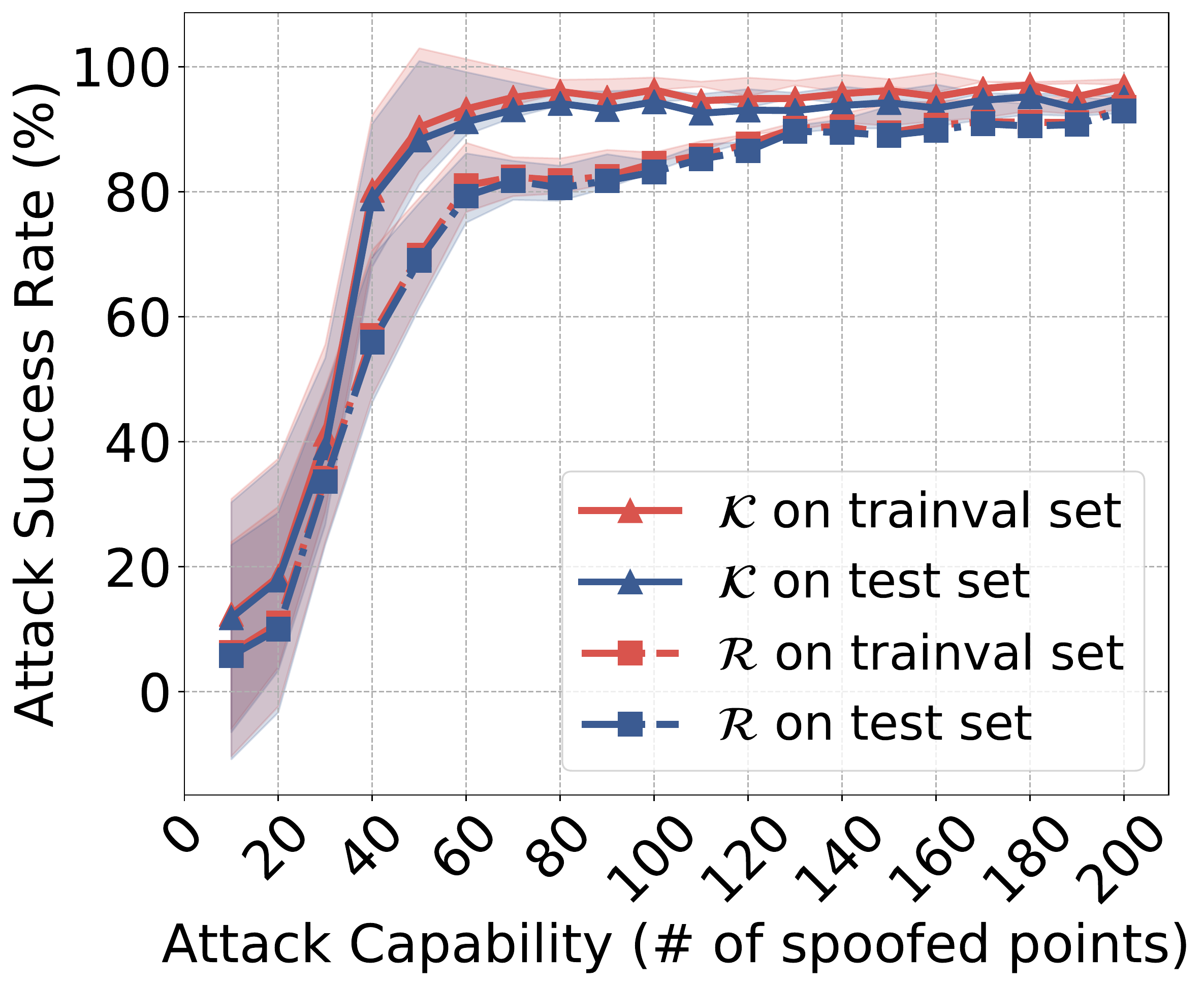}
\end{minipage}%
}%
\centering
\vspace{-0.2cm}
\vspace{-0.1in}
\caption{Attack success rates ($ASR$s) of proposed black-box spoofing attacks on MV3D and MVF models.}
\label{fig:asr_fusion}
\vspace{-0.6cm}
\end{figure}

We then evaluate their robustness against our proposed black-box attack. The experimental setups are identical to the settings in \S\ref{attack_analysis}. Figure \ref{fig:asr_fusion} shows that the $ASR$ of reproduced models are as high as the original PointPillars. It indicates that existing view fusion-based architectures both cannot help with defending against LiDAR spoofing attacks, although they provide marginally gain on the AP. We further perform ablation studies and find that the BEV (or 3D) features dominate the model decisions (elaborated in Appendix~\ref{ap:ablation}). Since the identified vulnerability exists in 3D space, the models are still vulnerable to LiDAR spoofing attacks.


\nsubsection{Sequential View Fusion}
\label{svf}

The insights drawn from existing view fusion schemes show that existing fusion designs cannot provide better robustness compared to the original models. The 3D (or BEV) representation dominates the model leaving the FV representation not critical in the end-to-end architectures.  

\begin{figure}[tp]
\vspace{-0.2cm}
\centering
  \includegraphics[width=\linewidth]{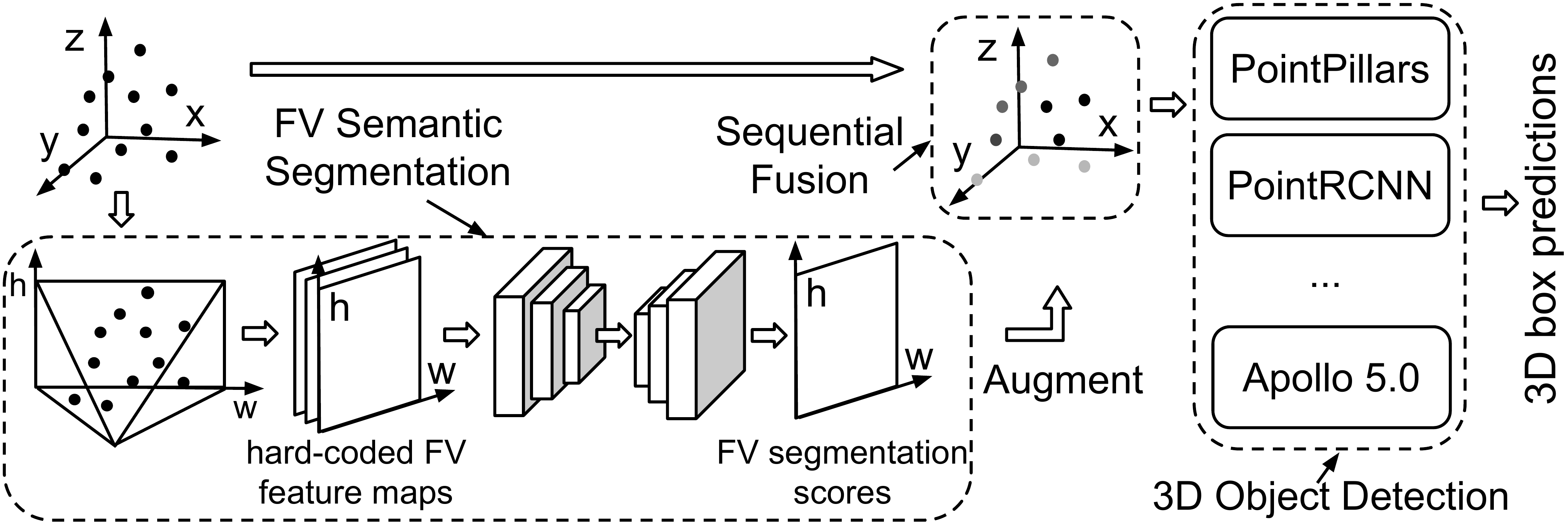}
  \vspace{-0.25in}
  \caption{Sequential view fusion (SVF) architecture.}
  \label{fig:svf}
  \vspace{-0.5cm}
\end{figure}

Based on the above understandings, we propose a new view fusion schema called sequential view fusion (SVF). SVF comprises of three modules (Figure \ref{fig:svf}), which are: 1) semantic segmentation: a semantic segmentation network that utilizes the FV representation to computes the point-wise confidence scores (\ie the probability that one point belongs to a vehicle). 2) view fusion: the 3D representation is augmented with semantic segmentation scores. 3) 3D object detection: a LiDAR-based object detection network that takes the augmented point clouds to predict bounding boxes. Instead of leaving the models to learn the importance of different representations by themselves, we attach a semantic segmentation network to the raw FV data. By doing so, we enforce the end-to-end learning to appreciate the FV features, so that the trained model will be resilient to LiDAR spoofing attacks.




\textbf{Semantic segmentation.} The semantic segmentation networks accept the FV represented point clouds and associate each point in FV with a probability score that it belongs to a vehicle. These scores provide aggregated information on the FV representation. Semantic segmentation over FV has several strengths. First, as mentioned before, the FV representation is noisy because of the nature of LiDAR. Compared to 3D object detection or instance segmentation, which is intractable over FV, semantic segmentation is an easier task as it does not need to estimate object-level output. Second, there are extensive studies on semantic segmentation over FV represented point clouds~\cite{wu2018squeezeseg,wang2018pointseg,biasutti2019lu}, and the segmentation networks achieve much more satisfactory results than the 3D object detection task over FV. 

In our implementation, we adopt the high-level design in LU-Net~\cite{biasutti2019lu}. It is worth noting that the end-to-end SVF architecture is agnostic to the semantic segmentation module. 

\textbf{View fusion.} The fusion module re-architects existing symmetric designs which integrate the 3D representation with the confidence scores generated by the semantic segmentation module. Specifically, we use Equation \ref{eq:mapping} for mapping between $\vec{p}=(x,y,z)$ and $\vec{p}_{FV}(r,c)$, and augment each $\vec{p}$ with the point-wise confidence score from its corresponding $\vec{p}_{FV}$. 

\textbf{3D object detection.} SVF is also agnostic to the 3D object detection module. In this paper, we utilize PointPillars and PointRCNN in our implementation. Most of the models introduced in \S\ref{background} can fit into the end-to-end SVF architecture.

\nsubsection{SVF Evaluation}
\label{svf_eval}

\textbf{Experimental setup.} We train SVF-PointPillars and SVF-PointRCNN on the KITTI training set. The setup of robustness analysis against LiDAR spoofing attacks is identical to the settings in \S\ref{attack_analysis}. We also try to evaluate SVF against Adv-LiDAR~\cite{cao2019adversarial} on Apollo 5.0 and the adaptive attacks.
 
\textbf{Evaluation metrics.} We evaluate the AP of SVF-PointPillars and SVF-PointRCNN on the KITTI validation set, and leverage $ASR$ to test their robustness against LiDAR spoofing attacks.

As shown in Table \ref{tb:svf_performance}, both SVF models achieve comparable AP compared to the original models. The marginal degradation comes from two-state training. More specifically, the distributions of the semantic segmentation outputs in the training and validation sets do not align well with each other. \CR{We find that the drop of AP will indeed cause a tiny amount of false negatives but will not influence the driving behaviors. Moreover, such degradation could be compensated with better training strategies (\eg finer-tuning of the parameters) since the capacity of SVF is larger than the original models.}

\begin{table}[!h]
    \captionsetup{font=small}
    \footnotesize
    \caption{SVF-PointPillars' and SVF-PointRCNN's APs (\%) of 3D car detection on the KITTI validation set.}
    \vspace{-0.1in}
    \centering
    \begin{tabular}{|c|c|c|c|}
    \hline
    \multirow{2}*{Model} &  \multicolumn{3}{ c |}{Car Detection} \\
    \cline{2-4}
     &  Easy & Moderate & Hard \\
    \hline
    \hline
    SVF-PointPillars & 85.93 & 74.12 & 70.19 \\
    \hline
    SVF-PointRCNN &  88.12 & 76.56 & 74.81 \\
    \hline
    \end{tabular}
    \label{tb:svf_performance}
    \vspace{-0.3cm}
\end{table}

\begin{figure}[tp]
\vspace{-0.3cm}
\centering
\subfigure[$ASR$ of SVF-PointPillars.]{
\begin{minipage}[t]{0.5\linewidth}
\centering
\includegraphics[width=0.95\linewidth]{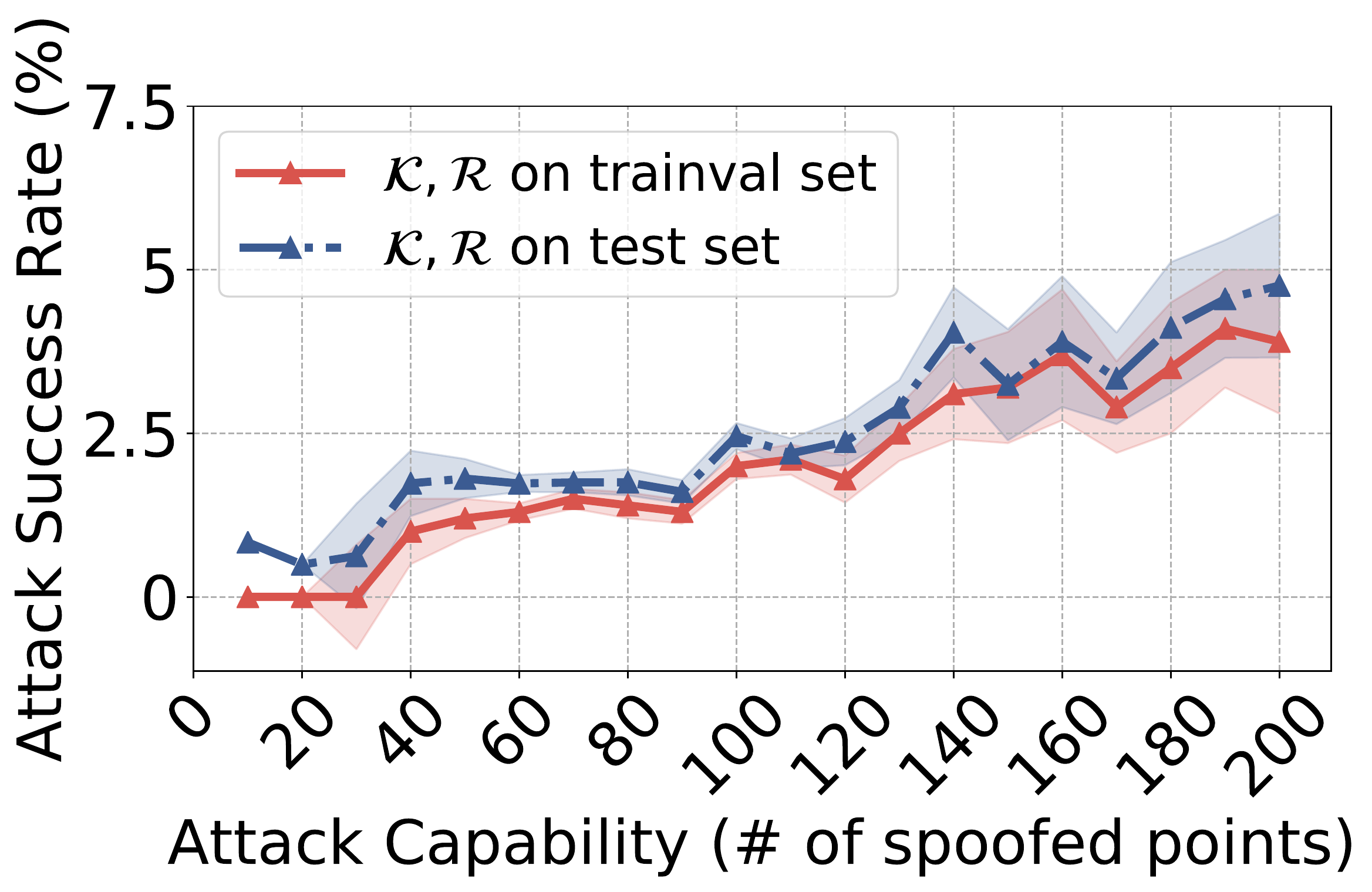}
\end{minipage}%
}%
\subfigure[$ASR$ of SVF-PointRCNN.]{
\begin{minipage}[t]{0.5\linewidth}
\centering
\includegraphics[width=0.95\linewidth]{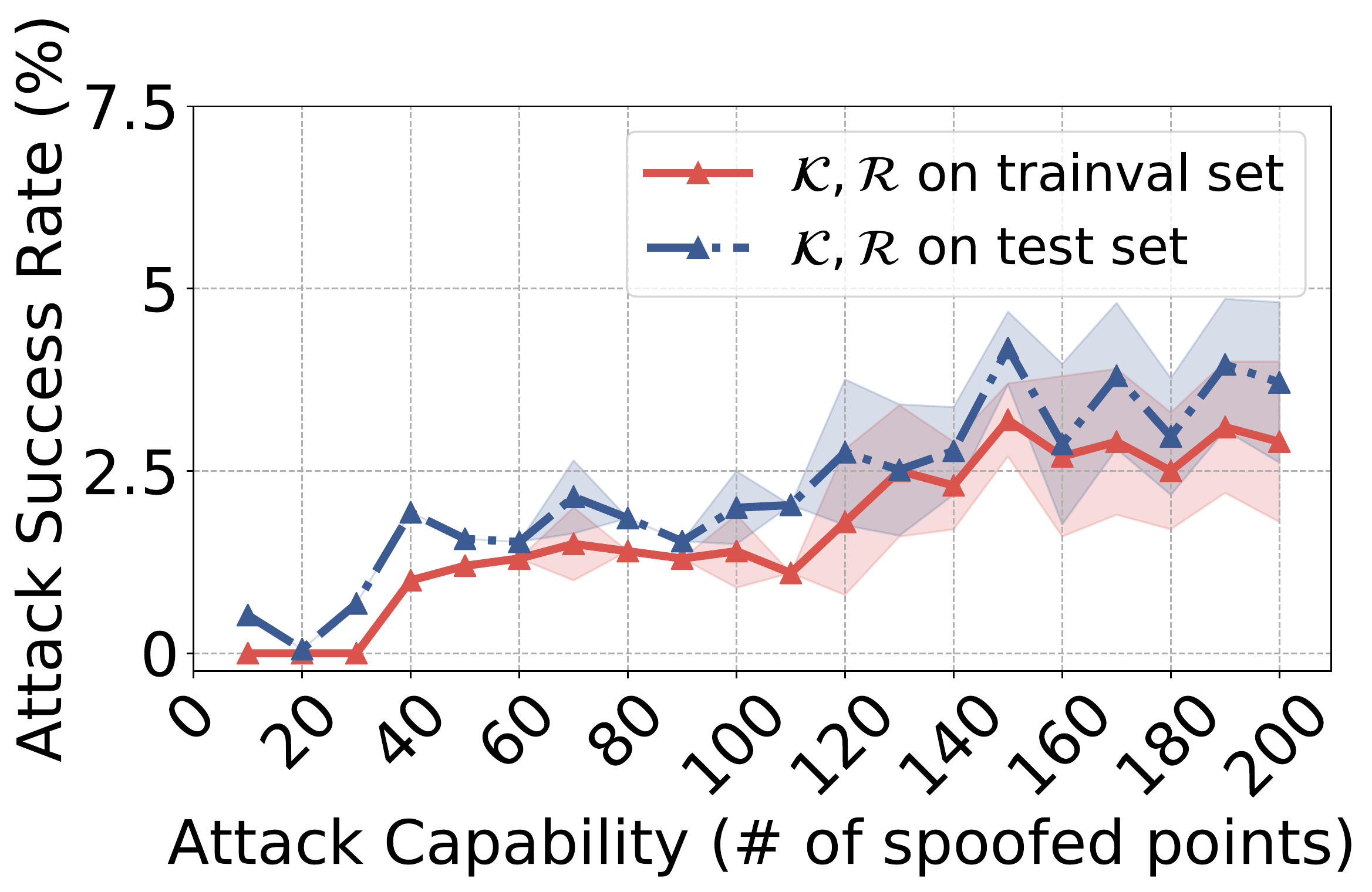}
\end{minipage}%
}%
\centering
\vspace{-0.4cm}
\caption{Attack success rates ($ASR$s) of proposed black-box spoofing attack on SVF models.}
\label{fig:asr_svf}
\vspace{-0.5cm}
\end{figure}

We then perform the robustness evaluation of SVF models. Figure \ref{fig:asr_svf} shows the $ASR$ of our proposed spoofing attacks. As shown, the attacks are no longer effective in SVF models. The $ASR$ reduces from more than 95\% (original models) to less than 4.5\% on both models with the maximum attack capability, which is also an around 2.2$\times$ improvement compared to CARLO-guarded models. The mean $ASR$ also drops from 80\% to around 2.3\%. We also perform ablation study on SVF, and demonstrate that the FV features are more important in SVF models (detailed in Appendix~\ref{ap:ablation}).

\nsubsubsection{Defense against White-box and Adaptive Attacks}
\label{ml_white_box}

\begin{figure}[!t]
 \vspace{-0.3cm}
  \begin{minipage}[t]{0.48\linewidth}
    \centering 
    \includegraphics[width=\linewidth]{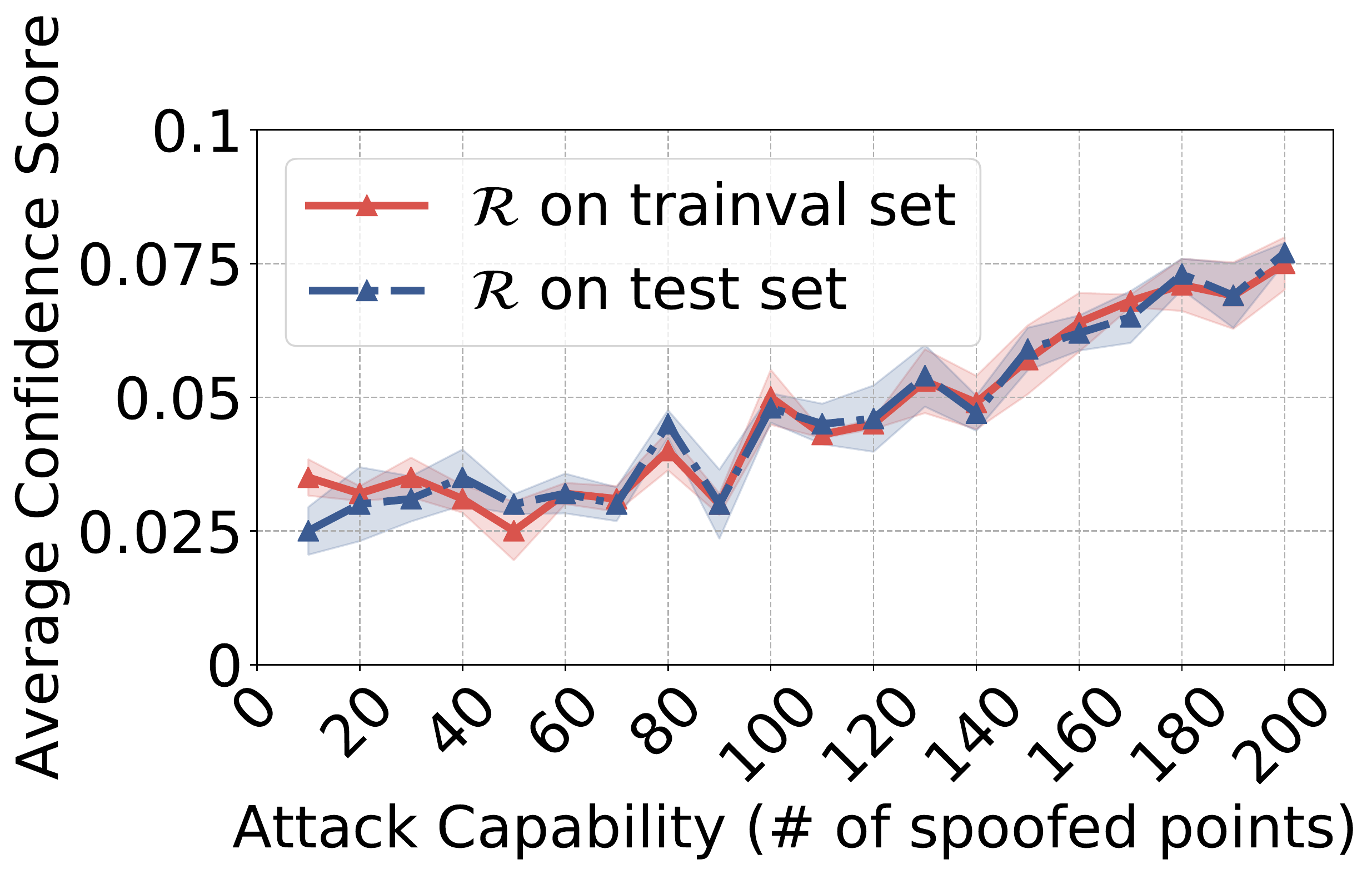} 
    \vspace{-0.6cm}
    \caption{Average confidence score of Adv-LiDAR on the segmentation network.} 
    \label{fig:white_box_svf}
  \end{minipage}%
  \hfill
  \begin{minipage}[t]{0.48\linewidth} 
    \centering 
    \includegraphics[width=\linewidth]{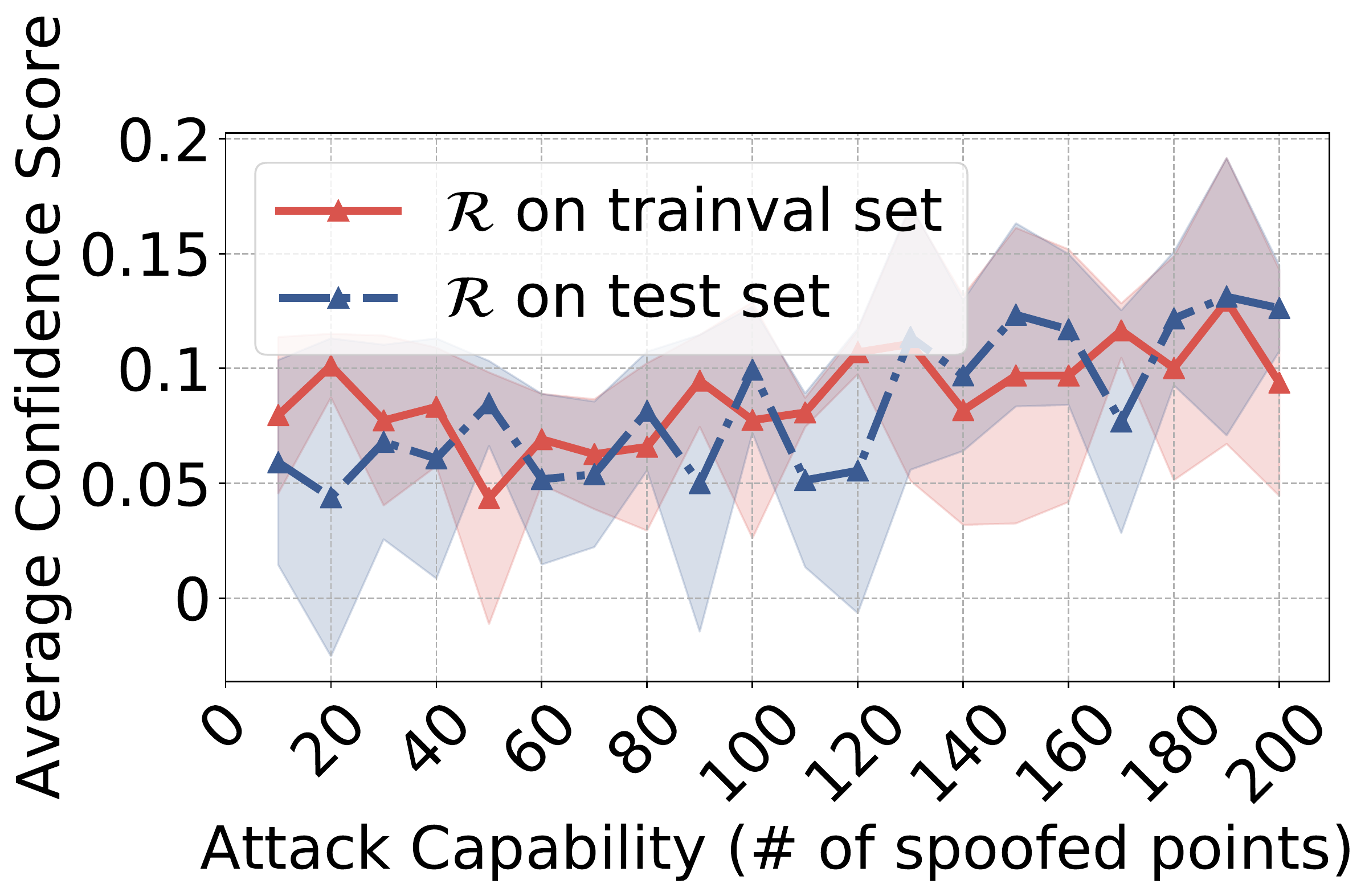} 
    \vspace{-0.6cm}
    \caption{Average confidence score of the adaptive attack on the segmentation network.}
    \label{fig:adaptive_svf}
  \end{minipage} 
  \vspace{-0.6cm}
\end{figure}

Since SVF requires re-training for the model, we cannot directly evaluate Adv-LiDAR on SVF-Apollo (\S\ref{background_kitti}). As a result, we decouple the problem to whether Adv-LiDAR can fool both the semantic segmentation and 3D object detection modules. We first directly apply the attack traces that successfully fool Apollo 5.0 to the segmentation network and record the mean confidence score of all the points belonging to the attack trace. Figure \ref{fig:white_box_svf} shows that the mean confidence scores are consistently below 0.08, which is too low to be classified as a valid vehicle with mean confidence scores of around 0.73 in our trained model.


Model-level defenses are usually vulnerable to simple adaptive attacks~\cite{athalye2018obfuscated,carlini2017adversarial}. To demonstrate the effectiveness of SVF against adaptive attack, we assume that the adversaries are aware of the SVF architecture. The attack goal is to both fool the semantic segmentation and 3D object detection modules. We also leverage the formulation in \cite{cao2019adversarial} to utilize the global transformation matrix $H(\theta,\tau)$ to control the spoofed points. 
\CR{
\begin{equation}
\begin{split}
    & \min_{\theta,\tau} \qquad - \gL_{seg} (x \odot  V \cdot H(\theta,\tau)^{T}) \qquad
\end{split}
\label{eq:adaptive_2}
\end{equation}
where $\odot$ represents the point cloud merge in the front view and $\gL_{seg}(\cdot)$ defines the average confidence score of the attack trace (\ie $ V \cdot H(\theta,\tau)^{T}$).} Figure \ref{fig:adaptive_svf} shows that none of the attack traces' average confidence score reaches 0.2 in the segmentation module, which is still far from the mean average confidence score of valid vehicle 0.73. Therefore, the adaptive attacks also cannot break the robustness of SVF.


\nsection{Discussion and Future Work}
\CR{In this section, we discuss the distinct features of our proposed black-box attack along with its practicality and completeness. We further discuss the comparisons between the presented defense strategies and their limitations, accordingly.}

\nsubsection{Attack Discussion}
\vspace{0.25cm}

\nsubsubsection{Comparison with Physical Adversarial Attacks}
\CR{First, we distinguish the spoofing attacks on LiDAR with extensive prior work on physical-world adversarial machine learning attacks in mainly three aspects:}

\CR{\textit{1. Different perturbation methods.} Images and point clouds have different data structures, which further lead to different perturbation methods applied. Images have compact and ordered structures. In contrast, point clouds are irregular, represented as $N \times C$, where $N$ is the number of points, and $C$ contains the location and intensity information (\ie~$xyz$-$i$)~\cite{kitti_3d}. Attackers are able to generate adversarial examples by modifying the RGB values in images. For attacks on LiDAR, however, attackers can directly shift the point in the 3D Euclidean space as long as it obeys the physics of LiDAR.}

\CR{\textit{2. Different perturbation capabilities.} Prior attacks on 2D images treat the whole target area as the attack surface since the threat model assumes that attackers have full controls over the target object (\eg attackers can potentially modify any area of a stop sign in~\cite{eykholt2018robust}). However, due to the characteristics of LiDAR spoofing attacks, the attack surface is limited by the sensor attack capability ($\mathcal{A}$) in \S\ref{threat}. Such a small attack surface also introduces difficulties in launching the attack.}

\CR{\textit{3. Different perturbation constraints.} Prior attacks on 2D images leverage $L_p$ norms as the main constraints for the formulated optimization problem~\cite{eykholt2018robust} whose goal is to minimize the perturbation to be stealthy. Such constraints do not apply to attack LiDAR because point clouds are not perceived by humans. Thus, stealthiness is not a focus in such attacks. Instead, the optimized attack traces must not exceed the sensor attack capability ($\mathcal{A}$) boundary, in which case $\mathcal{A}$ naturally becomes the primary constraint for attacking LiDAR.}

\CR{Second, the high-level methodology of our proposed attack is similar to replay attacks, in which adversaries playback the intercepted data to deceive target systems~\cite{replay}. However, different from existing replay attacks~\cite{nassi2020phantom} that retransmit \textit{logically correct} data to launch attacks, the limited sensor attack capability ($\mathcal{A}$) cannot support the injection of a \textit{physically valid} vehicle's trace into the LiDAR point cloud~\cite{cao2019adversarial}. Thus, the success of our black-box attack indeed relies on the identified vulnerability.}

\vspace{-0.2cm}
\nsubsubsection{Attack Practicality and Completeness}
\label{practi}

\CR{One major limitation of our proposed attack is that the presented results cannot directly demonstrate attack practicality in the physical world. First, due to the limitations of our delay component (\ie the function generator in our implementation), we can only control spoofed points at 10cm-level precision. Therefore, we only construct two fine-controlled physical attack traces for a proof-of-concept demonstration. The two attack traces contain 140 and 47 points. We evaluated them on the KITTI trainval set, and they achieve 87.68\%, 98.12\%, and 74.91\% $ASR$s on Apollo 5.0, PointPillars, and PointRCNN, respectively. Second, launching our black-box attack on a real AV requires accurate aiming of attack lasers at a target LiDAR, which is challenging to perform without real-world road tests and precision instruments~\cite{cao2019adversarial}. Since this paper aims to explore and expose the underlying vulnerability, we leave real-world testing as future work.}

\CR{Although we demonstrate that our proposed black-box attack achieves high attack success rates, the identified vulnerability does not provide \textit{completeness}. This means that there may exist other potential vulnerabilities hidden in the AD systems to be discovered and exploited. Future research may include verification of the AD models and comprehensive empirical studies to explore the underlying vulnerabilities.}


\vspace{-0.1cm}
\nsubsection{Defense Discussion}
\textbf{CARLO vs. SVF.} Both CARLO and SVF achieve satisfactory defense performance while maintaining comparable AP with the original model. In addition, both of them are model-agnostic so that they can be incorporated into most existing LiDAR-based perception systems. CARLO is a practical post-detection module. It does not require re-training the model, which can be quite labor-intensive. CARLO is also realistic because it does not assume that users have white-box access to the model. SVF, on the other hand, is a general architecture for ensuring robust LiDAR-based perception. SVF embeds physical information into model learning, which requires re-training. Compared to CARLO, SVF achieves better defense performance but suffers from a  slight drop in AP, indicating that it may require more training efforts.


\textbf{Limitations.} The main limitation of our mitigation strategies is the lack of guarantees. First, although both defenses can effectively defend against LiDAR spoofing attacks under the current sensor attack capability, our countermeasures may not work at some point with the increasing capability of sensor attacks. We argue that if attackers can spoof a set of points located in the distribution of physical invariants for valid vehicles (\eg injecting around 1500 points into the point cloud), there is arguably no way to distinguish them at the model level and it is safer for AVs to engage emergency brakes in that situation. \CR{Second, both defenses have a small portion of false alarms (\ie the 0.5\% false negatives in CARLO and the slight AP drop of SVF). However, we manually verify that they are not front-near vehicles; hence they would not impact the AV's driving behaviors, as mentioned before. Third, since the adaptive attacks are formulated with our efforts, future research may present more powerful attacks or advanced perturbation methods to break our defenses. In the future, we plan to improve SVF to provide guaranteed robustness by combining multiple sensors' inputs.}


\nsection{Related Work}
\label{related_work}

\textbf{Vehicular system security.} Extensive prior works explore security problems in vehicular systems and have identified vulnerabilities in in-vehicle networks of modern automobiles~\cite{vehicle-oakland10,vehicle-sec11,fault-vehicle-ccs16,247700}, in-vehicle cache side channels~\cite{247674}, and Connected Vehicle (CV)-based systems~\cite{cvattack-ndss18, trb:2018:yiheng:signalsecurity, wong2019trajectory}. Other studies try to provide robustness vehicular systems, such as secured in-vehicle communications~\cite{avatefipour2019intelligent,patsakis2014towards,woo2016practical}, secured in-vehicle payment transactions~\cite{gaddam2015mechanism}, and secured CV communications~\cite{rao2007secure}. In comparison, our work focuses on the emerging autonomous vehicle systems and specifically targets the robustness of LiDAR-based perception in AVs, which are under-explored in previous studies.  

\textbf{3D adversarial machine learning.} Adversarial attacks and defenses towards 3D deep learning have been increasingly explored recently. Point cloud classification models have been demonstrated vulnerable to adversarial perturbations~\cite{Xiang_2019_CVPR,tsairobust,wen2019geometry}. Xiao \etal generate adversarial examples for 3D mesh classification~\cite{Xiao_2019_CVPR}. Liu \etal and Yang \etal, on the other hand, leverage heuristics to detect the adversarial examples for point cloud classification~\cite{liu2019extending,yang2019adversarial}. In comparison, our work targets LiDAR-based 3D object detection in AVs. As introduced in \S\ref{background_spoof}, LiDAR point clouds only have measurements of the object's facing surface, which are different from full 3D point cloud data or meshes. Our attack method is motivated to generate adversarial examples in a black-box manner based on the discovered vulnerability. The mitigation strategies are designed to defend against current sensor attack capability, thus provide better robustness against both white- and black-box LiDAR spoofing attacks.

\nsection{Conclusion}
In this paper, we perform the first study to explore the general vulnerability of LiDAR-based perception architectures. We construct the first black-box spoofing attack based on the identified vulnerability, which universally achieves an 80\% mean success rate on target models. We further perform the first defense study, proposing CARLO to accurately detect spoofing attacks which reduce their success rate to 5.5\%. Lastly, we present SVF, the first general architecture for robust LiDAR-based perception that reduces the mean spoofing attack success rate to 2.3\%. 

\nsection{Acknowledgements}
We appreciate our shepherds, Xiaoyu Ji and Wenyuan Xu, and the anonymous reviewers for their insightful comments. We thank Xiao Zhu, Shengtuo Hu, Jiwon Joung, and Xumiao Zhang for proofreading our paper. This project is partially supported by Mcity and NSF under the grants CNS-1930041, CNS-1850533, CNS-1929771, and CNS-1932464.

\vspace{-0.2cm}

\renewcommand{\bibfont}{\footnotesize}
\bibliographystyle{abbrv}
\bibliography{refs}
\appendix

\vspace{0.3cm}
{\noindent\large\bf Appendices}

\nsection{Spoofing Attack Details}
\label{ap:Spoof}
\vspace{0.1cm}

\begin{figure}[!t]
  \begin{minipage}[t]{0.48\linewidth}
  \vspace{0pt}
    \centering 
    \includegraphics[width=\linewidth,height=\linewidth]{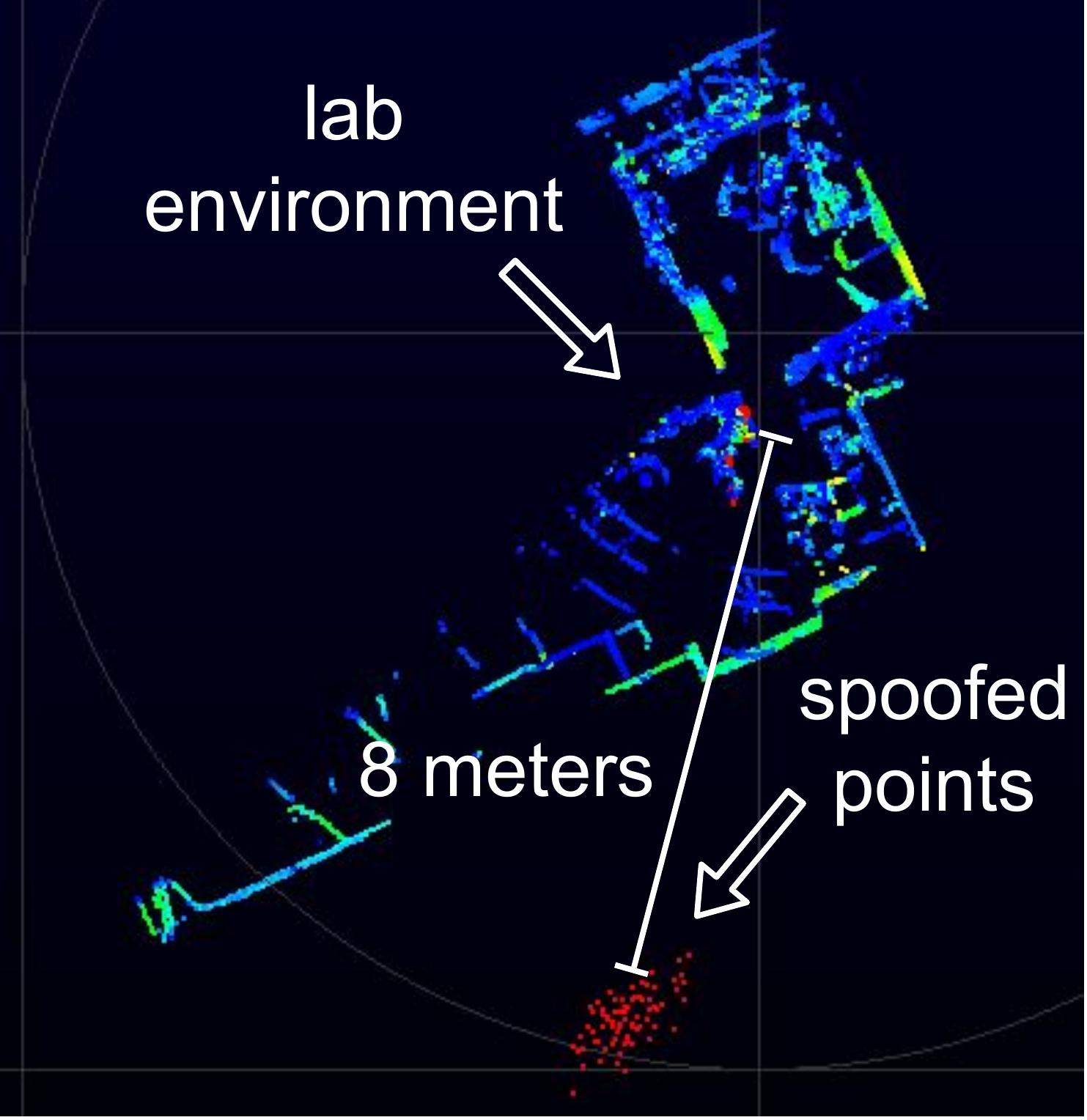}
  \caption{Physical spoofing in in-lab environments.}
  \label{fig:lab}
  \end{minipage}%
  \hfill
  \begin{minipage}[t]{0.48\linewidth} 
  \vspace{0pt}
    \centering 
    \includegraphics[width=\linewidth]{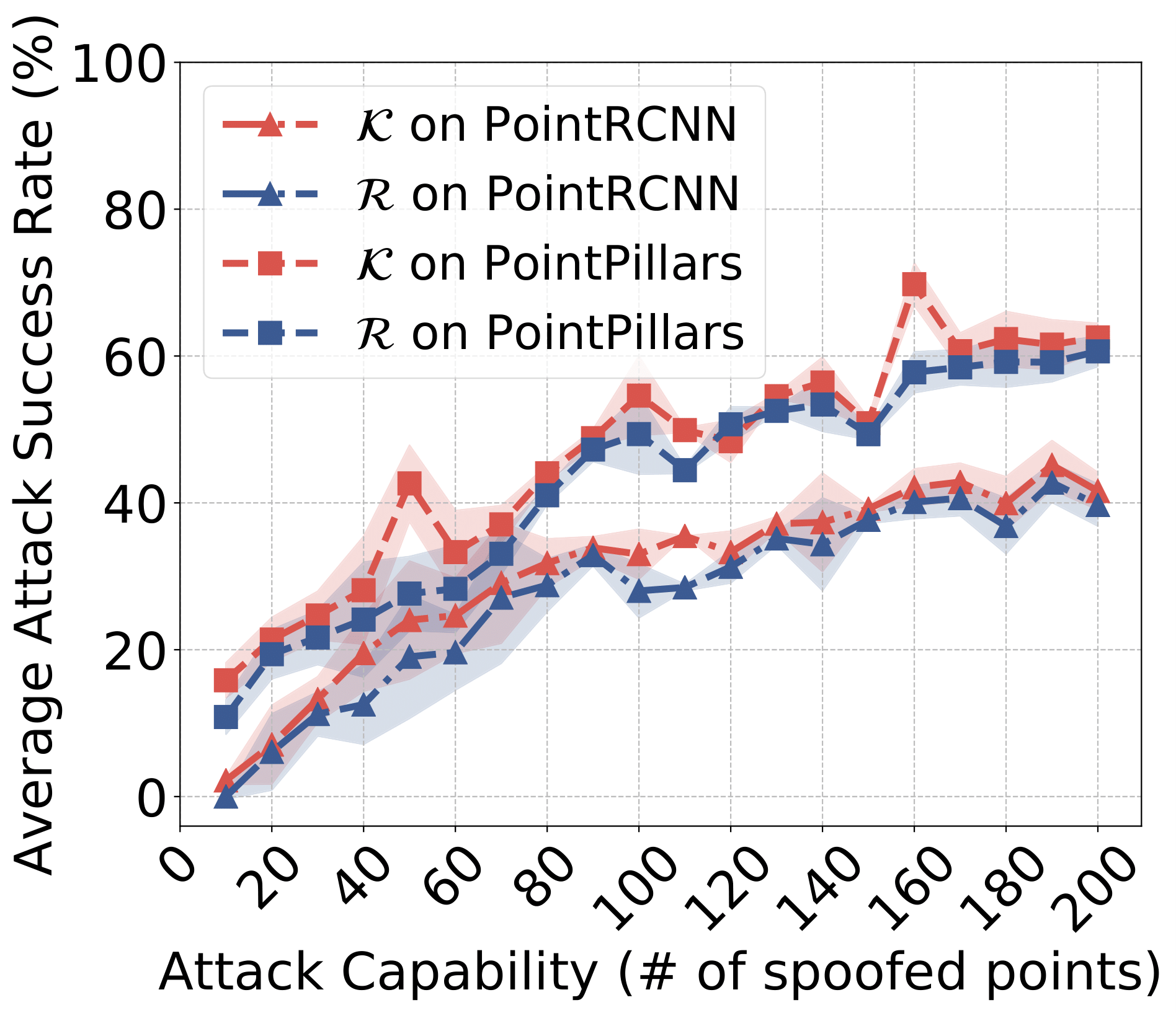}
    \vspace{-0.6cm}
      \caption{Average attack success rates ($A^2SR$s) of proposed black-box attack on PointPillars and PointRCNN.}
      \label{fig:aasr}
  \end{minipage} 
  \vspace{-0.5cm}
\end{figure}

The attack consists of three modules: a photodiode, a delay component, and an infrared laser~\cite{shin2017illusion}. The photodiode functions as a synchronizer that triggers the delay component whenever it captures laser signals from the victim LiDAR sensor. The delay component triggers the laser module after a configurable time delay to attack the following firing cycles of the victim LiDAR sensor. The attack can be programmatically controlled so that an adversary can target different locations and angles in the point cloud. \CR{Specifically, we use an OSRAM SFH 213 FA as the photodiode, a Tektronix AFG3251 function generator as the delay component, and a PCO-7114 laser driver that drives the attack laser OSRAM SPL PL90 in our setups. Figure \ref{fig:lab} shows the physical spoofing attack conducted in a controlled environment.}



\nsection{Supplementary Attack Evaluation}
\label{ap:attack_eval}

We define a new metric for general evaluations on the object detection-based attacks called average attack success rate ($A^2SR$). As mentioned before, the default thresholds are empirically set. Thus, evaluations of $ASR$ provide limited insights. Similar to AP defined in PASCAL \cite{Everingham10} criteria, we average the $ASR$ in 11 recall intervals to better understand the impact of the proposed attacks and the characteristics of different architectures:
\begin{equation}
    A^2SR = \frac{1}{11} \sum_{r \in \{0.0,0.1,\ldots,1.0\}} ASR_{t_{r}}
    \label{eq:a2sr}
\end{equation}
where $t_{r}$ represents the threshold that makes the recall of the target model at $r$. The evaluation of recall follows the description of the Moderate category in \S\ref{background_kitti}. In this paper, we test $A^2SR$ on PointPillars and PointRCNN since Apollo model is not designed to be evaluated on KITTI (\S\ref{background_kitti}).


Figure \ref{fig:aasr} shows that the $A^2SR$ of PointPillars is generally higher than PointRCNN, which means the spoofed points can achieve higher confidences in PointPillars. Such results are expected since point-wise features contain more detailed information than voxel-based features; hence point-wise features could be more resilient to spoofing attacks.

\begin{figure}[h!]
\centering
  \vspace{-0.25cm}
  \includegraphics[width=\linewidth]{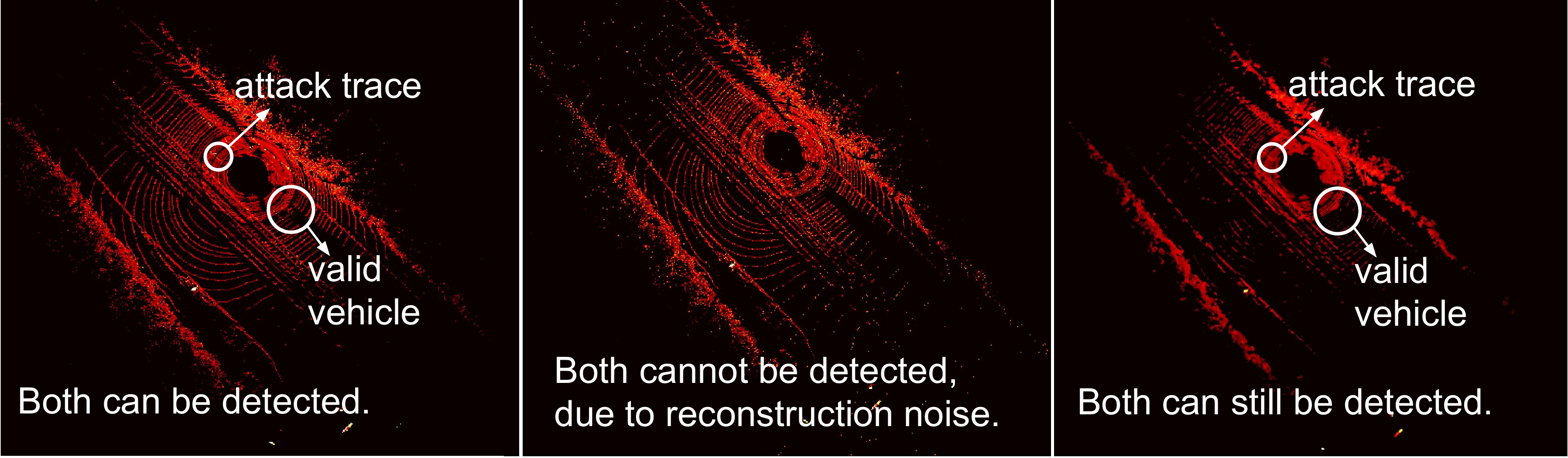}
  \caption{Illustrative example: the left figure is the original feature map of a point cloud sample from Apollo 5.0; the middle one is the feature map after ME reconstruction; and the right one is the feature map after squeezing.}
  \vspace{-0.3cm}
  \label{fig:fs}
\end{figure}

We also leverage feature squeezing~\cite{xu2017feature} and ME-Net~\cite{yang2019menet} to evaluate our proposed attack on Apollo 5.0. We utilize median smoothing as the method for feature squeezing, and follow general settings in~\cite{yang2019menet} for matrix estimation. We perform evaluations on 100 samples from the KITTI validation set. Results show that ME-Net can eliminate the fake vehicle but introduce new false negatives, which will lead to more severe safety issues. In contrast, feature squeezing cannot effectively eliminate fake vehicles, as shown in Figure~\ref{fig:fs}.


\nsection{CARLO Algorithm Details}
\label{ap:algorithm}

Algorithm \ref{alg1} shows the detailed CARLO algorithm combined with its two building blocks: FSD and LPD. Especially, to estimate the free space inside a bounding box $\mathsf{B}$, we first extract all the laser fires that have chances to hit $\mathsf{B}$, which form a frustum in the 3D space. We then grid the 3D Euclidean space of the frustum into small 3D cells and initialize all the cells as occluded cells in the beginning. For each laser, we use 3D Bresenham's line algorithm~\cite{bresenham1977linear} to compute the cells it traversed from the origin of the laser beam (\ie the LiDAR sensor) to the end (\ie the hit point). If a cell is traversed by a laser beam, we label it as a free cell because it does not belong to a solid object. Finally, we union the free cells for all the possible laser rays to get the total free cells in the frustum.

\nsection{Ablation Study of View Fusion Models}
\label{ap:ablation}

\begin{figure}[!t]
  \begin{minipage}[t]{0.45\linewidth}
    \centering 
    \includegraphics[width=4cm,height=2.1cm]{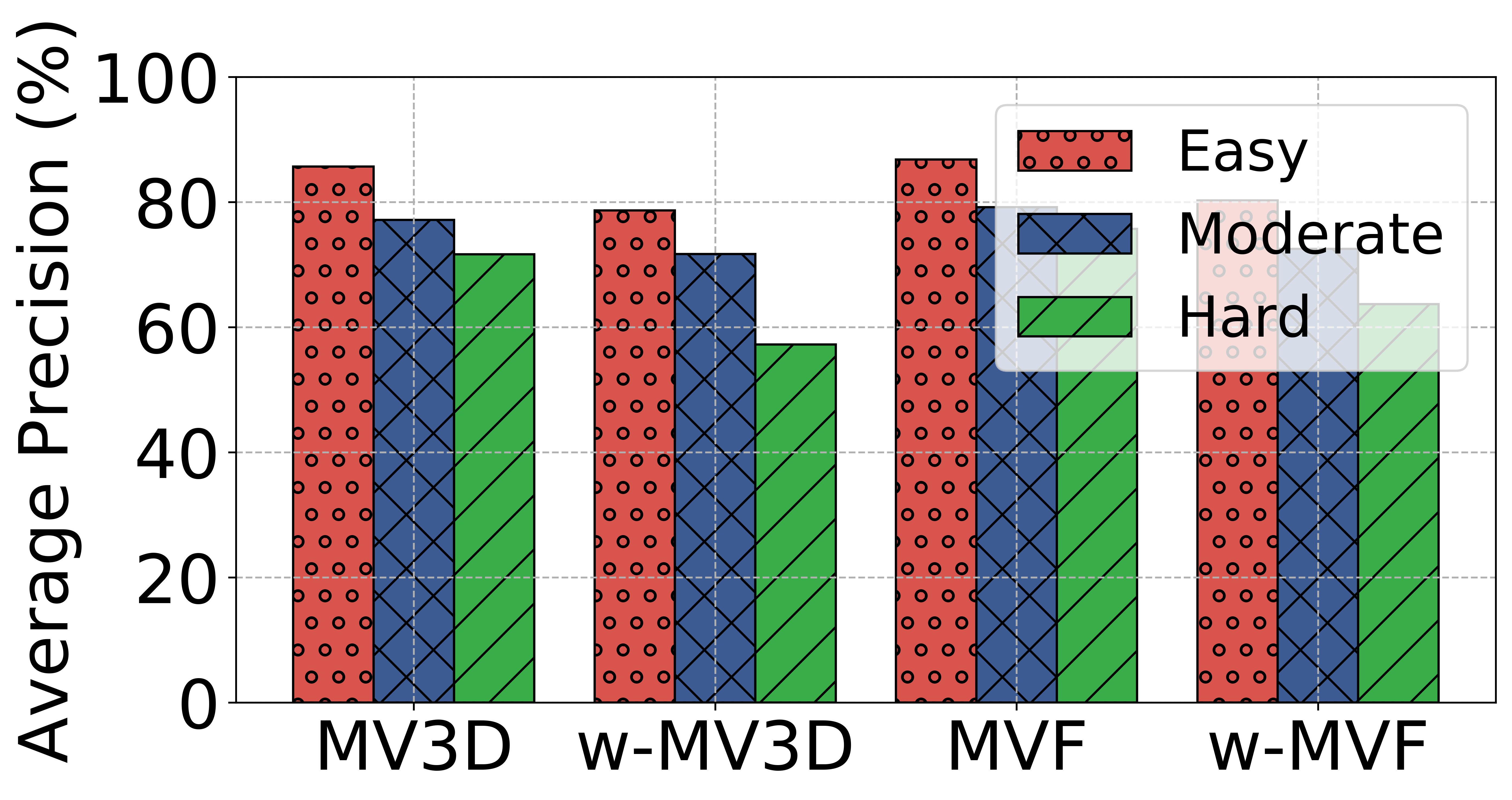} 
    \caption{APs of weakened view fusion-based models (w-: weakened models).} 
    \label{fig:mv3d_weak}
  \end{minipage}%
  \hspace{0.05\linewidth}
  \begin{minipage}[t]{0.45\linewidth} 
    \centering 
    \includegraphics[width=4cm,height=2.1cm]{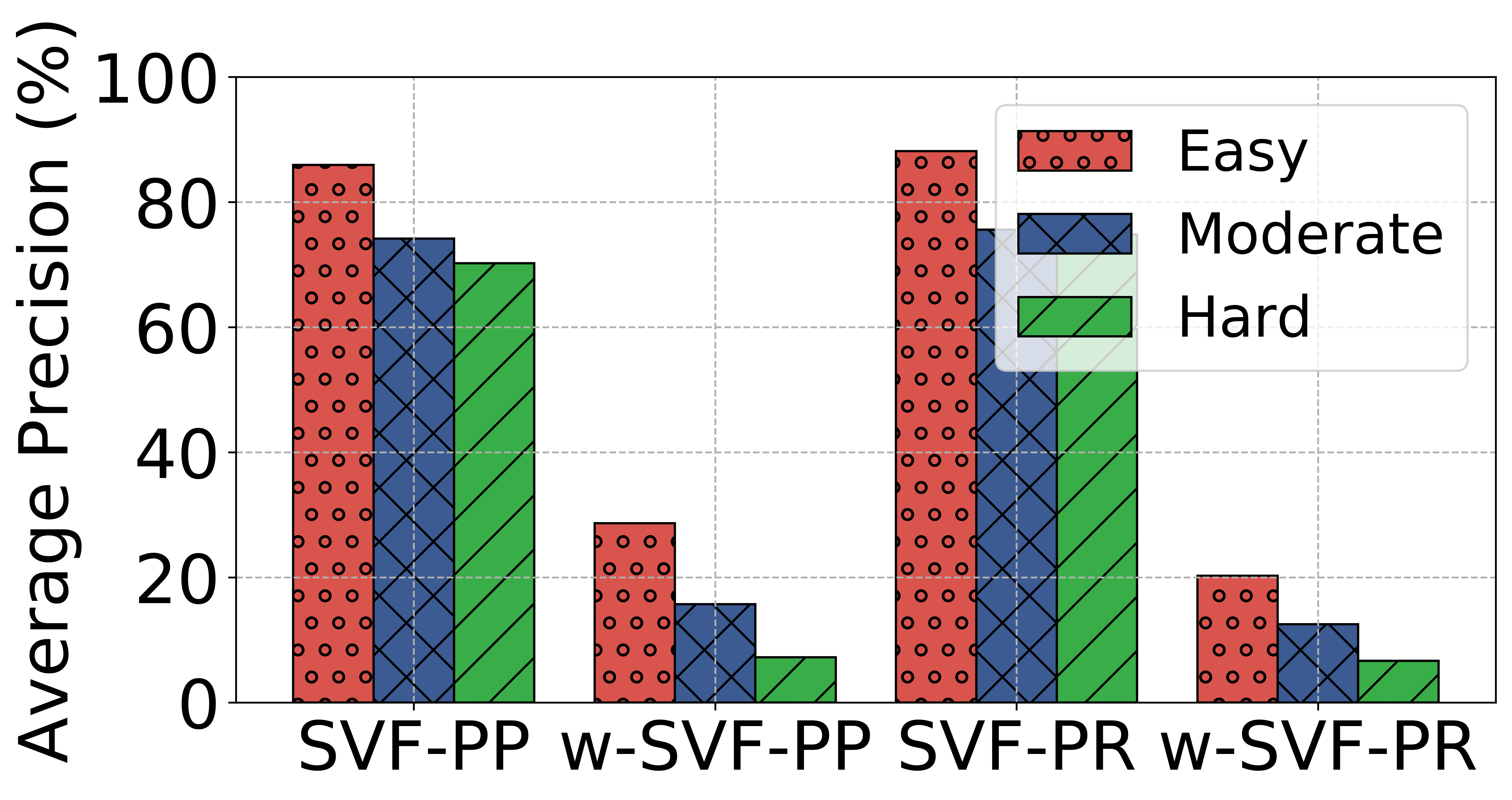} 
    \caption{APs of weakened SVF models (PP: PointPillars; PR: PointRCNN).} 
    \label{fig:svf_weak}
  \end{minipage} 
  \vspace{-0.5cm}
\end{figure}

We perform ablation studies to explore the reasons behind the results shown in Figure \ref{fig:asr_fusion}. In particular, we find most of the existing fusion-based models utilize a symmetric design where the FV and 3D (or BEV) features are fed into similar modules for learning, and the learned features are simply stacked or averaged for later stages (Figure \ref{fig:existing_fusion}). We design experiments to study the effectiveness of such a design empirically. Explicitly, we zero out the features from FV to measure how much the FV representation contributes to the final detection. As shown in Figure \ref{fig:mv3d_weak}, the APs only have relatively small degradation, which implies that BEV (or 3D) features dominate the model decisions. Kim \etal also empirically demonstrates that current sensor fusion-based models are vulnerable to single-source perturbations~\cite{kim2019single}. Similarly, we showcase that current view fusion-based models are vulnerable to the perturbation represented in the dominated view.

To better understand why SVF models can provide better robustness, we analyze how the FV representation helps in SVF models. Similarly, we zero out the augmented confidence score features and evaluate the AP. Figure \ref{fig:svf_weak} shows the weakened models' performance, which empirically demonstrates that the features from FV account more in SVF models.

\nsection{Supplementary Figures}
\label{ap:figure}

\begin{figure}[!t]
  \begin{minipage}[t]{0.48\linewidth}
  \vspace{0pt}
    \centering 
    \includegraphics[width=\linewidth]{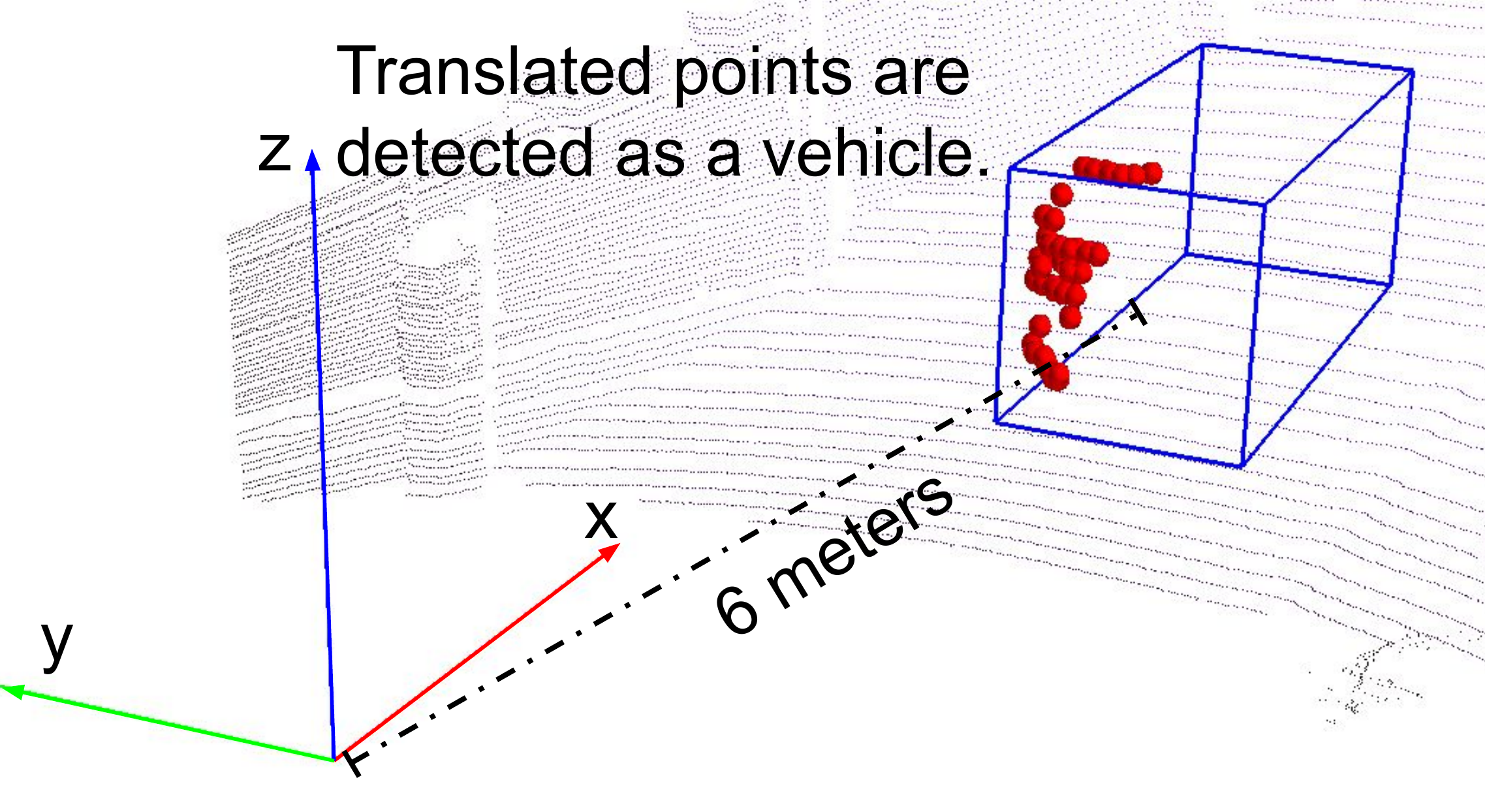}
  \caption{Translated points from Figure \ref{fig:c1-3} are detected as a valid vehicle by PointRCNN.}
  \label{fig:sup_1}
  \end{minipage}%
  \hfill
  \begin{minipage}[t]{0.48\linewidth} 
  \vspace{0pt}
    \centering 
    \includegraphics[width=\linewidth]{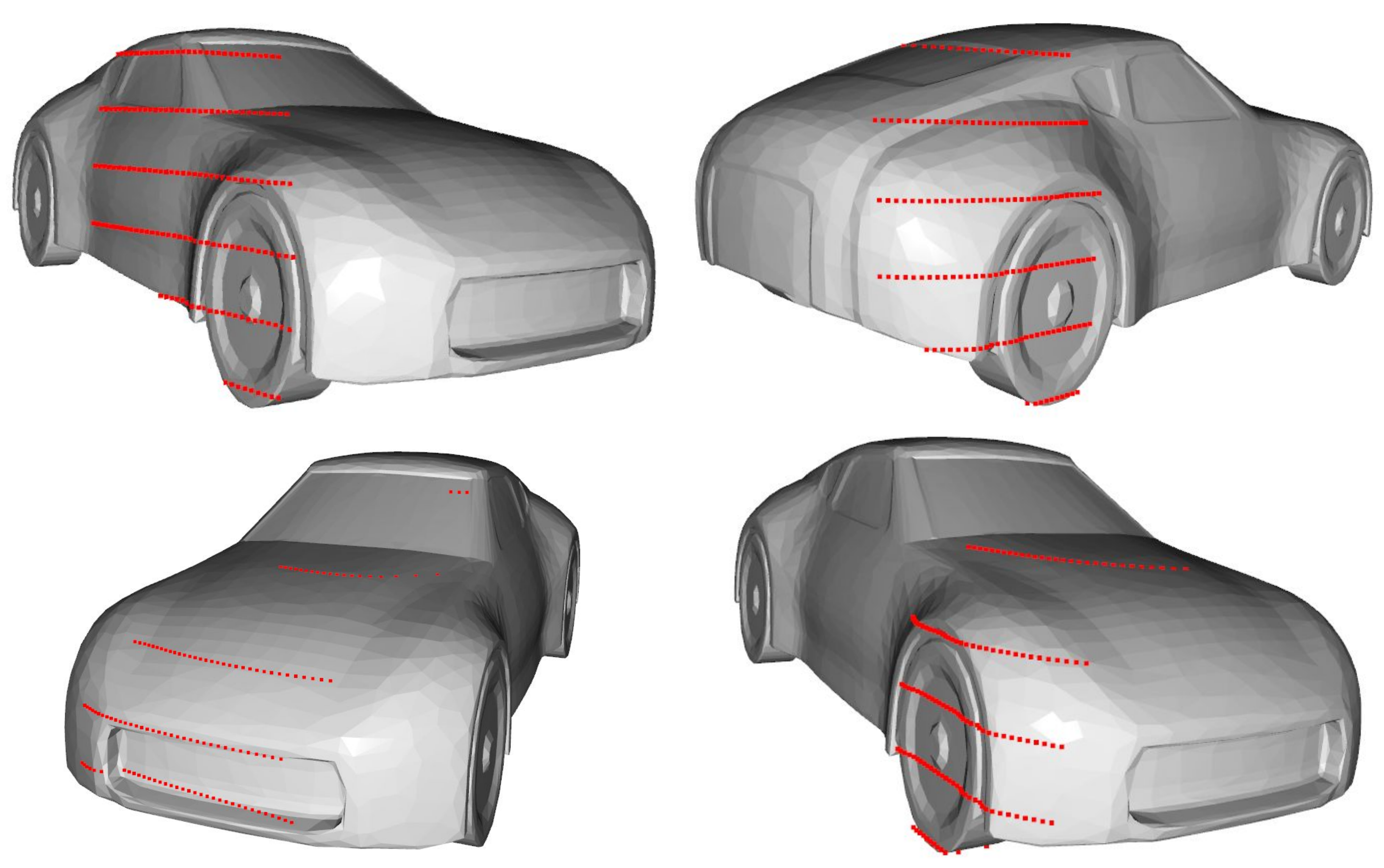}
  \caption{More examples of our rendered traces with occlusions.}
  \label{fig:sup_2}
  \end{minipage} 
  \vspace{-0.5cm}
\end{figure}

Figure \ref{fig:sup_1} shows an illustrative example that translated points from Figure \ref{fig:c1-3} can be detected as a valid vehicle in PointRCNN. Figure \ref{fig:sup_2} shows more rendered original attack traces.



\begin{algorithm}[h!]
\small
\SetAlgoLined
\SetKwInput{Input}{input}
\SetKwInput{Output}{output}
\SetKwInput{Init}{Initialization}
\SetKwInput{Blank}{}
\SetKwInput{Ret}{Return}
\SetKwComment{Comment}{ $triangleright$ }{} 

\tabcolsep=0pt
\begin{tabular}{@{}ll}
    \Input{}&Detected bounding boxes $\boldsymbol{B}=\{\mathsf{B}\}$\;\\
&  LiDAR laser ray directions $\boldsymbol{L}=\{\mathsf{L}\}$\;\\
&  3D point cloud $\boldsymbol{X}=\{\vec{p}\}$  \;\\
&  Threshold of FSD $\frac{a+b}{2}$  \;\\
&  Thresholds of LPD $b'+\epsilon$, $a'-\epsilon$ \;\\
    \Output{}&Valid bounding boxes $\boldsymbol{B}_{\mathrm{valid}}=\{\mathsf{B}\}$\;\\
    &Adversarial bounding boxes $\boldsymbol{B}_{\mathrm{adv}} =\{\mathsf{B}\}$\;\\
\end{tabular}
\BlankLine
$\V{Initialization:}$ $\boldsymbol{B}_{\mathrm{valid}} \gets \emptyset$, $\boldsymbol{B}_{\mathrm{adv}} \gets \emptyset$, $g \gets 0$, $f \gets 0$\;
 \tcc{Initiate parameters.}
 \For{$\mathsf{B} \in \boldsymbol{B}$}{
 \tcc{Initiate parameters, where $FS(\cdot)$ is the free space and $\mathsf{F}_\mathsf{B}$ is the frstum of $\mathsf{B}$.}
 $\mathsf{F}_\mathsf{B} \gets \emptyset$, $FS(\cdot) \gets \emptyset$;\\
 \For{$\mathsf{L} \in \boldsymbol{L}$}{
   \tcc*[h]{Predict whether $\mathsf{L}$ will intersect with $\mathsf{B}$.}\\
   \uIf{$\mathsf{L} \cap \mathsf{B}$}{
    $\vec{p}_{\mathsf{L}} \gets \mathsf{L}$\;
    $\mathsf{F}_\mathsf{B}$.append($[\,\mathsf{L},\vec{p}_{\mathsf{L}}\,]$)\;
  }
  \tcc*[h]{Extract the frustum $\mathsf{F}_\mathsf{B}$ of  $\mathsf{B}$.}
  }
  $g \gets$ Equation \ref{eq:lpd}\;
  \tcc*[h]{Calculate $g$ by $\mathsf{F}_\mathsf{B}$ for $\mathsf{B}$ (LPD).}\\
  \uIf{$g < a'-\epsilon$}{
    $\boldsymbol{B}_{\mathrm{valid}}$.append($\mathsf{B}$)\;
    \tcc*[h]{Certainly valid vehicles.}
  }
  \uElseIf{$g > b'+ \epsilon$}{
    $\boldsymbol{B}_{\mathrm{adv}}$.append($\mathsf{B}$)\;
    \tcc*[h]{Certainly spoofed vehicles.}
  }
  \uElse{
    \tcc*[h]{Calculate$f$ by $\mathsf{F}_\mathsf{B}$ for $\mathsf{B}$ (FSD).}\\
    \For{$[\mathsf{L},\vec{p}_{\mathsf{L}}] \in \mathsf{F}_{\mathsf{B}}$}{
        $FS(\mathsf{L}) \gets$ Bresenham$([\mathsf{L},\vec{p}_{\mathsf{L}}])$\cite{bresenham1977linear}\;
        $FS(\mathsf{B}) \gets FS(\mathsf{B}) \cup FS(\mathsf{L})$ \;
    }
    $f \gets$ Equation \ref{eq:fsd}\; 
    \uIf{$f<\frac{a+b}{2}$}{
        $\boldsymbol{B}_{\mathrm{valid}}$.append($\mathsf{B}$)\;
    }
    \uElse{
        $\boldsymbol{B}_{\mathrm{adv}}$.append($\mathsf{B}$)\;
    }
  }
}

$\V{Return:}$ $\boldsymbol{B}_{\mathrm{valid}}$, $\boldsymbol{B}_{\mathrm{adv}}$\;
 \caption{CARLO}
 \label{alg1}
\end{algorithm}

\end{document}